\documentclass[12pt, a4paper]{article}
\usepackage[utf8]{inputenc} 
\usepackage{chet}
\usepackage{authblk}
\usepackage{slashed}
\usepackage{indentfirst}
\usepackage{amsmath}
\usepackage{amsfonts}
\usepackage{amssymb}
\usepackage{courier}
\usepackage{hyperref}

\newcommand{\B}[1]{\overline{#1}}
\newcommand{\HH}[1]{\widehat{#1}}
\newcommand{\WT}[1]{\widetilde{#1}}
\let\ampersand\&
\renewcommand*\&{and}

\usepackage[T1]{fontenc} 
\usepackage{xcolor}

\def \p{\partial}
\def \pb{\overline\partial}

\def \Tb{\overline T}
\def \a{\alpha}
\def\ab{\overline\alpha}
\def \b{\beta}
\def\bb{\overline\beta}

\def \d{\delta}

\def \g{\gamma}
\def\gb{\overline\gamma}

\def \L{\Lambda}
\def \Lb{\overline\Lambda}

\def \ob{\bar\omega}
\def \O{\Omega}
\def\Oh{\widehat\Omega}

\def \N{\nabla}
\def\Nh{\widehat\nabla}
\def \Nb{\overline\nabla}
\def\ua{\underline{a}}
\def \r{\rho}
\def\rb{\overline\rho}

\def \s{\sigma}

\def\Pib{\overline\Pi}

\def\Th{\widehat T}
\def\Rh{\widehat R}
\def\psib{\overline\psi}
\def\dd{\overline d}
\def\wb{\overline w}
\def\Ch{\widehat C}
\def\Yh{\widehat Y}
\def\Dh{\widehat D}
\def\Uh{\widehat U}
\def\Vh{\widehat V}
\def\Zh{\widehat Z}
\def\Eh{\widehat E}
\def\Fh{\widehat F}
\def\Gh{\widehat G}
\def\Ih{\widehat I}
\def\Jh{\widehat J}
\def\Lh{\widehat L}

\def\taub{\overline\tau}
\def\Qh{\widehat Q}

\newcommand{\halpha}{\overline{\alpha}}
\newcommand{\hbeta}{\overline{\beta}}
\newcommand{\hgamma}{\overline{\gamma}}
\newcommand{\hnabla}{\overline{\nabla}}
\newcommand{\ub}{{\underline{b}}}
\newcommand{\uc}{{\underline{c}}}

\newcommand{\ue}{{\underline{e}}}
\newcommand{\hC}{\widehat{C}}
\newcommand{\hD}{\widehat{D}}
\newcommand{\hR}{\widehat{R}}
\newcommand{\hT}{\widehat{T}}
\newcommand{\hY}{\widehat{Y}}

\newcommand{\hV}{\widehat{V}}
\newcommand{\hN}{\widehat{N}}
\newcommand{\hG}{\widehat{G}}

\newcommand{\hlambda}{\overline{\lambda}}
\newcommand{\hdelta}{\overline{\delta}}
\newcommand{\hOmega}{\widehat{\Omega}}
\newcommand{\homega}{\widehat{\omega}}

\newcommand{\hp}{\widehat{p}}

\newcommand{\hZ}{\overline{Z}}
\newcommand{\hd}{\overline{d}}
\newcommand{\hpsi}{\overline{\psi}}

\newcommand{\hrho}{\overline{\rho}}

\newcommand{\hsigma}{\overline{\sigma}}

\newcommand{\htheta}{\overline{\theta}}

\newcommand{\hLambda}{\overline{\Lambda}}
\newcommand{\hw}{\overline{w}}

\title{B-RNS-GSS Type II superstring in Ramond-Ramond backgrounds}
\author{Osvaldo Chandia$^a$\footnote[1]{\texttt{ochandiaq@gmail.com}} and João Gomide$^b$\footnote[2]{\texttt{jpr.gomide@unesp.br}}}
\affil{$^a$\textit{Departamento de Ciencias, Facultad de Artes Liberales}\\
\textit{Universidad Adolfo Ib\'a\~nez, Santiago, Chile.\\ $^b$\textit{ICTP South American Institute for
Fundamental Research}\\
\textit{Instituto de Física Teórica, UNESP - Univ. Estadual Paulista}\\
\textit{Rua Dr. Bento Teobaldo Ferraz 271, 01140-070, São Paulo, SP, Brazil}}}

\abstract{The B-RNS-GSS formalism containing both RNS and GS worldsheet variables is extended for the description of the Type II superstring in curved backgrounds. The BRST charge of the theory is expressed in terms of $\mathcal{N}=(1,1)$ superconformal generators $(T,G; \Th, \Gh)$, and the classical anticommutation relations $\{G,G\}=-2T$ and $\{\Gh,\Gh\}=-2\Th$ are shown to imply the Type II supergravity equations of motion. The particular case of the Type IIB superstring in an $AdS_5\times S^5$ background is discussed and shown to be classically integrable.}

\begin{document} 
\maketitle
\flushbottom
\tableofcontents

\section{Introduction} \label{intro}
The Ramond-Neveu-Schwarz (RNS) formalism of the superstring has an elegant structure based on $\mathcal{N}=(1,1)$ worldsheet supersymmetry. Despite its elegance, vertex operators for spacetime fermions and the description of Ramond-Ramond backgrounds are complicated in this formalism \cite{Friedan:1985ge} because of the lack of manifest spacetime supersymmetry. A new formalism for the superstring with both worldsheet and spacetime supersymmetry was recently proposed in \cite{Berkovits:2021xwh}, where a set of extra worldsheet variables was added to the worldsheet action in such a way that simple vertex operators for spacetime fermions could be constructed. It was named the B-RNS-GSS formalism since it involves features of the pure spinor \cite{Berkovits:2000fe}, RNS and Green-Schwarz-Siegel \cite{Siegel:1985xj} formalisms of the superstring. Consequently, it is expected to play a useful role in resolving the open problem of proving the equivalence of the pure spinor and RNS formalisms \cite{Berkovits:2022qhc}. 

The quantization of the B-RNS-GSS model is performed with a BRST charge that resembles the BRST charge of the RNS formalism. It has the standard form for an $\mathcal{N}=1$ worldsheet superconformal field theory $Q=\oint\left(cT+\g G+\g^2b-bc\p c\right)$, where $T$ is the stress-energy tensor, $G$ is the generator of worldsheet supersymmetry and $(b,c,\beta,\g)$ are the usual superconformal ghosts. The fact that $G$ and $T$ generate an $\mathcal{N}=1$ superconformal algebra implies nilpotence and holomorphicity of $Q$. 

This new formalism is also suitable to describe curved backgrounds. Indeed, it was recently shown \cite{Berkovits:2022dbm} that, at the classical level, nilpotence and holomorphicity of $Q$ in the context of the B-RNS-GSS string in a curved heterotic background imply the $N=1$ $D=10$ supergravity and super-Yang-Mills constraints for the background superfields as formulated in \cite{Berkovits:2001ue}, using an analysis similar to the case of the pure spinor formalism. In this paper, the B-RNS-GSS formalism will be defined in curved Type II supergravity backgrounds, and nilpotence and holomorphicity of the BRST charge at the classical level will imply the $N=2$ $D=10$ Type II supergravity constraints of \cite{Berkovits:2001ue}. It should be stressed that this includes backgrounds with finite Ramond-Ramond flux, and the sigma-model for the B-RNS-GSS string in an $AdS_5\times S^5$ background will be constructed at the end of the paper to illustrate this fact.

The paper is organized as follows: in Section \ref{flat} we review the construction of the Type II version of \cite{Berkovits:2021xwh} in a flat background. In Section \ref{CG} we construct the Type II superstring in a curved background. In Section \ref{GGisT} we show that the equations $\{G,G\}=-2T$ and $\{\Gh,\Gh\}=-2\Th$ at the classical level impose the constraints of Type II supergravity. In Section \ref{GGhis0} we prove that the constraints from Section \ref{GGisT} imply $\{G,\Gh\}=0$. In Section \ref{HOL} we prove the holomorphicity of $G$ and the antiholomorphicity of $\Gh$, consequently showing the BRST invariance of the action for Type II supergravity. In Section \ref{AdS} we specialize our results for the $AdS_5\times S^5$ background of the Type IIB superstring and establish the classical integrability of the corresponding sigma-model. We conclude in Section \ref{conclusion} by commenting on possible applications of these results. The appendices are organized as follows: Appendix \ref{app0} gathers some gauge transformations of the action and BRST charge, Appendix \ref{app1} contains the results for the necessary Poisson brackets between different worldsheet fields, Appendix \ref{app2} gathers the equations of motion derived from the worldsheet action, and Appendix \ref{app3} gives the conventions for the $AdS_5\times S^5$ section.

\section{Review of the B-RNS-GSS formalism}\label{flat}
The B-RNS-GSS formalism can be constructed by adding the spacetime spinor Green-Schwarz-Siegel variables $(\theta^{\alpha},p_{\alpha})$ and $(\htheta^{\halpha},\hp_{\halpha})$, as well as the unconstrained spacetime spinor variables $(\Lambda^{\alpha},w_{\alpha})$ and $(\hLambda^{\halpha},\hw_{\halpha})$ for $\alpha,\halpha=1$ to $16$, to the RNS formalism worldsheet variables $(x^m,\psi^m,\hpsi^m,b,c,\beta,\gamma,\B{b},\B{c},\B{\beta},\B{\gamma})$ for $m=0$ to $9$. These variables can be organized in terms of worldsheet superfields such that the $\mathcal{N}=(1,1)$ superconformal symmetry is manifest
\begin{equation}
    X^m=x^m+\kappa\psi^m+\B{\kappa}\hpsi^m+\kappa\B{\kappa}f^m,
\end{equation}
\begin{equation}
    \Theta^{\alpha}=\theta^{\alpha}-\kappa\Lambda^{\alpha}-\B{\kappa}h^{\alpha}+\kappa\B{\kappa}f^{\alpha},\quad \HH{\Theta}^{\halpha}=\htheta^{\halpha}-\kappa\HH{h}^{\halpha}-\B{\kappa}\hLambda^{\halpha}+\kappa\B{\kappa}\HH{f}^{\halpha},
\end{equation}
\begin{equation}
    \Phi_{\alpha}=w_{\alpha}-\kappa p_{\alpha}-\B{\kappa}h_{\alpha}+\kappa\B{\kappa}f_{\alpha},\quad \HH{\Phi}_{\halpha}=\hw_{\halpha}-\kappa\HH{h}_{\halpha}-\B{\kappa}\HH{p}_{\halpha}+\kappa\B{\kappa}\HH{f}_{\halpha},
\end{equation}
where $f$'s and $h$'s are auxiliary variables. The worldsheet action can be written in the manifestly $\mathcal{N}=(1,1)$ worldsheet superconformal form
\begin{equation}
    S=-\int d^2z\int d^2\kappa\Big[\frac{1}{2}DX^m\B{D}X_m-\Phi_{\alpha}\B{D}\Theta^{\alpha}+\HH{\Phi}_{\halpha}D\HH{\Theta}^{\halpha}\Big],
\end{equation}
where $D=\partial_{\kappa}-\kappa\partial_z$ and $\B{D}=\partial_{\B{\kappa}}-\B{\kappa}\B{\partial}_{\B{z}}$ denote the fermionic $\mathcal{N}=(1,1)$ derivatives. After integrating out the auxiliary variables, the component form of the action is simply
\begin{equation} \label{flataction}
    S=S_{RNS}+\int d^2z\Big(p_{\alpha}\B{\partial}\theta^{\alpha}+w_{\alpha}\B{\partial}\Lambda^{\alpha}+\hp_{\halpha}\partial\htheta^{\halpha}+\hw_{\halpha}\partial\hLambda^{\halpha}\Big).
\end{equation}
This formalism can also be made manifestly spacetime supersymmetric at the expense of breaking this manifest $\mathcal{N}=(1,1)$ worldsheet supersymmetry generated by the superconformal currents
\begin{equation}
    G=\psi^m\partial x_m+\Lambda^{\alpha}p_{\alpha}+w_{\alpha}\partial\theta^{\alpha},\quad \hG=\hpsi^m\B{\partial}x_m+\hLambda^{\halpha}\HH{p}_{\halpha}+\hw_{\halpha}\B{\partial}\htheta^{\halpha}.
\end{equation}
In order to do so, one must perform the similarity transformation $\mathcal{O}\to e^{R}\mathcal{O}e^{-R}$ with
\begin{equation}
    R=\frac{1}{2}\oint \psi^m\big(\Lambda\gamma^m\theta\big)+\frac{1}{2}\oint\hpsi^m\big(\hLambda\gamma^m\htheta\big)
\end{equation}
on all operators $\mathcal{O}$. Then, the theory is manifestly invariant under the transformations generated by the usual Green-Schwarz-Siegel supersymmetry generator
\begin{equation}
    q_{\alpha}=\oint\Big[p_{\alpha}+\frac{1}{2}\partial x^m(\gamma_m\theta)_{\alpha}+\frac{1}{24}(\theta\gamma^m\partial\theta)(\gamma_m\theta)_\alpha\Big]
\end{equation}
and similarly for $\HH{q}_{\halpha}$. Note that the similarity transformation takes the superconformal current $G$ to
\begin{equation}
    G=\frac{1}{2}\big(\Lambda\gamma^m\Lambda\big)\psi_m+\Lambda^{\alpha}d_{\alpha}+\psi^m\Pi_m+w_{\alpha}\partial\theta^{\alpha}
\end{equation}
and similarly for $\hG$. The variables
\begin{equation}
    d_{\alpha}\equiv p_{\alpha}-\frac{1}{2}\Big(\partial x^m-\frac{1}{4}\partial\theta\gamma^m\theta\Big)\big(\gamma_m\theta\big)_{\alpha},\quad \Pi^m\equiv\partial x^m-\frac{1}{2}\big(\partial\theta\gamma^m\theta\big)
\end{equation}
are defined as in the Green-Schwarz-Siegel formalism.

The BRST operator of the B-RNS-GSS formalism has the standard $\mathcal{N}=(1,1)$ worldsheet superconformal form, which for the left-moving charge means
\begin{equation}
    Q_{B-RNS-GSS}=\oint\Big[cT+\gamma G-bc\partial c+\gamma^2 b\Big],
\end{equation}
where $T$ is the holomorphic stress-energy tensor obtained from the action (\ref{flataction}) and the spinors $(\Lambda^{\alpha},w_{\alpha})$ carry holomorphic conformal weight $(\frac{1}{2},\frac{1}{2})$. Analogous statements follow for the right-moving BRST charge.

The presence of extra fields implies that the spectrum of the B-RNS-GSS string is larger than that of the RNS string. Physical vertex operators for the usual GSO(+) projected RNS open superstring spectrum have been constructed recently \cite{Berkovits:2021xwh,Berkovits:2013eqa}. These operators are in the cohomology of $Q_{B-RNS-GSS}$ and satisfy the extra condition that they carry non-positive charge with respect to the U(1) generators
\begin{equation} \label{flatU1}
    J\equiv\xi\eta-\Lambda w,\quad\B{J}\equiv\B{\xi}\B{\eta}-\B{\Lambda}\B{w}.
\end{equation}
To understand the non-positive charge requirement on the vertex operators, consider twisting\footnote{In order for the conformal anomaly to cancel such that the ``untwisted'' theory presented here is consistent at the quantum level, one must also add non-minimal bosonic $\big(\WT{\Lambda}_{\alpha},\WT{w}^{\alpha}\big)$ and fermionic $\big(R_{\alpha},S^{\alpha}\big)$ left-moving variables of conformal weight $(0,1)$ and $(\frac{1}{2},\frac{1}{2})$ respectively, as well as their right-moving counterparts. These variables are not needed for the construction of massless vertex operators, and will therefore be ignored in the remainder of this paper. For more details on their role in the B-RNS-GSS formalism, see \cite{Berkovits:2022dbm,Berkovits:2022qhc}.} the $(\Lambda^{\alpha},w_{\alpha},\hLambda^{\halpha},\hw_{\halpha})$ variables by
\begin{equation}
    \Lambda^{\alpha}\to\frac{1}{\gamma}\Lambda^{\alpha},\quad w_{\alpha}\to \gamma w_{\alpha},\quad \hLambda^{\halpha}\to \frac{1}{\hgamma}\hLambda^{\halpha},\quad \hw_{\halpha}\to \hgamma\hw_{\halpha}.
\end{equation}
After performing the twist, the bosonic spinor variables carry integer worldsheet spin as in the pure spinor formalism.
But due to the inclusion of inverse powers of $\gamma$, vertex operators with positive charge with respect to (\ref{flatU1}) would no longer be in the small Hilbert space with respect to the twisted worldsheet variables. So the additional requirement of non-positive charge is needed to guarantee that the vertex operators remain in the small Hilbert space after performing the twist. 

The integrated left-moving massless open string vertex is given by
\begin{equation}
\int dz~U_{open}=\int dz\Bigg(\partial\theta^{\alpha}A_{\alpha}+\Pi^mA_m+d_{\alpha}W^{\alpha}+\frac{1}{2}N^{mn}F_{mn}-\psi^mw_{\alpha}\partial_mW^{\alpha}\Bigg),
\end{equation}
with $N_{mn}=-\psi^m\psi^n+\frac{1}{2}\Lambda\gamma^{mn}w$, 
and a similar expression holds for the right-moving vertex. The closed string B-RNS-GSS integrated massless vertex operator can then be obtained by the left-right product of the left and right-moving open vertices, which yields
\begin{equation} \label{closedvertex}
\begin{split}
   \int d^2z\Bigg(&\partial\theta^{\alpha}\B{\partial}\htheta^{\hbeta}A_{\alpha\hbeta}+\partial\theta^{\alpha}\hd_{\hbeta}E_{\alpha}\,^{\hbeta}+d_{\alpha}\B{\partial}\htheta^{\hbeta}E_{\hbeta}\,^{\alpha}+\Pi^m\B{\Pi}^nA_{mn}+N_{mn}\HH{N}_{pq}S^{mnpq}\\
   &+\partial\theta^{\alpha}\B{\Pi}^nA_{\alpha n}+\Pi^m\B{\partial}\htheta^{\hbeta}A_{m\hbeta}+\Pi^m\hd_{\hbeta}E_m\,^{\hbeta}+d_{\alpha}\B{\Pi}^nE_n\,^{\alpha}+d_{\alpha}\hd_{\hbeta}P^{\alpha\hbeta}\\
   &+\partial\theta^{\alpha}\HH{N}_{mn}\HH{\Omega}_{\alpha}\,^{mn}+N_{mn}\B{\partial}\htheta^{\hbeta}\Omega_{\hbeta}\,^{mn}+N_{mn}\hd_{\hbeta}D^{\hbeta mn}+d_{\alpha}\HH{N}_{mn}D^{\alpha mn}\\
   &+\Pi^m\HH{N}_{np}\HH{\Omega}_m\,^{np}+N_{mn}\B{\Pi}^p\Omega_p\,^{mn}+\Pi^m\B{\psi}^n\hw_{\hbeta}C_{mn}\,^{\hbeta}+\psi^mw_{\alpha}\B{\Pi}^nC_{nm}\,^{\alpha}\\
   &+\partial\theta^{\alpha}\B{\psi}^n\hw_{\hbeta}C_{\alpha n}\,^{\hbeta}+\psi^mw_{\alpha}\B{\partial}\htheta^{\hbeta}C_{\hbeta m}\,^{\alpha}+\psi^mw_{\alpha}\hd_{\hbeta}I_m\,^{\alpha\hbeta}+d_{\alpha}\B{\psi}^n\hw_{\hbeta}I_n\,^{\alpha\hbeta}\\
   &+N_{mn}\B{\psi}^p\hw_{\hbeta}J_p\,^{mn\hbeta}+\psi^mw_{\alpha}\HH{N}_{np}J_m\,^{np\alpha}+\psi^mw_{\alpha}\B{\psi}^n\hw_{\hbeta}K_{mn}\,^{\alpha\hbeta}\Bigg).
\end{split}
\end{equation}

\section{Curved background}\label{CG}
The closed vertex (\ref{closedvertex}) does not include all possible terms with non-positive U(1) charge, as terms with $-2$ U(1) charge (i.e. with factors of $w_{\alpha}w_{\beta}$ and/or $\hw_{\halpha}\hw_{\hbeta}$) can clearly be written. The lessons learned in the analysis of the heterotic string \cite{Berkovits:2022dbm} show that such terms must be considered and that they describe non-linear deformations around flat-space. After including these terms and covariantizing the flat action (\ref{flataction}), we arrive at the most general action satisfying the U(1)-charge/small Hilbert space condition
\begin{align} \label{action}
S=\int d^2\s ~\Big[&  \frac12 \Pi_a \Pib^a + \frac12 \Pi^A \Pib^B B_{BA} + \frac12 \psi_a \Nb \psi^a + \frac12 \overline\psi_a \Nh \overline\psi^a + w_\a \overline\nabla \Lambda^\alpha + \overline{w}_{\overline\alpha} \nabla \overline\Lambda^{\overline\alpha}  \cr
&+ d_\a \Pib^\a + \overline{d}_{\ab} \Pi^{\ab}+ d_\a {\overline d}_{\ab} P^{\a\ab}+b\pb c + \b \pb \g + {\overline b}\p {\overline c} + \bb \p \gb  \cr
&+  \psi^a w_\alpha \Pib^A C_{Aa}{}^\a + \overline{\psi}^a {\overline w}_{\overline\alpha} \Pi^A {\widehat C}_{Aa}{}^{\ab} + w_\a w_\b \Pib^A Y_A{}^{\a\b} + {\overline w}_{\ab} {\overline w}_{\bb} \Pi^A {\widehat Y}_A{}^{\ab\bb}  \cr
&+ d_\a \Lb^{\ab} {\overline w}_{\bb} D_{\ab}{}^{\bb\a} + {\overline d}_{\ab} \L^\a w_\b {\widehat D}_\a{}^{\b\ab} + d_\a \psib^a \psib^b \Uh_{ab}{}^\a + {\overline d}_{\ab} \psi^a \psi^b U_{ab}{}^{\ab} \cr 
&+ d_\a {\overline\psi}^a {\overline w}_{\ab} V_a{}^{\a\ab} + {\overline d}_{\ab} \psi^a w_\a {\widehat V}_a{}^{\ab\a} + d_\a {\overline w}_{\ab} {\overline w}_{\bb} Z^{\a\ab\bb} + {\overline d}_{\ab} w_\a w_\b {\widehat Z}^{\ab\a\b}\cr 
&+ \L^\a w_\b \Lb^{\ab} {\overline w}_{\bb} S_{\a\ab}{}^{\b\bb} + \L^\a w_\b \psib^a \psib^b E_{ab\a}{}^\b + \Lb^{\ab} \wb_{\bb} \psi^a \psi^b \Eh_{ab\ab}{}^{\bb}+ \psi^a \psi^b \psib^c \psib^d H_{abcd}  \cr
&+ \L^\a w_\b \psib^a \wb_{\ab} F_{a\a}{}^{\b\ab} + \Lb^{\ab} \wb_{\bb} \psi^a w_\a \Fh_{a\ab}{}^{\bb\a} + \psi^a \psi^b \psib^c \wb_{\ab}I_{abc}{}^{\ab} + \psib^a \psib^b \psi^c w_\a \Ih_{abc}{}^\a \cr 
& + \L^\a w_\b \wb_{\ab}\wb_{\bb}G_\a{}^{\b\ab\bb} + \Lb^{\ab}\wb_{\bb} w_\a w_\b \Gh_{\ab}{}^{\bb\a\b}+ \psi^a \psi^b \wb_{\ab} \wb_{\bb} J_{ab}{}^{\ab\bb}\cr
&+ \psib^a \psib^b w_\a w_\b \Jh_{ab}{}^{\a\b} +  \psi^a w_\a \psib^b \wb_{\ab} K_{ab}{}^{\a\ab} + \psi^a w_\a \wb_{\ab} \wb_{\bb} L_a{}^{\a\ab\bb}\cr 
&+ \psib^a \ob_{\ab}w_\a w_\b \Lh_a{}^{\ab\a\b} + w_\a w_\b \wb_{\ab} \wb_{\bb} M^{\a\b\ab\bb} \Big] ,
\end{align}
where $\Pi^A$ and $\Pib^A$ are given by
\begin{align}
    \Pi^A=\p Z^M E_M{}^A,\quad \Pib^A=\pb Z^M E_M{}^A ,
\end{align}
and $E^A=dZ^M E_M{}^A$ is the supervielbein one-form. The worldsheet fields $d_\a$ and $\dd_{\ab}$ generate translations in the fermionic directions of superspace. The covariant derivatives are defined by\footnote{We define $\p_0=\frac12(\pb+\p)$ and $\p_1=\frac12(\pb-\p)$, where $\partial_0$ denotes derivation in the $\tau$-direction and $\partial_1$ denotes derivation in the $\sigma$-direction of the worldsheet.}
\begin{align} \label{covs}
    &\Nb\L^\a=\B{\partial}\L^\a+\L^\b\pb Z^M \O_{M\b}{}^\a,\quad\N\Lb^{\ab}=\partial\hLambda^{\halpha}+\Lb^{\bb}\p Z^M \Oh_{M\bb}{}^{\ab} ,\cr
    &\Nb\psi^a=\B{\partial}\psi^a+\psi^b\pb Z^M \O_{Mb}{}^a,\quad\HH{\N}\psib^a=\p\psib^a+\psib^b\p Z^M \Oh_{Mb}{}^a.
\end{align}
The $\O$ and $\Oh$ superfields and the remaining background superfields $C,\dots, M$ will be constrained by BRST invariance. 

The BRST symmetry is generated by the left and right-moving BRST charges which retain their standard $\mathcal{N}=(1,1)$ worldsheet supersymmetric form
\begin{align}
Q=\oint d\s \Big( cT + cT_{\beta\gamma} + bc\p c + \g^2 b + \g G \Big), \quad \Qh = \oint d\s \Big( {\overline c} \Th + \B{c}\hT_{\hbeta\hgamma} + {\overline b} {\overline c} \pb {\overline c} +\gb^2  {\overline b}+ \gb \Gh \Big) ,
\label{BRST}
\end{align}
where $T$ and $\Th$ are the stress-energy tensors given by
\begin{align}
    T=-\frac12 \Pi_a \Pi^a -\frac12 \psi_a \N \psi^a - d_\a \Pi^\a - \frac12 w_\a \N \L^\a +\frac12 \L^\a \N w_\a - \psi^a w_\a \Pi^A C_{Aa}{}^\a - w_\a w_\b \Pi^A Y_A{}^{\a\b},
\end{align}
and
\begin{align}
    \Th = -\frac12 \Pib_a \Pib^a - \frac12 \psib_a {\overline{\widehat{\nabla}}} \psib^a - \dd_{\ab} \Pib^{\ab} - \frac12 \wb_{\ab} \Nb \Lb^{\ab} + \frac12 \Lb^{\ab} \Nb \wb_{\ab} - \psib^a \wb_{\ab} \Pib^A \Ch_{Aa}{}^{\ab} - \wb_{\ab} \wb_{\bb} \Pib^A \Yh_A{}^{\ab\bb}.
\end{align}
The supercurrents $G$ and $\hG$ are generalized to
\begin{align}
G =&~ \frac12\psi^a(\L\g_a\L) + \L^\a d_\a + \psi^a \Pi_a + w_\a \Pi^\a \cr 
&+ \frac12 \L^\a \L^\b w_\g G_{\a\b}{}^\g + \frac12 \L^\a \psi^a \psi^b G_{\a ab} +\frac16 \psi^a \psi^b \psi^c G_{abc} +\psi^a \L^\a w_\b G_{a\a}{}^\b \cr 
&+ \frac12 \psi^a \psi^b w_\a G_{ab}{}^\a + \frac12 \L^\a w_\b w_\g G_\a{}^{\b\g}+\frac12 \psi^a w_\a w_\b G_a{}^{\a\b} + \frac16 w_\a w_\b w_\g G^{\a\b\g}  ,
\label{Ggeneral}
\end{align}
\begin{align}
\Gh =&~ \frac12\psib^a(\Lb\g_a\Lb) + \Lb^{\ab} \dd_{\ab} + \psib^a \Pib_a + \wb_{\ab} \Pib^{\ab} \cr 
&+ \frac12 \Lb^{\ab} \Lb^{\bb} \wb_{\gb} \Gh_{\ab\bb}{}^{\gb} + \frac12 \Lb^{\ab} \psib^a \psib^b \Gh_{\ab ab} +\frac16 \psib^a \psib^b \psib^c \Gh_{abc} +\psib^a \Lb^{\ab} \wb_{\bb} \Gh_{a\ab}{}^{\bb} \cr 
&+ \frac12 \psib^a \psib^b \wb_{\ab} \Gh_{ab}{}^{\ab} + \frac12 \Lb^{\ab} \wb_{\bb} \wb_{\gb} \Gh_{\ab}{}^{\bb\gb}+ \frac12 \psib^a \wb_{\ab} \wb_{\bb} \Gh_a{}^{\ab\bb} + \frac16 \wb_{\ab} \wb_{\bb} \wb_{\gb} \Gh^{\ab\bb\gb}.
\label{Ghgeneral}
\end{align}
Note that these expressions for the supercurrents also satisfy the U(1)-charge/small Hilbert space condition on the deformations.

The background superfields that have been introduced above parametrize the deformations away from the flat background. As in \cite{Berkovits:2022dbm}, the physical deformations are the ones such that the BRST charge is nilpotent. The BRST charges are nilpotent
\begin{align}
Q^2=\Qh^2=\{Q,\Qh\}=0 ,
\label{main}
\end{align}
provided that the supercurrents $G,\hG$ satisfy
\begin{align}
\{G,G\}=-2T,\quad \{\Gh,\Gh\}=-2\Th,\quad \{G,\Gh\}=0,
\label{Gmain}
\end{align}
where $\{,\}$ denotes the Poisson bracket. These requirements will constrain the background superfields. As we will see, the constraints are such that the supercurrent $G,\hG$ are respectively holomorphic and anti-holomorphic, and will imply that the background superfields describe Type II supergravity backgrounds.

Before ending this section, the torsion and curvature components for the background geometry will be defined because the constraints of Type II supergravity are expressed in terms of them. The vielbein $E^A$ one-form and connections $\O_\a{}^\b, \Oh_{\ab}{}^{\bb}$ one-forms allow the definition of torsion and curvature two-forms. Note that up to this point, the $\O, \Oh$ connections can be decomposed as
\begin{align} \label{Odec}
    &\O_\a{}^\b=\d_\a^\b\O+\frac14(\g^{ab})_\a{}^\b\O_{ab}+(\g^{abcd})_\a{}^\b\O_{abcd},\cr
    &\Oh_{\ab}{}^{\bb}=\d_{\ab}^{\bb}\Oh+\frac14(\g^{ab})_{\ab}{}^{\bb}\Oh_{ab}+(\g^{abcd})_{\ab}{}^{\bb}\Oh_{abcd}.
\end{align}
Our calculation will show that $\O_{abcd}=\Oh_{abcd}=0$ and the torsion and curvatures will satisfy the constraints of \cite{Berkovits:2001ue}. The torsions one-forms $T^\a$ and $T^{\ab}$ are defined as
\begin{align} \label{torsionA}
    T^\a=dE^\a+E^\b\O_\b{}^\a,\quad T^{\ab}=dE^{\ab}+E^{\bb}\Oh_{\bb}{}^{\ab}.
\end{align}
Since we have $\O_{ab}$ and $\Oh_{ab}$, we can define the torsions
\begin{align}\label{torsion}
    T^a=dE^a+E^b\O_b{}^a,\quad \Th^a=dE^a+E^b\Oh_b{}^a ,
\end{align}
where the indices $a,b,\dots$ are raised or lowered with the Minkowski metric $\eta^{ab}$ or $\eta_{ab}$. The torsion components will be constrained by imposing (\ref{Gmain}). Similarly, the curvatures are defined as
\begin{align}
    &R_\a{}^\b=d\O_\a{}^\b+\O_\a{}^\g\O_\g{}^\b,\quad \Rh_{\ab}{}^{\bb}=d\Oh_{\ab}{}^{\bb}+\Oh_{\ab}{}^{\gb}\Oh_{\gb}{}^{\bb},\cr
    &R_{ab}=d\O_{ab}+\O_a{}^c\O_{cb},\quad \Rh_{ab}=d\Oh_{ab}+\Oh_a{}^c\Oh_{cb}.
\end{align}
The product between forms is the wedge product. We are going to need the Bianchi identities
\begin{align}\label{Bianchis}
    &\N T^\a = T^\b R_\b{}^\a,\quad \N T^{\ab} = \Th^{\bb} R_{\bb}{}^{\ab},\cr
    &\N T^a = T^b R_b{}^a,\quad \Nh\Th^a=\Th^b\Rh_b{}^a .
\end{align}
Finally, we define two types of covariant derivatives depending on which of the $\O_{ab}$ or $\Oh_{ab}$ connections one uses\footnote{It will be shown in this paper that the constraints imposed by BRST invariance relate the connections $\Oh_{ab}$ and $\O_{ab}$ such that $(\hOmega_{abc}-\Omega_{abc})\propto \hT_{abc},~(\hOmega_{\halpha bc}-\Omega_{\halpha bc})\propto\hT_{\halpha bc}$ and $(\Omega_{\alpha bc}-\hOmega_{\alpha bc})\propto T_{\alpha bc}$.}. For example, the covariant derivatives on a zero-form $\Psi_{a\a\bb}$ are
\begin{align}\label{ccovs}
    &\N_A\Psi_{a\a\bb}=\p_A\Psi_{a\a\bb}-\O_{Aa}{}^b\Psi_{b\a\bb}-\O_{A\a}{}^\g\Psi_{a\g\bb}-\Oh_{A\bb}{}^{\gb}\Psi_{a\a\gb},\cr
    &\Nh_A\Psi_{a\a\bb}=\p_A\Psi_{a\a\bb}-\Oh_{Aa}{}^b\Psi_{b\a\bb}-\O_{A\a}{}^\g\Psi_{a\g\bb}-\Oh_{A\bb}{}^{\gb}\Psi_{a\a\gb}.
\end{align}
Type II supergravity also contains the components of the three-form superfield $H=dB$ in its spectrum, where $B$ is the two-form of the action.

It is important to note that the action (\ref{action}) and BRST charges (\ref{BRST}) are invariant under several gauge symmetries associated to relabelling background fields, which are gathered in Appendix \ref{app0}. This behavior has already been encountered during the analysis of the heterotic string case. Before moving on to the Poisson brackets, we will already make use of the gauge symmetries parameterized by $\omega_{\alpha\beta}\,^{\gamma}$ and $\homega_{\halpha\hbeta}\,^{\hgamma}$ to gauge
\begin{equation}
    \Omega_{\alpha}\,^{bcde}=\hOmega_{\halpha}\,^{bcde}=0,
\end{equation}
\begin{equation}
    G_{\alpha\beta}\,^{\gamma}=\WT{F}^{1440}_{abcd\delta}(\gamma_e)^{\delta\gamma}(\gamma^{abcde})_{\alpha\beta},
\end{equation}
\begin{equation}
    \HH{G}_{\halpha\hbeta}\,^{\hgamma}=\HH{\WT{F}}^{1440}_{abcd\hdelta}(\gamma_e)^{\hdelta\hgamma}(\gamma^{abcde})_{\halpha\hbeta},
\end{equation}
where the $1440$ superscript denotes the dimension of the irreducible representation to which $\WT{F}^{1440}_{abcd\delta}$ and $\HH{\WT{F}}^{1440}_{abcd\hdelta}$ belong. Recall that the 2176 components of $G_{\alpha\beta}\,^{\gamma}$ split up into the following irreducible representations
\begin{equation}
    2176=16\oplus16\oplus144\oplus560\oplus1440,
\end{equation}
\begin{equation}
    G_{\alpha\beta}\,^{\gamma}=G_a\,^{\gamma}\big(\gamma^a\big)_{\alpha\beta}+F_{abcde}\,^{\gamma}\big(\gamma^{abcde}\big)_{\alpha\beta},
\end{equation}
\begin{equation}
    G_a\,^{\gamma}=\widetilde{G}^{144}_a\,^{\gamma}+G^{16}_{\delta}\big(\gamma_a\big)\,^{\delta\gamma},
\end{equation}
\begin{equation}
    F_{abcde}\,^{\gamma}=\widetilde{F}^{1440}_{[abcd|\delta|}\big(\gamma_{e]}\big)^{\delta\gamma}+\widetilde{F}^{560}_{[ab|\delta|}\big(\gamma_{cde]}\big)^{\delta\gamma}+F^{16}_{\delta}\big(\gamma_{abcde}\big)^{\delta\gamma},
\end{equation}
and analogously for $\hG_{\halpha\hbeta}\,^{\hgamma}$.

\section{Computation of $\{G,G\}=-2T$ and $\{\Gh,\Gh\}=-2\Th$}\label{GGisT}
In this section, we prove the claim that imposing the conditions $\big\{G,G\big\}=-2T$ and $\big\{\hG,\hG\big\}=-2\hT$ implies the $N=2$ $D=10$ Type II supergravity constraints of \cite{Berkovits:2001ue}, as well as determines the background fields in the action and in the supercurrents in terms of supergravity superfields. The canonical Poisson brackets for the fundamental variables in the action (\ref{action}) are  
\begin{align} \label{brack0}
    &\big[Z^M(\s), P_N(\s')\big\} = \d_N^M \d(\s-\s') ,\cr  
    &\big\{\psi^a(\s),\psi^b(\s')\big\} = \eta^{ab} \d(\s-\s'),\quad \big\{\psib^a(\s),\psib^b(\s')\big\} = \eta^{ab} \d(\s-\s') ,\cr
    &\big[\L^\a(\s),w_\b(\s')\big] = \d_\b^\a \d(\s-\s'),\quad \big[\Lb^{\ab}(\sigma),\wb_{\bb}(\sigma')\big] = \d_{\bb}^{\ab} \d(\s-\s'),
\end{align}
where the momentum $P_M\equiv\frac{\delta S}{\delta (\partial_{0}Z^M)}$ is given by
\begin{align}
(-1)^M P_M =&~ \p_0 Z^N G_{NM} + \p_1 Z^N B_{NM} - E_M{}^\a d_\a - E_M{}^{\ab} \dd_{\ab} \cr
& - \frac12 \psi^a \psi^b \O_{Mab} - \frac12 \psib^a \psib^b \Oh_{Mab} + \L^\a w_\b \O_{M\a}{}^\b + \Lb^{\ab} \wb_{\bb} \Oh_{M\ab}{}^\b \cr 
&+ (-1)^M \psi^a w_\a C_{Ma}{}^\a + (-1)^M \psib^a \wb_{\ab} \Ch_{Ma}{}^{\ab} + w_\a w_\b Y_M{}^{\a\b} + \wb_{\ab} \wb_{\bb} \Yh_M{}^{\ab\bb}.
\label{PM}
\end{align}
The last equation allows us to express $d_a, d_\a$ and $\dd_{\ab}$ in terms of canonical variables. Namely,
\begin{align}
d_a \equiv \p_0 Z^M E_{Ma} =& (-1)^M E_a{}^M P_M - \p_1 Z^M B_{Ma} \cr 
&+ \frac12 \psi^b \psi^c \O_{abc} + \frac12 \psib^b \psib^c \Oh_{abc} - \L^\a w_\b \O_{a\a}{}^\b - \hLambda^{\ab} \wb_{\bb} \Oh_{a\ab}{}^{\bb} \cr
&-\psi^b w_\a C_{ab}{}^\a - \psib^b \wb_{\ab} \Ch_{ab}{}^{\ab} - w_\a w_\b Y_a{}^{\a\b} - \wb_{\ab} \wb_{\bb} \Yh_a{}^{\ab\bb},
\label{p0is}
\end{align}
\begin{align}
d_\a =& (-1)^{M+1}  E_\a{}^M  P_M+ \p_1 Z^M B_{M\a} - \frac12 \psi^a \psi^b \O_{\a ab} - \frac12 \psib^a \psib^b \Oh_{\a ab} + \L^\b w_\g \O_{\a\b}{}^\g + \Lb^{\bb} \wb_{\gb} \Oh_{\a\bb}{}^{\gb} \cr
&-\psi^a w_\b C_{\a a}{}^\b - \psib^a \wb_{\bb} \Ch_{\a a}{}^{\bb} + w_\b w_\g Y_\a{}^{\b\g} + \wb_{\bb}\wb_{\gb} \Yh_\a{}^{\bb\gb} ,
\label{dis}
\end{align}
\begin{align}
\dd_{\ab} =& (-1)^{M+1}  E_{\ab}{}^M  P_M+ \p_1 Z^M B_{M\ab} - \frac12 \psi^a \psi^b \O_{\ab ab} - \frac12 \psib^a \psib^b \Oh_{\ab ab} + \L^\b w_\g \O_{\ab\b}{}^\g + \Lb^{\bb} \wb_{\gb} \Oh_{\ab\bb}{}^{\gb} \cr
&-\psi^a w_\b C_{\ab a}{}^\b - \psib^a \wb_{\bb} \Ch_{\ab a}{}^{\bb} + w_\b w_\g Y_{\ab}{}^{\b\g} + \wb_{\bb}\wb_{\gb} \Yh_{\ab}{}^{\bb\gb} .
\label{dbis}
\end{align}
From these definitions, the Poisson brackets necessary for the calculation of (\ref{Gmain}) can be computed. They have been listed in Appendix \ref{app1}. The computation of (\ref{Gmain}) will also make use of the equations of motion derived from (\ref{action}). The variation of the action with respect to $d_{\alpha}$ and $\dd_{\halpha}$ gives rise to
\begin{align}
&\Pib^\a = -\dd_{\bb} P^{\a\bb} - \Lb^{\bb} \wb_{\gb} D_{\bb}{}^{\gb\a} - \psib^a \psib^b \Uh_{ab}{}^\a - \psib^a \wb_{\bb} V_a{}^{\a\bb} - \wb_{\bb} \wb_{\gb} Z^{\a\bb\gb} ,\cr
&\Pi^{\ab} = d_\b P^{\b\ab} - \L^\b w_\g \Dh_\b{}^{\g\ab} - \psi^a \psi^b U_{ab}{}^{\ab} - \psi^a w_\b \Vh_a{}^{\ab\b} - w_\b w_\g \Zh^{\ab\b\g},
\label{eom0}
\end{align}
and the remaining equations of motion are listed in Appendix \ref{app2}.

Explicitly, we compute
\begin{align}
    \big\{G,G\big\}=\oint d\s \oint d\s' \{G(\s'),G(\s)\}=-\oint d\s ~2T(\s),
\end{align}
\begin{align}
    \big\{\hG,\hG\big\}=\oint d\s \oint d\s' \{\hG(\s'),\hG(\s)\}=-\oint d\s ~2\hT(\s),
\end{align}
that is, these are same-time commutators and the spatial coordinates which are not written explicitly are understood to be integrated over.
The computation is long and there are many contributions coming from the different necessary Poisson brackets. However, a pattern appears which is helpful when it comes to analysing the final result. The resulting terms are of two types:
\begin{equation}
    \varphi^A\varphi^B\mathcal{C}_{AB}\,^C\Delta_C\quad \mathrm{or}\quad 
    \varphi^A\varphi^B\varphi^C\varphi^D\mathcal{I}_{ABCD} ,
\end{equation}
where $ \varphi\in\big\{\Lambda^{\alpha},\psi^a,w_{\alpha},\hLambda^{\halpha},\hpsi^a,\hw_{\halpha}\big\}$ and $\Delta\in\big\{d_{\alpha},\hd_{\halpha},\Pi^{\alpha},\B{\Pi}^{\halpha},\Pi^a,\B{\Pi}^a\big\}$.
The notation $\mathcal{C}_{AB}\,^C$ and $\mathcal{I}_{ABCD}$ denotes functions of the background superfields in the action and in the supercurrents $(G,\hG)$, as well as of supergravity superfields. The first type of term will be analysed in the next subsection. These terms are the easiest, and the functions $\mathcal{C}_{AB}\,^C$ are either $N=2$ $D=10$ Type II supergravity constraints, or determine the superfields in the supercurrent $G$ and $\big\{C,Y,D,U,V,Z\big\}$ which appear in the action. The second type of term will be analysed later, as they are more complicated. The functions $\mathcal{I}_{ABCD}$ will be equivalent to Bianchi identities, which vanish identically, or will determine the remaining background fields $\big\{S,E,\HH{E},F,\HH{F},G,\HH{G},H,I,\HH{I},J,\HH{J},K,L,\HH{L},M\big\}$ which appear in the action.

\subsection{Three worldsheet fields}
We will now focus on the contributions with three worldsheet fields coming from the $\big\{G,G\big\}=-2T$ Poisson bracket computation. We start by considering the contributions involving a factor of $\Pib^a$. The terms with $\L^\a\L^\b\Pib^a$ imply the constraint
\begin{align}\label{GG1}
    T_{\a\b a}-H_{\a\b a}=0.
\end{align}
The terms with $\psi^a\L^\a\Pib^b$ lead to
\begin{align}\label{GG2}
    T_{\a(ab)}=H_{\a ab}=0.
\end{align}
The terms with $\psi^a\psi^b\Pib^c$ lead to the equation
\begin{align}\label{GG3}
    T_{abc}-T_{c[ab]}-H_{abc}=0.
\end{align}
Symmetrizing $bc$ in this equation gives the constraint $T_{a(bc)}=0$, whereas  antisimmetrizing $bc$ implies the constraint $T_{cba}+H_{cba}=0$. From now on, the torsion component $T_{abc}$ is totally antisymmetric. The terms with $\L^\a w_\b\Pib^a$ imply the constraint
\begin{align}\label{GG4}
    T_{a\a}{}^\b-\frac12 P^{\b\gb}\left(T_{\gb\a a}-H_{\gb\a a}\right)=0.
\end{align}
The terms with $\psi^a w_\a\Pib^b$ lead to defining equations for the $C_{ab}\,^{\gamma}$ superfield
\begin{align}\label{GG5}
    &C_{(ab)}{}^\a - P^{\a\bb} T_{\bb(ab)} = 0 ,\cr 
    &2T_{ab}{}^\a - C_{[ab]}{}^\a + P^{\a\bb} H_{\bb ab}=0. 
\end{align}
The terms with $w_\a w_\b\Pib^a$ lead to the equation
\begin{align}\label{GG6}
    Y_a{}^{\a\b} - \frac12 P^{(\a\gb} T_{a\gb}{}^{\b)} + \frac14 P^{\a\gb} P^{\b\rb} \left( T_{\gb\rb a} - H_{\gb\rb a} \right) = 0.
\end{align}

Consider now the terms involving $\Pi^a$ in $\{G,G\}$. The terms with $\L^\a\L^\b\Pi^a$ give the constraint
\begin{align}\label{GG7}
    \frac12 \left(T_{\a\b a}+H_{\a\b a}\right) + (\g_a)_{\a\b} = 0 ,
\end{align}
which together with (\ref{GG1}) determines the Type II supergravity constraint
\begin{equation}
    T_{\a\b a}=H_{\a\b a}=-(\g_a)_{\a\b}.
\end{equation}
The terms with $\L^\a w_\b\Pi^a$ determine $G_{a\a}{}^\b$ of the supercurrent $G$ (\ref{Ggeneral}) as
\begin{align}\label{GG8}
    G_{a\a}{}^\b=C_{\a a}{}^\b+T_{a\a}{}^\b+\frac12 P^{\b\gb} \left( T_{\gb\a a} + H_{\gb\a a} \right).
\end{align}
The terms with $\psi^a\psi^b\Pi^c$ determine $G_{abc}$ from $G$ as
\begin{align}\label{GG9}
    G_{abc} = -T_{abc}.
\end{align}
The terms with $\psi^a\L^\a\Pi^b$ determine $G_{\a ab}$ of $G$ according to
\begin{align}\label{GG10}
    G_{\a ab} = \frac12 ( T_{\a[ab]} + H_{\a ab} ) = T_{\a ab},
\end{align}
where (\ref{GG2}) was used. The terms with $\psi^a w_\a\Pi^b$ determine $G_{ab}{}^\a$ of $G$ as
\begin{align}\label{GG11}
    G_{ab}{}^\a = -T_{ab}{}^\a + C_{[ab]}{}^\a  - \frac12 P^{\a\bb} ( T_{\bb[ab]} + H_{\bb ab} ) .
\end{align}
Finally, the terms with $w_\a w_\b\Pi^a$ lead to
\begin{align}\label{GG12}
     G_a{}^{\a\b} &= 2 Y_a{}^{\a\b} - P^{(\a\gb} C_{\gb a}{}^{\b)}  - P^{(\a\gb} T_{a\gb}{}^{\b)} -\frac12 P^{\a\gb} P^{\b\rb} ( T_{\gb\rb a} + H_{\gb\rb a} ) \cr 
    &= -P^{(\a\gb} C_{\gb a}{}^{\b)} -P^{\a\gb} P^{\b\rb}  T_{\gb\rb a} ,
\end{align}
where (\ref{GG6}) was used. 

Take now the terms with $\dd_{\halpha}$ in $\{G,G\}$. The terms with $\L^\a\L^\b\dd_{\gb }$ lead to the constraint
\begin{align}\label{GG13}
    T_{\a\b}{}^{\gb} + \frac12 P^{\r\gb} H_{\r\a\b} = 0 .
\end{align}
The terms with $\psi^a\L^\a\dd_{\bb}$ give the constraint
\begin{align}\label{GG14}
    T_{a\a}{}^{\bb} = - \frac12 P^{\g\bb} ( T_{\g\a a} + H_{\g\a a} ) = (\g_a)_{\a\g} P^{\g\bb} ,
\end{align}
where $T_{\a\b a}=H_{\a\b a}=-(\g_a)_{\a\b}$ was used. The terms with $\psi^a\psi^b\dd_{\ab}$ determine $U_{ab}\,^{\hgamma}$ by
\begin{align}\label{GG15}
    2 U_{ab}{}^{\ab} + T_{ab}{}^{\ab} + P^{\b\ab} T_{\b ab} = 0 ,
\end{align}
where (\ref{GG2}) was used. The terms with $\L^\a w_\b\dd_{\gb}$ lead to
\begin{align}\label{GG16}
    \Dh_{\a}{}^{\b\gb} + \N_\a P^{\b\gb} - P^{\b\rb} T_{\rb\a}{}^{\gb}  - \frac12 P^{\b\rb} P^{\s\gb} H_{\rb\s\a} = 0 . 
\end{align}
The terms with $\psi^a w_\a\dd_{\bb}$ give
\begin{align}\label{GG17}
    \Vh_a{}^{\bb\a} - \N_a P^{\a\bb} + P^{\a\gb} T_{a\gb}{}^{\bb} + P^{\g\bb} T_{a\g}{}^\a + P^{\g\bb} C_{\g a}{}^\a + \frac12 P^{\a\gb} P^{\r\bb}( T_{\r\gb a} + H_{\r\gb a} ) = 0 .
\end{align}
Finally, the terms with $w_\a w_\b\dd_{\gb}$ determine the background superfield $\hZ^{\gamma\beta\alpha}$
\begin{align}\label{GG18}
    \Zh^{\gb\a\b} - \frac12 P^{(\a\rb} \N_{\rb} P^{\b)\gb} + \frac12 P^{\r\gb} P^{(\a\overline{\sigma}} T_{\r\overline{\sigma}}{}^{\b)}+ \frac14 P^{\a\rb}P^{\b\overline{\sigma}} P^{\tau\gb}H_{\tau\rb\overline{\sigma}} - P^{\r\gb} Y_\r{}^{\a\b} = 0 .
\end{align}

Consider now the terms with $d_{\alpha}$ in $\{G,G\}$. The terms with $\L^\a\L^\b d_\g$ give
\begin{align}\label{GG19}
    G_{\a\b}{}^\g=-T_{\a\b}{}^\g-\frac12P^{\g\rb}H_{\rb\a\b} .
\end{align}
The terms with $\psi^a\L^\a d_\g$ lead to (\ref{GG8}). Similarly, the terms with $\psi^a\psi^bd_\a$ lead to (\ref{GG11}). The terms with $\L^\a w_\b d_\g$ imply
\begin{align}\label{GG20}
    G_\a{}^{\b\g} = -2 Y_\a{}^{\b\g} + P^{(\b\rb} T_{\rb\a}{}^{\g)} + \frac12 P^{\b\rb} P^{\g\overline{\sigma}} H_{\a\rb\overline{\sigma}} .
\end{align}
The terms with $\psi^a w_\a d_\b$ lead to equation (\ref{GG12}). Finally, the terms with $w_\a w_\b d_\g$ give
\begin{align}\label{GG21}
    G^{\a\b\g} = 2 P^{(\a\rb} Y_{\rb}{}^{\b\g)} - P^{(\a\rb} P^{\b\overline{\sigma}} T_{\rb\overline{\sigma}}{}^{\g)} - \frac12 P^{\a\rb} P^{\b\overline{\sigma}} P^{\g\taub} H_{\rb\overline{\sigma}\taub}.
\end{align}

Focus now on the terms with factor of $\Pib^{\ab}$ in $\{G,G\}$. The term with $\L^\a\L^\b\Pib^{\gb}$ gives the constraint
\begin{align}\label{GG22}
     H_{\a\b\gb} = 0 .
\end{align}
The contributions with $\psi^a\L^\a\Pib^{\bb}$ give the equation
\begin{align}\label{GG23}
    T_{\a\bb a} + H_{\a\bb a} = 0 .
\end{align}
The terms with $\L^\a w_\b\Pib^{\gb}$ lead to the constraint
\begin{align}\label{GG24}
    T_{\gb\a}{}^\b + \frac12 P^{\b\rb} H_{\rb\gb\a} = 0 .
\end{align}
The terms with $\psi^a\psi^b\Pib^{\ab}$ imply
\begin{align}\label{GG25}
    T_{\ab[ab]} + H_{\ab ab} = 0 .
\end{align}
The terms with $\psi^a w_\a\Pib^{\bb}$ lead to
\begin{align}\label{GG26}
    T_{a\bb}{}^\a + C_{\bb a}{}^\a + \frac12 P^{\a\gb} ( T_{\gb\bb a} + H_{\gb\bb a} ) = 0 .
\end{align}
Finally, the contributions with $w_\a w_\b \Pib^{\gb}$ imply
\begin{align}\label{GG27}
    Y_{\gb}{}^{\a\b} - \frac12 P^{(\a\rb} T_{\rb\gb}{}^{\b)} - \frac14 P^{\a\rb} P^{\b\overline{\sigma}}H_{\rb\overline{\sigma}\gb} = 0.
\end{align}

The last contributions we must consider are the ones with a factor of $\Pi^{\a}$ in $\{G,G\}$. The term with $\L^\a\L^\b\Pi^\g$ leads to the constraint
\begin{align}\label{GG28}
    H_{\a\b\g} = 0 .
\end{align}
The terms with $\psi^a\L^\a\Pi^\b$ vanish because of the constraint $T_{\a\b a}=H_{\a\b a}=-(\g_a)_{\a\b}$. The terms with $\L^\a w_\b\Pi^\g$ imply (\ref{GG19}). The terms with $\psi^a\psi^b\Pi^\a$ are equivalent to the conditions (\ref{GG2}) and (\ref{GG10}). The terms with $\psi^a w_\a\Pi^\b$ lead to (\ref{GG8}). Finally, the terms with $w_\a w_\b\Pi^\g$ are equivalent to (\ref{GG20}).

The analysis of the terms with three worldsheet fields coming from the $\big\{\hG,\hG\big\}=-2\hT$ Poisson bracket is analogous to the one we just covered and gives the ``hatted'' version of the constraints and definitions above (note that some of the signs are different, though). After gathering all the contributions and simplifying all the expressions with the constraints coming from both Poisson brackets, the analysis of the terms with three worldsheet fields in the commutators $\big\{G,G\big\}=-2T$ and $\big\{\hG,\hG\big\}=-2\hT$ leads to the constraints:
\begin{align}\label{TH}
    &H_{\a\b\g} = H_{\a\b\gb} = H_{\a\bb\gb} = H_{\ab\bb\gb} =T_{\a\b}{}^{\gb} = T_{\a\bb}{}^{\gb} = T_{\ab\bb}{}^\g = T_{\ab\b}{}^\g =T_{\a\b}{}^\g=T_{\ab\bb}{}^{\gb} =0 ,\cr
    &T_{\a\b a} = H_{\a\b a} = -(\g_a)_{\a\b},\quad T_{\ab\bb a} = - H_{\ab\bb a} = -(\g_a)_{\ab\bb},\quad T_{\a\bb a} = H_{\a\bb a} = 0 ,\cr 
    &T_{\a(ab)} = H_{\a ab} = \Th_{\ab(ab)} = H_{\ab ab} = T_{\ab ab} = \Th_{\a ab} = 0 ,\cr 
    &T_{a\a}{}^\b = T_{a\ab}{}^{\bb} = 0,\quad T_{a\a}{}^{\bb} = (\g_a)_{\a\g} P^{\g\bb},\quad T_{a\ab}{}^\b = -(\g_a)_{\ab\gb} P^{\b\gb} ,\cr 
    &T_{abc} + H_{abc} = 0,\quad \Th_{abc} - H_{abc} = 0 , 
\end{align}
which are indeed the $N=2$ $D=10$ Type II supergravity constraints of \cite{Berkovits:2001ue}. We would like to emphasize that the computation described here yields both the physical and the conventional constraints. This is different from what happens in the pure spinor superstring case, where physical constraints come from the nilpotency of the BRST charge and conventional constraints come from the holomorphicity of the BRST current.

Note also that the constraints $T_{\halpha bc}=\hT_{\alpha bc}=0$ and $\hT_{abc}=-T_{abc}=H_{abc}$ imply
\begin{equation} \label{Omegabc}
    \Omega_{\alpha bc}-\hOmega_{\alpha bc}=T_{\alpha bc},\quad \hOmega_{\halpha bc}-\Omega_{\halpha bc}=\hT_{\halpha bc},\quad \hOmega_{abc}-\Omega_{abc}=\hT_{abc},
\end{equation}
as mentioned previously. Furthermore, note that in writing these expressions we have made use of the fact that the constraints
\begin{equation}
    G_{\alpha\beta}\,^{\gamma}=-T_{\alpha\beta}\,^{\gamma}=\WT{F}^{1440}_{abcd\delta}(\gamma_e)^{\delta\gamma}(\gamma^{abcde})_{\alpha\beta},
\end{equation}
\begin{equation}
    \HH{G}_{\halpha\hbeta}\,^{\hgamma}=-\hT_{\halpha\hbeta}\,^{\hgamma}=\HH{\WT{F}}^{1440}_{abcd\hdelta}(\gamma_e)^{\hdelta\hgamma}(\gamma^{abcde})_{\halpha\hbeta},
\end{equation} 
imply that $T_{\alpha\beta}\,^{\gamma}=\hT_{\halpha\hbeta}\,^{\hgamma}=0$. Indeed, following the analysis done for the heterotic string, the Bianchi identities
\begin{equation}
    \nabla_{(\alpha}T_{\beta\gamma)}\,^{d}+T_{(\alpha\beta}\,^ET_{|E|\gamma)}\,^{d}=0,
\end{equation}
\begin{equation}
    \HH{\nabla}_{(\halpha}\hT_{\hbeta\hgamma)}\,^{d}+\hT_{(\halpha\hbeta}\,^E\hT_{|E|\hgamma)}\,^{d}=0,
\end{equation}
imply that $\WT{F}^{1440}_{abcd\delta}$ and $\HH{\WT{F}}^{1440}_{abcd\hdelta}$ must be zero. To see why, note that these components are not in the kernel of the linear maps $T_{(\alpha\beta}\,^{\rho}\gamma_{\gamma)\rho}^d=0$ and $\hT_{(\halpha\hbeta}\,^{\hrho}\gamma_{\hgamma)\hrho}^d=0$. Furthermore, since we have the constraints $\hT_{\alpha\beta}\,^{\hgamma}=T_{\halpha\hbeta}\,^{\gamma}=T_{\alpha\hbeta}\,^c=0$, there are no other 1440 irrep contributions in the Bianchi identities above. Consequently, we do conclude that $\WT{F}^{1440}_{abcd\delta}=\HH{\WT{F}}^{1440}_{abcd\hdelta}=0$ and therefore
\begin{equation}
    T_{\alpha\beta}\,^{\gamma}=\hT_{\halpha\hbeta}\,^{\hgamma}=0.
\end{equation}

Finally, we stress the fact that imposing the U(1)-charge/small Hilbert space condition in addition to the anti-commutators was essential to achieve the constraints of (\ref{TH}). This has already been observed in the heterotic string case. Indeed, were we allowed to deform the supercurrents with terms with positive U(1) charge (or equivalently, in the large Hilbert space), we would lose, for example, the essential constraints
\begin{equation}
    T_{\alpha\beta}\,^c=-(\gamma^c)_{\alpha\beta},\quad T_{\halpha\hbeta}\,^c=-(\gamma^c)_{\halpha\hbeta},\quad H_{\alpha\beta\gamma}=H_{\halpha\hbeta\hgamma}=0.
\end{equation}

\subsection{Four worldsheet fields}
Now we move on to the contributions with four worldsheet fields. Again, we start by focusing on the contributions coming from the $\big\{G,G\big\}=-2T$ Poisson bracket. As mentioned above, this analysis will give the remaining background superfields of the action (\ref{action}) in terms of the supergravity superfields. It will also determine constraints on the curvature.

Consider first the terms with $\L^{\alpha}\L^{\beta} w_{\gamma}w_{\delta}$. These are
\begin{align}\label{GG29}
    \L^\a \L^\b w_\g w_\r P^{\g\overline{\sigma}} \left( \N_{\overline{\sigma}} T_{\a\b}{}^\r - \g^a_{\a\b} T_{a\overline{\sigma}}{}^\r - R_{\overline{\sigma}(\a\b)}{}^\r \right).
\end{align}
 This contribution vanishes because of the Bianchi identity involving $R_{(\overline{\sigma}\a\b)}{}^\r$. The terms with $\L^{\alpha}\L^{\beta}\wb_{\hgamma}\wb_{\hdelta}$ lead to
 \begin{align}\label{GG30}
    \L^\a \L^\b \wb_{\gb} \wb_{\rb} P^{\s\gb}P^{\tau\rb} (\g^a)_{(\a\b} (\g_a)_{\s)\tau},
\end{align}
which vanishes due to the Fierz identity $(\g^a)_{(\a\b} (\g_a)_{\s)\tau}=0$. The terms with $\psi^a\psi^b\L^{\gamma}\L^{\delta}$ give
\begin{align}\label{GG31}
    &\psi^a\psi^b\L^\a\L^\b \Big( \O_{[a} (\g_{b]})_{\a\b} + \O_{[a}{}^{cdef} (\g_{b]cdef})_{\a\b} \Big) \cr 
    &+\frac12\psi^a\psi^b \L^\a \L^\b \left( \N_{(\a} T_{\b)ab} + T_{(\a a}{}^c T_{\b)cb} + T_{\a\b}{}^A T_{Aab} - R_{\a\b ab} \right).
\end{align}
Using the Bianchi identity involving $R_{[\a\b a]b}$ in the second line, we obtain the equation
\begin{align}\label{GG32}
    \O_{[a} (\g_{b]})_{\a\b} + \O_{[a}{}^{cdef} (\g_{b]cdef})_{\a\b} = 0 ,
\end{align}
which implies $\O_a=0$ and $\O_a{}^{cdef}=0$. The vanishing of the term with $\psib^a\psib^b\L^{\gamma}\L^{\delta}$ implies the constraint for the curvature
\begin{align}\label{GG33}
    \Rh_{\a\b ab}=0.
\end{align}
The terms with $\L^{\alpha}\L^{\beta}\L^{\gamma} w_{\delta}$ are proportional to
\begin{align}\label{GG34}
    \L^\a\L^\b\L^\g w_\r R_{\a\b\g}{}^\r ,
\end{align}
which is zero due to the Bianchi identity involving $R_{(\a\b\g)}{}^\r$. The term with $\L^{\alpha}\L^{\beta}\Lb^{\hgamma}\wb_{\hdelta}$ leads to the constraint for the curvature
\begin{align}\label{GG35}
    \Rh_{\a\b\gb}{}^{\rb}=0.
\end{align}
The terms with $\psi^a\L^{\beta}\L^{\gamma} w_{\delta}$ give 
\begin{align}\label{GG36}
    &\psi^a \L^\a \L^\b w_\g \left( \g^b_{\a\b} T_{ab}{}^\g - R_{a(\a\b)}{}^\g \right) \cr 
    &-2\psi^a \L^\a \L^\b w_\g P^{\g\rb} \left( \O_{\rb} (\g_a)_{\a\b} + \O_{\rb}{}^{bcde} (\g_{abcde})_{\a\b} \right) .
\end{align}
The first line vanishes because of the Bianchi identity involving $R_{[a\a\b]}{}^\g$ and the vanishing of the second line implies $\O_{\rb}=0$ and $\O_{\rb}{}^{bcde}=0$. The terms with $\psib^a\L^{\beta}\L^{\gamma}\wb_{\hdelta}$ are
\begin{align}\label{GG37}
   \psib^a \L^\a \L^\b \wb_{\gb} \left( \Nh_{(\a} T_{\b)a}{}^{\gb} + \g^b_{\a\b} T_{ab}{}^{\gb} \right) ,
\end{align}
which is zero because of the Bianchi identity involving $\Rh_{[a\a\b]}{}^{\gb}$. The terms with $\psi^a\psi^b\psi^c\psi^d$ are
\begin{align}\label{GG38}
    \psi^a \psi^b \psi^c \psi^d \left( \frac13 \N_a T_{bcd} + \frac12 T_{ab}{}^\a T_{\a cd} + \frac14 T_{ab}{}^e T_{ecd} - \frac12 R_{abcd} \right).
\end{align}
This expression is zero after combining the Bianchi identity involving $R_{[abc]d}$ with the Bianchi identity involving $\N_{[a} H_{bcd]}$, together with the constraint $T_{abc}+H_{abc}=0$. The terms with $\psi^a\psi^b\psib^c\psib^d$ determines the background field $H_{abcd}$ of the action as
\begin{align}\label{GG39}
    H_{abcd} = -\frac12 \Uh_{cd}{}^\a T_{\a ab} + \frac14 \Rh_{abcd}.
\end{align}
The terms with $\psi^a\psi^b\L^{\gamma} w_{\delta}$ are
\begin{align}\label{GG40}
    &\psi^a\psi^b \L^\a w_\b \left( R_{ab\a}{}^\b - \N_\a T_{ab}{}^\b + T_{\a[a}{}^c T_{b]c}{}^\b + T_{\a[a}{}^{\gb} T_{b]\gb}{}^\b \right) \cr 
    &+ \psi^a \psi^b \L^\a w_\b P^{\b\gb} \left( R_{\a\gb ab} - \N_{\gb} T_{\a ab} + 2(\g_a)_{\a\r} T_{\gb b}{}^\r \right).
\end{align}
The first line vanishes because of the Bianchi identity involving $R_{[ab\a]}{}^\b$ and the second line vanishes due to the Bianchi identity involving $R_{[\a\gb a]b}$. The terms with $\psi^a\psi^b\Lb^{\hgamma}\wb_{\hdelta}$ determine the background superfield $\Eh_{ab\ab}{}^{\bb}$ of the action as
\begin{align}\label{GG41}
    \Eh_{ab\ab}{}^{\bb}=- \frac12 \left( \Rh_{ab\ab}{}^{\bb} + D_{\ab}{}^{\bb\g} T_{\g ab} \right) .
\end{align}
The terms with $\psi^a\psi^b w_{\gamma}w_{\delta}$ are
\begin{align}\label{GG42}
    &\psi^a \psi^b w_\a w_\b P^{\a\gb} \left( \N_{\gb} T_{ab}{}^\b + \N_{[a} T_{b]\gb}{}^\b + T_{ab}{}^c T_{c\gb}{}^\b \right) \cr 
    &-\frac12 \psi^a \psi^b w_\a w_\b P^{\a\gb} P^{\b\rb} R_{\gb\rb ab} .
\end{align}
The first line is zero because of the Bianchi identity involving $R_{[\gb ab]}{}^\b$ and the second line vanishes due to the ``hatted'' version of the curvature constraint (\ref{GG33}) that is obtained from the $\big\{\hG,\hG\big\}=-2\hT$ bracket. The terms with $\psi^a\psi^b\wb_{\hgamma}\wb_{\hdelta}$ determine the background field $J_{ab}{}^{\ab\bb}$ as
\begin{align}\label{GG43}
    J_{ab}{}^{\ab\bb}=\frac12 \left(\frac{1}{2} T_{ac}{}^{(\ab} T_{bd}{}^{\bb)} \eta^{cd} - Z^{\g\ab\bb} T_{\g ab} - \Nh_{[a} \Yh_{b]}{}^{\ab\bb} -  \Th_{ab}{}^c \Yh_c{}^{\ab\bb} - T_{ab}{}^{\gb} \Yh_{\gb}{}^{\ab\bb} \right).
\end{align}
The terms with $\psi^a\psi^b\psi^c w_{\delta}$ are
\begin{align}\label{GG44}
    &\psi^a\psi^b \psi^c w_\a \left( \N_a T_{bc}{}^\a + T_{ab}{}^d T_{dc}{}^\a + T_{ab}{}^{\bb} T_{\bb c}{}^\a \right) \cr 
    &+\psi^a\psi^b \psi^c w_\a P^{\a\bb} \left( R_{\bb abc} - \frac13 \N_{\bb} T_{abc} - T_{\bb a}{}^\g T_{\g bc} \right).
\end{align}
The first line vanishes because of the Bianchi identity involving $R_{[abc]}{}^\a$. The second line is zero after using the Bianchi identities involving $R_{[\bb ab]c}$ and $\N_{[\bb} H_{abc]}$ together with the constraint $H_{abc}=-T_{abc}$. The terms with $\psi^a\psi^b\psib^c\wb_{\hdelta}$ determine the background superfield $I_{abc}{}^{\ab}$ as
\begin{align}\label{GG45}
    I_{abc}{}^{\ab}=-\frac12 \left( V_c{}^{\b\ab} T_{\b ab} + \Th_{ab}{}^d T_{dc}{}^{\ab} + T_{ab}{}^\b T_{\b c}{}^{\ab} + T_{ab}{}^{\bb} \Ch_{\bb c}{}^{\ab} + \Nh_{[a} T_{b]c}{}^{\ab} \right).
\end{align}
The terms with $w_{\alpha}w_{\beta}w_{\gamma}w_{\delta}$ are proportional to
\begin{align}\label{GG46}
    w_\a w_\b w_\g w_\r P^{\a\overline{\sigma}} P^{\b\taub} P^{\g\overline\kappa} P^{\r\overline\lambda} (\g^a)_{\overline{\sigma}\taub} (\g_a)_{\overline\kappa\overline\lambda} ,
\end{align}
which is zero due to the $D=10$ Fierz identity $(\g^a)_{(\overline{\sigma}\taub} (\g_a)_{\overline\kappa)\overline\lambda}=0$. The terms with $w_{\alpha}w_{\beta}\wb_{\hgamma}\wb_{\hdelta}$ determine the background superfield $M^{\a\b\gb\rb}$ of the action to be
\begin{align}\label{GG47}
    M^{\a\b\gb\rb}=& -\frac12 P^{\a\overline{\sigma}} P^{\b\taub} \left( \N_{(\overline{\sigma}} \Yh_{\taub)}{}^{\gb\rb} + T_{\overline{\sigma}\taub}{}^A \Yh_A{}^{\gb\rb} \right) - \frac12
    P^{(\a|\overline{\sigma}} \N_{\overline{\sigma}} Z^{\b)\gb\rb} + Z^{\s\gb\rb} Y_\s{}^{\a\b} \cr 
    &-\frac12 P^{\a\overline{\sigma}} P^{\b\taub} \Ch_{\overline{\sigma} a}{}^{\gb} \Ch_{\taub b}{}^{\rb} \eta^{ab} - \frac12 V_a{}^{\a\gb} V_b{}^{\b\rb} \eta^{ab} +  P^{(\a\overline{\sigma}} \Yh_{\overline{\sigma}}{}^{\gb\taub}D_{\taub}{}^{\rb\b)} \cr 
    &+\frac14 P^{(\a|\overline{\sigma}} V_a{}^{\b)(\gb} \Ch_{\overline{\sigma} b}{}^{\rb)} \eta^{ab} - \frac{1}{2}D_{\overline{\sigma}}{}^{(\gb\a} Z^{\b\rb)\overline{\sigma}}.
\end{align}
The terms with $\psib^a\psib^b w_{\gamma}w_{\delta}$ determine the background superfied $\Jh_{ab}{}^{\a\b}$ as
\begin{align}\label{GG48}
    \Jh_{ab}{}^{\a\b}&= -\frac12 P^{(\a|\gb} \Nh_{\gb} \Uh_{ab}{}^{\b)} + \Uh_{ab}{}^\g Y_\g{}^{\a\b} + \Uh_{ac}{}^{(\a} \Uh_{bd}{}^{\b)} \eta^{cd} + \frac14 P^{\a\gb} P^{\b\rb} \Rh_{\gb\rb ab}.
\end{align}
The terms with $\L^{\alpha} w_{\beta}w_{\gamma}w_{\delta}$ amount to
\begin{align}\label{GG49}
    \L^\a w_\b w_\g w_\r P^{\b\overline{\sigma}} P^{\g\taub}R_{\overline{\sigma}\taub\a}{}^{\rho} , 
\end{align}
which vanishes because of the ``hatted'' version of the curvature constraint (\ref{GG35}). The terms with $w_{\alpha}w_{\beta}\Lb^{\hgamma}\wb_{\hdelta}$ determine the background superfield $\Gh_{\ab}{}^{\bb\g\r}$ to be
\begin{align}\label{GG50}
    \Gh_{\ab}{}^{\bb\g\r}= -\frac12 P^{(\g|\overline{\sigma}} \N_{\overline{\sigma}} D_{\ab}{}^{\bb|\r)} + \frac12 D_{\ab}{}^{\overline{\sigma}(\g} D_{\overline{\sigma}}{}^{\bb|\r)} - \frac12 P^{\g\overline{\sigma}} P^{\r\taub} \Rh_{\overline{\sigma}\taub\ab}{}^{\bb} + D_{\ab}{}^{\bb\s} Y_\s{}^{\g\r}.
\end{align}
The terms with $\psi^a w_{\beta}w_{\gamma}w_{\delta}$ are proportional to
\begin{align}\label{GG51}
    \psi^a w_\a w_\b w_\g P^{\a\rb} P^{\b\overline{\sigma}} \left( \N_{(\rb} T_{\overline{\sigma})a}{}^\g - (\g^b)_{\rb\overline{\sigma}} T_{ba}{}^\g \right), 
\end{align}
which is zero because of the Bianchi identity involving $R_{[a\rb\overline{\sigma}]}{}^\g$. The terms with $\psib^a\wb_{\hbeta} w_{\gamma}w_{\delta}$ give the background superfield $\Lh_a{}^{\gb\a\b}$ of the action as
\begin{align}\label{GG52}
    \Lh_a{}^{\gb\a\b}=&-\frac12 P^{\a\rb} P^{\b\overline{\sigma}} \left( \Nh_{(\rb} \Ch_{\overline{\sigma})a}{}^{\gb} + T_{\rb\overline{\sigma}}{}^A T_{Aa}{}^{\gb} \right) + \frac12 P^{(\a|\rb} \Nh_{\rb} V_a{}^{\b)\gb} + \Uh_{ab}{}^{(\a} V_c{}^{\b)\gb} \eta^{bc} \cr 
    &-\frac12 P^{(\a|\rb} D_{\overline{\sigma}}{}^{\gb|\b)} \Ch_{\rb a}{}^{\overline{\sigma}} + P^{(\a|\rb} \Uh_{ca}{}^{\b)} \Ch_{\rb b}{}^{\gb} \eta^{bc} + V_a{}^{\r\gb} Y_\r{}^{\a\b} +\frac12 D_{\rb}{}^{\gb(\a} V_a{}^{\b)\rb}.
\end{align}
The terms with $\psi^a\L^{\beta} w_{\gamma}w_{\delta}$ are proportional to
\begin{align}\label{GG53}
    \psi^a \L^\a w_\b w_\g P^{\b\rb} \left( \N_\a T_{\rb a}{}^\g - T_{\a a}{}^b T_{b\rb}{}^\g -R_{a\rb\a}{}^\g  \right) ,
\end{align}
which vanishes because of the Bianchi identity involving $R_{[a\rb\a]}{}^\g$. The terms with $\psi^ a\L^{\beta}\wb_{\hgamma}\wb_{\hdelta}$ are proportional to  
\begin{align}\label{GG54}
    \psi^a \L^\a \wb_{\bb} \wb_{\gb} \left( \Nh_\a \Yh_a{}^{\bb\gb} + T_{\a a}{}^{\rb} \Yh_{\rb}{}^{\bb\gb} - (\g_a)_{\a\r} Z^{\r\bb\gb} - T_{ab}{}^{\bb} T_{\a c}{}^{\gb} \eta^{bc} \right) .
\end{align}
After using the equations for $\Yh_a{}^{\bb\gb}$ and $Z^{\r\bb\gb}$ obtained from the $\big\{\hG,\hG\big\}=-2\hT$ bracket, as well as $\Omega_{\alpha bc}-\hOmega_{\alpha bc}=T_{\alpha bc}$ and the Bianchi identities involving $\HH{\nabla}_{[\alpha}T_{\beta c]}\,^{\hdelta}$ and $\nabla_{[\alpha}T_{\beta\gamma]}\,^d$, this contribution vanishes.
The terms with $\psi^a\psi^b\psi^c\L^{\delta}$ are
\begin{align}\label{GG55}
    \psi^a \psi^b \psi^c \L^\a \left( -R_{\a abc} + \frac13 \N_\a T_{abc} - \N_a T_{\a bc} - (\g_a)_{\a\b} T_{bc}{}^\b + T_{\a a}{}^d T_{dbc} \right) , 
\end{align}
which is zero because of the Bianchi for $R_{[\a ab]c}$, the Bianchi identity for $\N_{[\a} H_{abc]}$ and the constraint $T_{abc}+H_{abc}=0$.
The terms with $\psi^{a}\psib^b\psib^c\L^{\delta}$ are proportional to
\begin{align}\label{GG56}
    \psi^a \psib^b \psib^c \L^\a \left( \Rh_{\a abc} + 2 (\g_a)_{\a\b} \Uh_{bc}{}^\b \right) .
\end{align}
Since the connections $\Oh^{ab}$ and $\O^{ab}$ differ by torsion components as in Equation (\ref{Omegabc}), the two-forms $\Rh_{ab}$ and $R_{ab}$ also differ by torsion components. In particular
\begin{align}
    \Rh_{\a abc}=R_{\a abc}-\N_{[\a}T_{a]bc}+T_{\a a}{}^d T_{dbc}-T_{\a b}{}^d T_{adc}+T_{ab}{}^d T_{\a dc}+T_{\a a}{}^{\bb} \Th_{\bb bc}.
\end{align}
Recall that $R_{\a abc}$ is determined by the Bianchi identity involving $R_{[\a ab]c}$, the Bianchi identity involving $\N_{[\a}H_{abc]}$ and the constraint $H_{abc}+T_{abc}=0$. Using this and the definition of $\Uh_{bc}{}^\b$ it follows that the expression (\ref{GG56}) vanishes. The terms with $\psi^a\L^{\beta}\Lb^{\hgamma}\wb_{\hdelta}$ are proportional to
\begin{align}\label{GG57}
    \psi^a\L^\a\Lb^{\bb}\wb_{\gb}\left(\Rh_{a\a\bb}{}^{\gb}+(\g_a)_{\a\r}D_{\bb}{}^{\gb\r}\right),
\end{align}
which, after using the definition of $D_{\bb}{}^{\gb\r}$, gives the constraint for the curvature
\begin{align}\label{curvGG}
    \Rh_{a\a\bb}{}^{\gb} = -(\g_a)_{\a\r}\N_{\bb}P^{\r\gb}.
\end{align}
The terms with $\psi^a\psib^b\L^{\gamma}\wb_{\hdelta}$ are proportional to
\begin{align}\label{GG58}
    \psi^a \psib^b \L^\a \wb_{\bb} \left( \Nh_a T_{\a b}{}^{\bb} - \Nh_\a T_{ab}{}^{\bb} + T_{a\a}{}^{\gb} \Ch_{\gb b}{}^{\bb} - (\g_a)_{\a\g} V_b{}^{\g\bb} \right) ,
\end{align}
which vanishes after using the expression for $V_b\,^{\gamma\hbeta}$ in terms of supergravity superfields, together with $\hOmega_{abc}-\Omega_{abc}=\hT_{abc}$ and the Bianchi identity involving $\HH{\nabla}_{[\alpha}T_{ab]}\,^{\hbeta}$. The terms with $\L^{\alpha} w_{\beta}\Lb^{\hgamma}\wb_{\hdelta}$ determine $S_{\a\gb}{}^{\b\rb}$ as
\begin{align}\label{GG59}
    S_{\a\gb}{}^{\b\rb} &= \N_\a D_{\gb}{}^{\rb\b} + P^{\b\overline{\sigma}} \Rh_{\a\overline{\sigma}\gb}{}^{\rb}.
\end{align}
The terms with $\psib^a\psib^b\L^{\gamma} w_{\delta}$ determine the background superfied $E_{ab\a}{}^\b$ as
\begin{align}\label{GG60}
    E_{ab\a}{}^\b=\Nh_\a \Uh_{ab}{}^\b - \frac12 P^{\b\gb} \Rh_{\a\gb ab}.
\end{align}
The terms with $\L^{\alpha} w_{\beta}\psib^c\wb_{\hdelta}$ give the background superfield $F_{a\a}{}^{\b\gb}$ according to
\begin{align}\label{GG61}
    F_{a\a}{}^{\b\gb}=-\Nh_\a V_a{}^{\b\gb} + P^{\b\rb} \left( \Nh_{\rb} T_{\a a}{}^{\gb} + \Nh_\a \Ch_{\rb a}{}^{\gb} \right) + T_{\a a}{}^{\rb} D_{\rb}{}^{\gb\b} + 2 U_{ab}{}^\b T_{\a c}{}^{\gb} \eta^{bc}.
\end{align}
The terms with $\L^{\alpha} w_{\beta}\wb_{\hgamma}\wb_{\hdelta}$ define the background superfield $G_\a{}^{\b\gb\rb}$ as
\begin{align}\label{GG62}
    G_\a{}^{\b\gb\rb}= \N_\a Z^{\b\gb\rb} + P^{\b\overline{\sigma}} \N_\a \Yh_{\overline{\sigma}}{}^{\gb\rb} - T_{\a a}{}^{\gb} V_b{}^{\b\rb} \eta^{ab} + P^{\b\overline{\sigma}} T_{\a a}{}^{\gb}\Ch_{\overline{\sigma} b}{}^{\rb} \eta^{ab}.
\end{align}
The terms with $\psi^a w_{\beta}\Lb^{\hgamma}\wb_{\hdelta}$ determine the background superfield $\Fh_{a\bb}{}^{\gb\a}$ to be
\begin{align}\label{GG63}
    \Fh_{a\bb}{}^{\gb\a}=-\N_a D_{\bb}{}^{\gb\a} + D_{\bb}{}^{\gb\r} C_{\r a}{}^\a - P^{\a\rb}\Rh_{a\rb\bb}{}^{\gb}.
\end{align}
The terms with $\psi^a w_{\beta}\psib^c\psib^d$ give the background superfield $\Ih_{bca}{}^\a$ as
\begin{align}\label{GG64}
    \Ih_{bca}{}^\a=-\Nh_a \Uh_{bc}{}^\a + \Uh_{bc}{}^\b C_{\b a}{}^\a - \frac12 P^{\a\bb} \Rh_{\bb abc}.
\end{align}
The terms with $\psi^a w_{\beta}\psib^c\wb_{\hdelta}$ determine the background superfield $K_{ab}{}^{\a\bb}$ as
\begin{align}\label{GG65}
    K_{ab}{}^{\a\bb}=& -\Nh_a V_b{}^{\a\bb} + V_b{}^{\g\bb} C_{\g a}{}^\a + P^{\a\gb} \left( \Nh_a \Ch_{\gb b}{}^{\bb} - \Nh_{\gb} T_{ab}{}^{\bb} + T_{a\gb}{}^c T_{cb}{}^{\bb} + T_{a\gb}{}^\r T_{\r b}{}^{\bb} \right) \cr 
    &+ T_{ab}{}^{\gb} D_{\gb}{}^{\bb\a} - 2 T_{ac}{}^{\bb} U_{db}{}^\a \eta^{cd} .
\end{align}
The terms with $\psi^a w_{\beta}\wb_{\hgamma}\wb_{\hdelta}$ give the background superfield $L_a{}^{\a\bb\gb}$ as
\begin{align}\label{GG66}
    L_a{}^{\a\bb\gb}=& -\N_a Z^{\a\bb\gb} + Z^{\r\bb\gb} C_{\r a}{}^\a + P^{\a\rb} \left( \Nh_{[\rb} \Yh_{a]}{}^{\bb\gb} + \Th_{\rb a}{}^b \Yh_b{}^{\bb\gb} - \frac12 T_{ab}{}^{(\bb} \Ch_{\rb c}{}^{\gb)} \eta^{bc} \right) \cr 
    & + \Yh_a{}^{\rb(\bb} D_{\rb}{}^{\gb)\a} + \frac12 V_b{}^{\a(\bb} T_{ac}{}^{\gb)} \eta^{bc} .
\end{align}
The terms with $\psi^a\L^{\beta}\L^{\gamma}\L^{\delta}$ are proportional to 
\begin{align}\label{GG67}
    \psi^a \L^\a \L^\b \L^\g \left( \N_\a T_{\b\g a} - T_{\a\b}{}^b T_{\g ab} \right) ,
\end{align}
which is zero because of the Bianchi identity involving $R_{(\a\b\g)a}$. Finally, the terms with $\Lambda^{\alpha}\Lambda^{\beta}\Lambda^{\gamma}\Lambda^{\delta}$ are proportional to
\begin{equation}
    \Lambda^{\alpha}\Lambda^{\beta}\Lambda^{\gamma}\Lambda^{\delta}(\gamma^e)_{\alpha\beta}(\gamma_e)_{\gamma\delta},
\end{equation}
which vanishes due to the Fierz identity $(\gamma^a)_{(\alpha\beta}(\gamma_a)_{\gamma)\delta}=0$.

Unsurprisingly, the analysis of the terms with four worldsheet fields coming from $\big\{\hG,\hG\big\}=-2\hT$ is analogous and gives the ``hatted'' version of the constraints and definitions above. Besides the constraints for the torsion and the $H$ components given in (\ref{TH}), we have also obtained the constraints for the curvature
\begin{align}\label{curvatures}
    &\Rh_{\a\b ab} = 0,\quad \Rh_{\a\b\gb}{}^{\rb} = 0,\quad R_{\ab\bb ab} = 0,\quad R_{\ab\bb\g}{}^\r = 0 ,\cr 
    &\Rh_{a\a\bb}{}^{\gb} = -(\g_a)_{\a\r} \N_{\bb}P^{\r\gb} ,\quad R_{a\ab\b}{}^\g = (\g_a)_{\ab\rb} \N_\b P^{\g\rb}.
\end{align}
Note that these equations are just consequences of Bianchi identities and the constraints in (\ref{TH}). We have also determined the  remaining background superfields of the action (\ref{action}). Note, however, that the superfields $\big\{S,E,\HH{E},F,\HH{F},G,\HH{G},H,I,\HH{I},J,\HH{J},K,L,\HH{L},M\big\}$ have two definitions, which come either from $\big\{G,G\big\}=-2T$ or $\big\{\hG,\hG\big\}=-2\hT$. Fortunately, the supergravity constraints (\ref{TH}), (\ref{curvatures}) and the definitions of the background superfields $\big\{C,Y,D,U,V,Z\big\}$ imply that both definitions are equivalent. For example, consider $\Eh_{ab\a}{}^{\bb}$ as given by the $\big\{\hG,\hG\big\}=-2\hT$ Poisson bracket
\begin{align}\label{GhGh60p}
    \Eh_{ab\ab}{}^{\bb}=\N_{\ab} U_{ab}{}^{\bb} + \frac12 P^{\g\bb} R_{\g\ab ab}.
\end{align}
Using the definitions of $U_{ab}{}^{\bb}$ and $D_{\ab}{}^{\bb\g}$ found previously, this expression becomes
\begin{align}\label{EEh}
    \Eh_{ab\ab}{}^{\bb}=-\frac12\N_{\ab} T_{ab}{}^{\bb} - \frac12 D_{\ab}{}^{\bb\g} T_{\g ab} - \frac12 P^{\g\bb} \left( \N_{\ab} T_{\g ab} + R_{\g\ab ab} \right).
\end{align}
Then, using the Bianchi identity involving $\Rh_{[\ab ab]}{}^{\bb}$ and the Bianchi identity involving $R_{[\g\ab a]b}$, one obtains
\begin{align}\label{GG41p}
    \Eh_{ab\ab}{}^{\bb}=- \frac12 \left( \Rh_{ab\ab}{}^{\bb} + D_{\ab}{}^{\bb\g} T_{\g ab} \right),
\end{align}
which corresponds to the alternative definition (\ref{GG41}) of $\Eh_{ab\ab}{}^{\bb}$ given by $\big\{G,G\big\}=-2T$. A similar analysis shows that the other background superfields of the action are uniquely defined.

\subsection{Simplifying constraints}
Before we end this section, we note that the expressions which determine the background superfields in the action and in the supercurrents found above can be simplified, since we have not fixed all of the gauge symmetries in (\ref{g1}) to (\ref{gf}). Indeed, we may use the transformations parameterized by $\rho_{\alpha b}\,^{\gamma},\hrho_{\halpha b}\,^{\hgamma},\WT{\rho}_{\alpha}\,^{\beta\gamma}$ and $\HH{\WT{\rho}}_{\halpha}\,^{\hbeta\hgamma}$ to gauge away the $C_{\alpha b}\,^{\gamma},\hC_{\halpha b}\,^{\hgamma},Y_{\alpha}\,^{\beta\gamma}$ and $\hY_{\halpha}\,^{\hbeta\hgamma}$ background superfields respectively. As a consequence, they drop out of all the constraints in which they appear, simplifying several expressions. 

We will now present the final expression for the background superfields. The background fields $\big\{U,D,V,Z,C,Y\big\}$ are determined by the equations
\begin{equation}
    U_{ab}{}^{\hgamma}=\frac{1}{2}P^{\gamma\hgamma}T_{\gamma ba}-\frac{1}{2}T_{ab}\,^{\hgamma},\quad \HH{U}_{ab}{}^{\gamma}=\frac{1}{2}P^{\gamma\hgamma}\hT_{\hgamma ab}-\frac{1}{2}T_{ab}\,^{\gamma},
\end{equation}
\begin{equation}
    \hD_{\alpha}{}^{\beta\hgamma}=-\nabla_{\alpha}P^{\beta\hgamma},\quad D_{\halpha}{}^{\gamma\hbeta}=\nabla_{\halpha}P^{\gamma\hbeta},
\end{equation}
\begin{equation}
    C_{\alpha b}\,^{\gamma}=\hC_{\halpha b}\,^{\hgamma}=0,
\end{equation}
\begin{equation}
    C_{ab}\,^{\gamma}=T_{ab}\,^{\gamma},\quad C_{\halpha b}\,^{\gamma}=T_{\halpha b}\,^{\gamma},
\end{equation}
\begin{equation}
    \hC_{ab}\,^{\hgamma}=T_{ab}\,^{\hgamma},\quad \hC_{\alpha b}\,^{\hgamma}=T_{\alpha b}\,^{\hgamma},
\end{equation}
\begin{equation}
    \hV_a{}^{\beta\hgamma}=\nabla_aP^{\beta\hgamma},\quad -V_{a}{}^{\gamma\hbeta}=\nabla_aP^{\gamma\hbeta},
\end{equation}
\begin{equation}
    \HH{Z}^{\alpha\beta\hgamma}=\frac{1}{2}P^{(\alpha|\hrho}\nabla_{\hrho}P^{\beta)\hgamma},\quad Z^{\alpha\hbeta\hgamma}=\frac{1}{2}P^{\rho(\hbeta}\nabla_{\rho}P^{\alpha|\hgamma)},
\end{equation}
\begin{equation}
    Y_{\alpha}\,^{\beta\gamma}=\hY_{\halpha}\,^{\hbeta\hgamma}=0,
\end{equation}
\begin{equation}
    Y_{\halpha}\,^{\beta\gamma}=\hY_{\alpha}\,^{\hbeta\hgamma}=0,
\end{equation}
\begin{equation}
    Y_{a}\,^{\beta\gamma}=\frac{1}{2}P^{\beta\hbeta}P^{\gamma\hgamma}T_{\hgamma\hbeta a},\quad \hY_{a}\,^{\hbeta\hgamma}=\frac{1}{2}P^{\beta\hbeta}P^{\gamma\hgamma}T_{\gamma\beta a}.
\end{equation}
The supercurrents $G,\hG$ are given by
\begin{align}
G =&~ \frac12\psi^a(\L\g_a\L) + \L^\a d_\a + \psi^a \Pi_a + w_\a \Pi^\a  + \frac12 \L^\a \psi^a \psi^b T_{\a ab} -\frac16 \psi^a \psi^b \psi^c T_{abc}\cr
&+ \frac12 \psi^a \psi^b w_\a T_{ab}{}^\a + \psi^a w_\a w_\b Y_a{}^{\a\b} ,
\label{GisR}
\end{align}
\begin{align}
\Gh =&~ \frac12\psib^a(\Lb\g_a\Lb) + \Lb^{\ab} \dd_{\ab} + \psib^a \Pib_a + \wb_{\ab} \Pib^{\ab} + \frac12 \Lb^{\ab} \psib^a \psib^b \Th_{\ab ab} -\frac16 \psib^a \psib^b \psib^c \Th_{abc} \cr
&+ \frac12 \psib^a \psib^b \wb_{\ab} T_{ab}{}^{\ab} + \psib^a \wb_{\ab} \wb_{\bb} \Yh_a{}^{\ab\bb}.
\label{GhisR}
\end{align}
Note that this result is the same as the one for the heterotic superstring, even though in this case the superfields $Y_a\,^{\beta\gamma}$ and $\hY_a\,^{\hbeta\hgamma}$ are not zero. Finally, the remaining background superfields in the action are determined by the following equations:
\begin{equation}
    S_{\a\gb}{}^{\b\rb}= \N_\a D_{\gb}{}^{\rb\b} + P^{\b\overline\sigma} \Rh_{\a\overline\sigma\gb}{}^{\rb},
\end{equation}
\begin{equation}
    H_{abcd} = -\frac12 \Uh_{cd}{}^\a T_{\a ab} + \frac14 \Rh_{abcd},
\end{equation}
\begin{equation}
    \Eh_{ab\ab}{}^{\bb}=- \frac12 \left( \Rh_{ab\ab}{}^{\bb} + D_{\ab}{}^{\bb\g} T_{\g ab} \right) ,
\end{equation}
\begin{equation}
    E_{ab\a}{}^\b=\Nh_\a \Uh_{ab}{}^\b - \frac12 P^{\b\gb} \Rh_{\a\gb ab},
\end{equation}
\begin{equation}
    F_{a\a}{}^{\b\gb}=-\Nh_\a V_a{}^{\b\gb} + P^{\b\rb} \Nh_{\rb} T_{\a a}{}^{\gb} + T_{\a a}{}^{\rb} D_{\rb}{}^{\gb\b} + 2 U_{ab}{}^\b T_{\a c}{}^{\gb} \eta^{bc},
\end{equation}
\begin{equation}
    \Fh_{a\bb}{}^{\gb\a}=-\N_a D_{\bb}{}^{\gb\a} - P^{\a\rb}\Rh_{a\rb\bb}{}^{\gb},
\end{equation}
\begin{equation}
    G_\a{}^{\b\gb\rb}= \N_\a Z^{\b\gb\rb} -\frac{1}{2} T_{\a a}{}^{(\gb} V_b{}^{\b|\rb)} \eta^{ab},
\end{equation}
\begin{equation}
    \Gh_{\ab}{}^{\bb\g\r}= -\frac12 P^{(\g|\overline\sigma} \N_{\overline\sigma} D_{\ab}{}^{\bb|\r)} + \frac12 D_{\ab}{}^{\overline\sigma(\g} D_{\overline\sigma}{}^{\bb|\r)} - \frac12 P^{\g\overline\sigma} P^{\r\taub} \Rh_{\overline\sigma\taub\ab}{}^{\bb},
\end{equation}
\begin{equation}
    I_{abc}{}^{\ab}=-\frac12 \left( V_c{}^{\b\ab} T_{\b ab} + \Th_{ab}{}^d T_{dc}{}^{\ab} + T_{ab}{}^\b T_{\b c}{}^{\ab} + \Nh_{[a} T_{b]c}{}^{\ab} \right),
\end{equation}
\begin{equation}
    \Ih_{bca}{}^\a=-\Nh_a \Uh_{bc}{}^\a - \frac12 P^{\a\bb} \Rh_{\bb abc},
\end{equation}
\begin{equation}
    J_{ab}{}^{\ab\bb}=\frac12 \left( \frac{1}{2}T_{ac}{}^{(\ab} T_{bd}{}^{\bb)} \eta^{cd} - Z^{\g\ab\bb} T_{\g ab} - \Nh_{[a} \Yh_{b]}{}^{\ab\bb} -  \Th_{ab}{}^c \Yh_c{}^{\ab\bb}\right),
\end{equation}
\begin{equation}
    \Jh_{ab}{}^{\a\b}= -\frac12 P^{(\a|\gb} \Nh_{\gb} \Uh_{ab}{}^{\b)} + \Uh_{ac}{}^{(\a} \Uh_{bd}{}^{\b)} \eta^{cd} + \frac14 P^{\a\gb} P^{\b\rb} \Rh_{\gb\rb ab},
\end{equation}
\begin{align}
    K_{ab}{}^{\a\bb}=& -\Nh_a V_b{}^{\a\bb} + P^{\a\gb} \left( T_{a\gb}{}^c T_{cb}{}^{\bb} + T_{a\gb}{}^\r T_{\r b}{}^{\bb} - \Nh_{\gb} T_{ab}{}^{\bb} \right) + T_{ab}{}^{\gb} D_{\gb}{}^{\bb\a} - 2 T_{ac}{}^{\bb} U_{db}{}^\a \eta^{cd} .
\end{align}
\begin{align}
    L_a{}^{\a\bb\gb}=& -\N_a Z^{\a\bb\gb} + P^{\a\rb} \left( \Nh_{\rb} \Yh_{a}{}^{\bb\gb} + \Th_{\rb a}{}^b \Yh_b{}^{\bb\gb} \right) + \Yh_a{}^{\rb(\bb} D_{\rb}{}^{\gb)\a} + \frac12 V_b{}^{\a(\bb} T_{ac}{}^{\gb)} \eta^{bc} .
\end{align}
\begin{align}
    \Lh_a{}^{\gb\a\b}=&-\frac12 P^{\a\rb} P^{\b\overline\sigma} T_{\rb\overline\sigma}{}^c T_{ca}{}^{\gb} + \frac12 P^{(\a|\rb} \Nh_{\rb} V_a{}^{\b)\gb} + \Uh_{ab}{}^{(\a} V_c{}^{\b)\gb} \eta^{bc} +\frac12 D_{\rb}{}^{\gb(\a} V_a{}^{\b)\rb}.
\end{align}
\begin{align}
    M^{\a\b\gb\rb}=& -\frac12 P^{\a\overline\sigma} P^{\b\taub} T_{\overline\sigma\taub}{}^c \Yh_c{}^{\gb\rb} - \frac12
    P^{(\a|\overline\sigma} \N_{\overline\sigma} Z^{\b)\gb\rb} - \frac12 V_a{}^{\a\gb} V_b{}^{\b\rb} \eta^{ab} - \frac{1}{2}D_{\overline{\sigma}}{}^{(\gb\a} Z^{\b\rb)\overline{\sigma}}.
\end{align}
This concludes our task of determining all the background superfields in the action and in the supercurrents $G,\hG$ in terms of supergravity superfields. We find a similar behavior to that found in the analysis of the heterotic string \cite{Berkovits:2022dbm} and in the pure spinor formalism \cite{Berkovits:2001ue}. More precisely, the superfields $S_{\alpha\hbeta}\,^{\gamma\hdelta}, \hD_{\alpha}\,^{\beta\hgamma}$ and $D_{\halpha}\,^{\gamma\hbeta}$ are determined in terms of the same supergravity fields as in the pure spinor formalism, where they also appear in the action. Furthermore, we find that the background fields associated with worldsheet field factors of U(1) charge $-2$ are indeed needed in the action and in the BRST charge, and that they describe non-linear deformations around flat space, as it had already been observed in the B-RNS-GSS heterotic string.

\section{Computation of $\{G,\Gh\}=0$}\label{GGhis0}
We now prove that  the constrains derived from $\{G,G\}=-2T$ and $\{\Gh,\Gh\}=-2\Th$ imply the vanishing of
$\{G,\Gh\}$. Explicitly, we compute
\begin{align}\label{GGh0}
    \oint d\s\oint d\s'\{G(\s'),\Gh(\s)\}=0.
\end{align}
with the commutators from Appendix \ref{app1}. Again, the resulting terms can be organized into a set with three worldsheet fields, and another with four worldsheet fields. Table \ref{table1} below shows what constraints or definitions coming from the $\{G,G\}=-2T$ and $\{\Gh,\Gh\}=-2\Tb$ Poisson brackets imply the vanishing of the terms with a given factor of three-worldsheet fields. The constraints related to the vanishing of the terms with factors of $\hLambda^{\halpha}w_{\beta},\hpsi^aw_{\beta}$ and $\hLambda^{\halpha}\psi_b$ are just the ``hatted'' version of the constraints associated to $\Lambda^{\alpha}\hw_{\hbeta},\psi^a\hw_{\hbeta}$ and $\Lambda^{\alpha}\hpsi_b$ from the table.

\begin{table}[]
\caption{}
\begin{center}
\begin{tabular}{||c|c||}
\hline
    Worldsheet field factor & Associated constraint or definition \\ 
    \hline    $\Lambda^{\alpha}\hLambda^{\hbeta}\Pi^{\gamma}$ & $H_{\alpha\beta\hgamma}=0$ \\ \hline   $\Lambda^{\alpha}\hLambda^{\hbeta}\B{\Pi}^{\hgamma}$ & $H_{\alpha\hbeta\hgamma}=0$\\ \hline
    $\Lambda^{\alpha}\hLambda^{\hbeta}d_{\gamma}$ & $T_{\alpha\hbeta}\,^{\gamma}=H_{\alpha\hbeta\hgamma}=0$ \\ \hline
    $\Lambda^{\alpha}\hLambda^{\hbeta}\hd_{\hgamma}$ & $\hT_{\alpha\hbeta}\,^{\hgamma}=H_{\alpha\hbeta\gamma}=0$ \\ \hline
    $\Lambda^{\alpha}\hLambda^{\hbeta}\Pi^c$ & $T_{\alpha\hbeta c}=H_{\alpha\hbeta c}=0$ \\ \hline
    $\Lambda^{\alpha}\hLambda^{\hbeta}\B{\Pi}^c$ & $T_{\alpha\hbeta c}=H_{\alpha\hbeta c}=0$ \\ \hline
    $\psi^a\hpsi^b\Pi^{\gamma}$ & $T_{\alpha(bc)}=H_{\alpha bc}=0$ \\ \hline
    $\psi^a\hpsi^b\B{E}^{\hgamma}$ & $T_{\halpha(bc)}=H_{\halpha bc}=0$ \\ \hline
    $\psi^a\hpsi^bd_{\gamma}$ & $C_{ab}\,^{\gamma}=T_{ab}\,^{\gamma}$ \\ \hline
    $\psi^a\hpsi^b\hd_{\hgamma}$ & $\hC_{ab}^{\hgamma}=\hT_{ab}\,^{\hgamma}$ \\ \hline
    $\psi^a\hpsi^b\Pi^c$ & $T_{abc}=H_{abc}$ \\ \hline
    $\psi^a\hpsi^b\B{\Pi}^c$ & $\hT_{abc}=H_{abc}$ \\ \hline
    $w_{\alpha}\hw_{\hbeta}\Pi^{\gamma}$ & $\hD_{\alpha}^{\beta\hgamma}$ definition \\ \hline
    $w_{\alpha}\hw_{\hbeta}\B{\Pi}^{\hgamma}$ & $D_{\halpha}^{\hbeta\gamma}$ definition \\ \hline
    $w_{\alpha}\hw_{\hbeta}d_{\gamma}$ & $\hZ^{\alpha\beta\hgamma}$ definition \\ \hline
    $w_{\alpha}\hw_{\hbeta}\hd_{\hgamma}$ & $Z^{\alpha\hbeta\hgamma}$ definition \\ \hline
    $w_{\alpha}\hw_{\hbeta}\Pi^c$ & $\hV_a^{\beta\hgamma}$ definition \\ \hline
    $w_{\alpha}\hw_{\hbeta}\B{\Pi}^c$ & $V_a^{\beta\hgamma}$ definition \\ \hline
    $\Lambda^{\alpha}\hpsi^b\Pi^{\gamma}$ & $T_{\alpha\beta c}=H_{\alpha\beta c}$ \\ \hline
    $\Lambda^{\alpha}\hpsi^c\B{\Pi}^{\hgamma}$ & $T_{\alpha\hbeta c}=H_{\alpha\hbeta c}$ \\ \hline
    $\Lambda^{\alpha}\hpsi^bd_{\gamma}$ & $T_{a\beta}\,^{\gamma}=0$ \\ \hline
    $\Lambda^{\alpha}\hpsi^b\hd_{\hgamma}$ & $\hC_{\alpha b}\,^{\hgamma}=T_{\alpha b}\,^{\hgamma}$ \\ \hline
    $\Lambda^{\alpha}\hpsi^b\Pi^c$ & $T_{\alpha (bc)}=H_{\alpha bc}=0$ \\ \hline
    $\Lambda^{\alpha}\hpsi^b\B{\Pi}^c$ & $\hT_{\alpha bc}=H_{\alpha bc}=0$ \\ \hline
    $\Lambda^{\alpha}\hw_{\hbeta}\Pi^{\gamma}$ & $T_{\alpha\beta}\,^{\hgamma}=H_{\alpha\beta\gamma}=0$ \\ \hline
    $\Lambda^{\alpha}\hw_{\hbeta}\B{\Pi}^{\hgamma}$ & $\hT_{\alpha\hbeta}\,^{\hgamma}=H_{\alpha\beta\hgamma}=0$ \\ \hline
    $\Lambda^{\alpha}\hw_{\hbeta}d_{\gamma}$ & $\hD_{\alpha}^{\beta\hgamma}$ definition \\ \hline
    $\Lambda^{\alpha}\hw_{\hbeta}\hd_{\hgamma}$ & $\hY_{\alpha}\,^{\hbeta\hgamma}=0$ \\ \hline
    $\Lambda^{\alpha}\hw_{\hbeta}\Pi^c$ & $T_{\alpha b}\,^{\hgamma}=P^{\gamma\hgamma}T_{\gamma\alpha b}$ \\ \hline
    $\Lambda^{\alpha}\hw_{\hbeta}\B{\Pi}^c$ & $\hC_{\alpha b}\,^{\hgamma}=T_{\alpha b}\,^{\hgamma}$ \\ \hline
    $\psi^a\hw_{\hbeta}\Pi^{\gamma}$ & $T_{\alpha b}\,^{\hgamma}=P^{\gamma\hgamma}T_{\gamma\alpha b}$ \\ \hline
    $\psi^a\hw_{\hbeta}\B{\Pi}^{\hgamma}$ & $\hT_{a\hbeta}\,^{\hgamma}=T_{\alpha\hbeta c}=H_{\alpha\hbeta c}=0$ \\ \hline
    $\psi^a\hw_{\hbeta}d_{\gamma}$ & $\hV_a^{\beta\hgamma}$ definition \\ \hline
    $\psi^a\hw_{\hbeta}\hd_{\hgamma}$ & $\hY_a\,^{\hbeta\hgamma}=\frac{1}{2}P^{\beta\hbeta}P^{\gamma\hgamma}T_{\gamma\beta a}$ \\ \hline
    $\psi^a\hw_{\hbeta}\Pi^c$ & $U_{ab}^{\hgamma}$ definition \\ \hline
    $\psi^a\hw_{\hbeta}\B{\Pi}^c$ & $\hC_{ab}\,^{\hgamma}=\hT_{ab}\,^{\hgamma}$ \\ \hline
\end{tabular}
\end{center}
    \label{table1}
\end{table}

It remains to prove that the  terms with four worldsheet fields in (\ref{GGh0}) also vanish. We use all the constraints found so far. The terms with $\L^{\alpha}\Lb^{\hbeta} w_{\gamma}w_{\delta}$ and with $\L^{\alpha}\Lb^{\hbeta}\wb_{\hgamma}\wb_{\hdelta}$ are zero trivially. The terms with $\psi^a\psi^b\L^{\gamma}\Lb^{\hdelta}$ are proportional to
\begin{align}\label{GGh9}
    \psi^a\psi^b\L^\a\Lb^{\bb} \left( \N_{\bb}T_{\a ab}-2(\g_a)_{\a\g}T_{\bb b}{}^\g-R_{\a\bb ab} \right) ,
\end{align}
which is zero due to the Bianchi identity involving $R_{[\a\bb a]b}$. The terms with $\L^{\alpha}\Lb^{\hbeta}\L^{\gamma} w_{\delta}$ give
\begin{align}\label{GGh10}
    \L^\a\Lb^{\bb}\L^\g w_\r\left(R_{\a\bb\g}{}^\r-\frac12(\g^a)_{\a\g}T_{\bb a}{}^\r \right) ,
\end{align}
which vanishes as implied by the Bianchi identity involving $R_{(\a\bb\g)}{}^\r$. The terms with $\psi^a\L^{\beta}\Lb^{\hgamma} w_{\delta}$ are
\begin{align}\label{GGh11}
    \psi^a\L^\a\Lb^{\bb} w_\g \left(\N_\a T_{\bb a}{}^\g - T_{\a a}{}^b T_{b\bb}{}^\g -R_{a\bb\a}{}^\g\right),
\end{align}
which vanishes due to the Bianchi identity involving $R_{[a\bb\a]}{}^\g$. The terms with $\psi^a\psib^b w_{\gamma}w_{\delta}$ are
\begin{align}\label{GGh12}
    \psi^a\psib^b w_\a w_\b\Big(\N_a Y_b{}^{\a\b} + T_{ab}{}^c Y_c{}^{\a\b}+P^{\a\gb}\left(\N_b T_{\gb a}{}^\b+\N_{\gb} T_{ab}{}^\b\right)\Big).
\end{align}
Using the Bianchi identity involving $R_{[\gb ab]}{}^\b$ in the term with a factor of $P^{\a\gb}$ and using the definition for $Y_a{}^{\a\b}$, this expression vanishes. The terms with $\psib^a\psi^b\psi^c\psi^d$ amount to
\begin{align}\label{GGh13}
    \psib^a\psi^b\psi^c\psi^d \left( \frac13\N_a T_{bcd}+T_{ab}{}^\a T_{\a cd}-R_{abcd}\right) ,
\end{align}
which is zero after using $R_{a[bc]d}=R_{[abc]d}+R_{bcda}$, the Bianchi identity involving $R_{[abc]d}$, the Bianchi identity involving $R_{[bcd]a}$, the Bianchi identity with $\N_{[a}H_{bcd]}$ and the constraint $H_{abc}+T_{abc}=0$. The terms with $\psi^a\psib^b\L^{\gamma} w_{\delta}$ are
\begin{align}\label{GGh14}
    \psi^a\psib^b\L^\a w_\b \left( R_{ab\a}{}^\b -\N_\a T_{ab}{}^\b + T_{\a[a}{}^cT_{b]c}{}^\b +T_{\a[a}{}^{\gb}T_{b]\gb}{}^\b \right) ,
\end{align}
which vanishes because of the Bianchi identity involving $R_{[ab\a]}{}^\b$. The terms with $\psib^a\psi^b\psi^c w_{\delta}$ are proportional to
\begin{align}\label{GGh15}
    &\psib^c\psi^a\psi^b w_\a \left( \N_{[a} T_{bc]}{}^\a + T_{[ab}{}^A T_{Ac]}{}^\a \right)\cr
    &+\psib^c\psi^a\psi^b w_\a \left( P^{\a\bb} R_{\bb cab}+2Y_c{}^{\a\b} T_{\b ab} -T_{ab}{}^{\bb} T_{\bb c}{}^\a \right).
\end{align}
The first line vanishes because of the Bianchi identity for $R_{[abc]}{}^\a$. The second line also vanishes after using the expression for $Y_a{}^{\a\b}$, the constraint $T_{\halpha b}\,^{\gamma}=-P^{\gamma\hgamma}T_{\hgamma\halpha b}$, as well as the Bianchi identity involving $R_{[\bb ca]b}$, the Bianchi identity involving $\N_{[\bb}H_{abc]}$ and the constraint $H_{abc}+T_{abc}=0$. 
After using the definitions for $D_{\ab}{}^{\bb\g}$,  $\Zh^{\ab\b\g}$, $\Vh_a{}^{\ab\b}$ and $Y_a{}^{\a\b}$, the contributions with $w_{\alpha}w_{\beta}w_{\gamma}\wb_{\hdelta}$ amount to
\begin{align}\label{GGh17}
    &w_\a w_\b w_\g \wb_{\rb} P^{\a\overline{\sigma}} P^{\b\taub} \left(\N_{\overline{\sigma}}\N_{\taub} P^{\g\rb} - \frac12 (\g^a)_{\overline{\sigma}\taub} \N_a P^{\g\rb} \right) \cr
    &=\frac12 w_\a w_\b w_\g \wb_{\rb} P^{\a\overline{\sigma}} P^{\b\taub}\left( \{\N_{\overline{\sigma}},\N_{\taub}\} P^{\g\rb} - (\g^a)_{\overline{\sigma}\taub} \N_a P^{\g\rb} \right) \cr
    &=\frac12 w_\a w_\b w_\g \wb_{\rb} P^{\a\overline{\sigma}} P^{\b\taub} P^{\g\overline\kappa} \Rh_{\overline{\sigma}\taub\overline\kappa}{}^{\rb} ,
\end{align}
which vanishes because of the Bianchi identity involving $\Rh_{(\overline{\sigma}\taub\overline\kappa)}{}^{\rb}$. Consider the terms with $\psi^a\psi^b w_{\gamma}\wb_{\hdelta}$. After trivial cancellations they become
\begin{align}\label{GGh18}
    &\psi^a\psi^bw_\a\wb_{\bb} \left( \N_a\N_b P^{\a\bb} + \frac12 P^{\a\gb}P^{\r\bb} R_{\r\gb ab} -\frac12 P^{\a\gb}\N_{\gb}T_{ab}{}^{\bb}-\frac12  P^{\a\gb}P^{\r\bb} \N_{\gb}T_{\r ab} \right.\cr
    &+\left.P^{\g\bb} \left( -\N_\g T_{ab}{}^\a+T_{a\g}{}^c T_{cb}{}^\a + T_{a\g}{}^{\rb} T_{\rb b}{}^\a \right) + \frac12 T_{ab}{}^A \N_A P^{\a\bb} + \frac12 P^{\g\bb} \N_\g T_{ab}{}^\a \right).
\end{align}
Commuting the derivatives in the first term leads to 
\begin{align}\label{GGh19}
    &\frac12\psi^a\psi^bw_\a\wb_{\bb} \left( P^{\g\bb} \left( R_{ab\g}{}^\a - \N_\g T_{ab}{}^\a + T_{\g[a}{}^c T_{b]c}{}^\a + T_{\g[a}{}^{\rb} T_{b]\rb}{}^\a \right) \right.\cr
    &+\left. P^{\a\gb} \left( \Rh_{ab \gb}{}^{\bb} - \N_{\gb} T_{ab}{}^{\bb} \right) + P^{\a\gb}P^{\r\bb} \left( R_{\r\gb ab}-\N_{\gb} T_{\r ab} \right) \right) .
\end{align}
The first line is zero due to the Bianchi identity involving $R_{[ab\g]}{}^\a$. The second line also vanishes after using the Bianchi identities involving $\Rh_{[ab\hgamma]}\,^{\hdelta}$ and $R_{[\alpha\hbeta c]d}$. The terms with $\L^{\alpha} w_{\beta}w_{\gamma}\wb_{\hdelta}$, after some trivial cancellations, amount to
\begin{align}\label{GGh20}
    &\L^\a w_\b w_\g\wb_{\rb} \left( P^{\b\overline{\sigma}} \{\N_\a,\N_{\overline{\sigma}}\} P^{\g\rb} + P^{\b\overline{\sigma}} P^{\tau\rb} \left( R_{\overline{\sigma}\tau\a}{}^\g + \frac12 (\g^a)_{\a\tau} T_{a\overline{\sigma}}{}^\g \right) \right). 
\end{align}
Computing the anticommutator one obtains
\begin{align}\label{GGh21}
    \L^\a w_\b w_\g\wb_{\rb} \left( P^{\b\overline{\sigma}} P^{\g\taub} \Rh_{\a(\overline{\sigma}\taub)}{}^{\rb} + P^{\b\overline{\sigma}} P^{\tau\rb} \left( R_{\overline{\sigma}(\a\tau)}{}^\g +  \frac12 (\g^a)_{\a\tau} T_{a\overline{\sigma}}{}^\g \right) \right).
\end{align}
The first and second terms vanish due to the Bianchi identities for $\Rh_{(\a\overline{\sigma}\taub)}{}^{\rb}$ and $R_{(\overline{\sigma}\a\tau)}{}^\g$ respectively. The terms with $\psi^a w_{\beta}w_{\gamma}\wb_{\hdelta}$ amount to
\begin{align}\label{GGh22}
    &\psi^a w_\a w_\b \wb_{\gb} \left( P^{\a\rb}[\N_a,\N_{\rb}] P^{\b\gb} + P^{\a\rb} T_{a\rb}{}^\s \N_\s P^{\b\gb} - P^{\a\rb}P^{\s\gb} \N_\s T_{\rb a}{}^\b \right. \cr
    &+\left. P^{\a\rb} P^{\s\gb} T_{\s a}{}^b T_{b\rb}{}^\b - P^{\a\rb} T_{\rb b}{}^\b T_{ac}{}^{\gb} \eta^{bc} - T_{ab}{}^{\gb} Y_c{}^{\a\b} \eta^{bc} \right) .
\end{align}
After computing the commutator we obtain
\begin{align}\label{GGh23}
    \psi^a w_\a w_\b \wb_{\gb} &\left( P^{\a\rb} P^{\s\gb}  \left( R_{a\rb\s}{}^\b - \N_\s T_{\rb a}{}^\b + T_{\sigma a}{}^b T_{b\rb}{}^\b \right) \right.\cr 
    &+\left.\frac12 P^{\a\rb} P^{\beta\overline{\sigma}} \left( \Rh_{a(\rb\overline{\sigma})}{}^{\gb} + (\g^b)_{\rb\overline{\sigma}} T_{ba}{}^{\gb} \right) \right).
\end{align}
The first and second lines vanish because of the Bianchi identities involving $R_{[a\rb\s]}{}^\b$ and $\Rh_{[a\rb\overline{\sigma}]}{}^{\gb}$ respectively. The terms with $\psib^a\L^{\beta} w_{\gamma}w_{\delta}$ are
\begin{align}\label{GGh24}
    \psib^a\L^\a w_\b w_\g \left( \N_\a Y_a{}^{\b\g} + T_{\a a}{}^b Y_b{}^{\b\g} + P^{\b\rb} R_{a\rb\a}{}^\g \right) .
\end{align}
Using the definition for $Y_a{}^{\b\g}$, as well as the Bianchi identity involving $R_{[a\rb\a]}{}^\g$, this expression vanishes. The terms with $\psib^a\L^{\beta}\wb_{\hgamma}\wb_{\hdelta}$ are
\begin{align}\label{GGh25}
    \psib^a\L^\a\wb_{\bb}\wb_{\gb}P^{\r\bb}\left( \Nh_{(\a} T_{\r)a}{}^{\gb} - (\g^b)_{\a\r} T_{ba}{}^{\gb} \right),  
\end{align}
which amounts zero as implied by the Bianchi identity involving $\Rh_{[a\a\r]}{}^{\gb}$. Consider the terms with $\psib^a\psi^b\psi^c\L^{\delta}$. They amount to
\begin{align}\label{GGh26}
    \psib^a\psi^b\psi^c\L^\a \left( R_{\a abc} + \N_a T_{\a bc} -2(\g_b)_{\a\b} T_{ac}{}^\b \right).
\end{align}
This expression is zero as implied by the Bianchi identity for $R_{[\a ab]c}$, the Bianchi identity for $\N_{[\a}H_{abc]}$ and the constraint $H_{abc}+T_{abc}=0$. Consider now the terms with $\psib^a\psib^b\psib^c\L^{\delta}$, which are proportional to
\begin{align}\label{GGh27}
    \psib^a\psib^b\psib^c\L^\a \left( \Rh_{\a abc} - \frac13 \Nh_\a\Th_{abc} - T_{\a a}{}^{\bb} \Th_{\bb bc} \right).
\end{align}
The Bianchi identity for $R_{[\a ab]c}$, the Bianchi identity for $\N_{[\a}H_{abc]}$ and the constraint $H_{abc}-\Th_{abc}=0$ imply that this contribution vanishes. Let us consider the terms with $\psib^a\L^{\beta}\L^{\gamma} w_{\delta}$ in (\ref{GGh0}). They amount to
\begin{align}\label{GGh28}
    \psib^a\L^\a\L^\b w_\g \left( R_{a(\a\b)}{}^\g - (\g^b)_{\a\b} T_{ab}{}^\g \right), 
\end{align}
which is zero as implied by the Bianchi identity involving $R_{[a\a\b]}{}^\g$. The terms with $\psib^a\psib^b\L^{\gamma}\wb_{\hdelta}$ lead to
\begin{align}\label{GGh29}
    \psib^a\psib^b\L^\a\wb_{\bb} \left( \Nh_{[a} T_{\a b]}{}^{\bb} - \Th_{ab}{}^c T_{c\a}{}^{\bb} \right). 
\end{align}
This contribution vanishes because of the Bianchi identity involving $\Rh_{[a\a b]}{}^{\bb}$. The terms with $\L^{\alpha}\wb_{\hbeta}\wb_{\hgamma}\wb_{\hdelta}$ sum up to
\begin{align}\label{GGh30}
    \L^\a\wb_{\bb}\wb_{\gb}\wb_{\rb} \left( P^{\s\bb} T_{\a a}{}^{\gb} T_{\s b}{}^{\rb} \eta^{ab} - P^{\s\bb} (\g^a)_{\a\s} \Yh_a{}^{\gb\rb} - T_{\a a}{}^{\bb} \Yh_b{}^{\gb\rb} \eta^{ab} \right). 
\end{align}
The definition for $\Yh_a{}^{\ab\bb}$ implies that this expression is equal to
\begin{align}\label{GGh31}
    \L^\a\wb_{\bb}\wb_{\gb}\wb_{\rb} P^{\s\bb} P^{\tau\gb} P^{\kappa\rb} (\g^a)_{\a\tau} (\g_a)_{\s\kappa} ,
\end{align}
which vanishes because of the Fierz identity $(\g^a)_{\a(\tau} (\g_a)_{\s\kappa)}=0$. The terms with the factor of $\psi^a\psi^b\L^{\gamma}\wb_{\hdelta}$ are proportional to
\begin{align}\label{GGh32}
    \psi^a\psi^b\L^\a\wb_{\bb} &\left( P^{\g\bb} \left( \N_{(\a} T_{\g)ab}-T_{\a a}{}^c T_{c\g b} + T_{\a b}{}^c T_{c\g a} - R_{\a\g ab} \right) \right.\cr
    &\quad+\left.\left( \N_{[\a} T_{ab]}{}^{\bb}+T_{\a a}{}^c T_{cb}{}^{\bb}-T_{\a b}{}^c T_{ca}{}^{\bb} \right) \right). 
\end{align}
Using the Bianchi identity involving $R_{[\a ab]}{}^{\bb}$ in the second line and combining the result with the first line leads to the Bianchi identity involving $R_{[\a\g a]b}$. This expression is therefore equal to zero. The terms with $\L^{\alpha}\L^{\beta} w_{\gamma}\wb_{\hdelta}$ are proportional to
\begin{align}\label{GGh32p}
    \L^\a\L^\b w_\g\wb_{\rb} \left( \{\N_\a,\N_\b\}P^{\g\rb}-(\g^a)_{\a\b}\N_a P^{\g\rb} -P^{\s\rb} R_{\s(\a\b)}{}^\g \right).
\end{align}
Computing the anti-commutator and imposing the curvature constraint for $\Rh_{a\alpha\hbeta}\,^{\hgamma}$ implies that this expression becomes
\begin{align}\label{GGh33}
    \L^\a\L^\b w_\g\wb_{\rb} P^{\s\rb} R_{(\s\a\b)}{}^\g,
\end{align}
which is zero because of the Bianchi identity involving $R_{(\s\a\b)}{}^\g$. The terms with $\psi^a\L^{\beta} w_{\gamma}\wb_{\hdelta}$, after some trivial cancellations, lead to
\begin{align}\label{GGh34}
    \psi^a\L^\a w_\b\wb_{\gb} &\Big( [\N_\a,\N_a]P^{\b\gb} + T_{\a a}{}^b \N_b P^{\b\gb} -P^{\r\gb} \left((\g^b)_{\a\r} T_{ba}{}^\b + R_{a\r\a}{}^\b\right)\cr
    &\quad-(\g_a)_{\a\s}\left( P^{\b\rb} \N_{\rb} P^{\s\gb}+P^{\s\rb}\N_{\rb}P^{\b\gb}\right) \Big).
\end{align}
After computing the commutator, this expression becomes
\begin{align}\label{GGh35}
    \psi^a\L^\a w_\b\wb_{\gb} &\Big( P^{\r\gb} \left( -R_{a(\a\r)}{}^\b-(\g^b)_{\a\r} T_{ba}{}^\b\right)\cr
    &\quad+ P^{\b\rb} \left( \Rh_{\a a\rb}{}^{\gb} - (\g_a)_{\a\s}\N_{\rb} P^{\s\gb} \right) \Big).
\end{align}
The first line vanishes because of the Bianchi identity involving $R_{[a\a\r]}{}^\b$ and the second line vanishes because of the curvature constraint for $\Rh_{\alpha a\hrho}\,^{\hgamma}$. The terms with a factor of $\psi^a\psi^b\psi^c\wb_{\hdelta}$ are 
\begin{align}\label{GGh36}
    \psi^a\psi^b\psi^c\wb_{\ab}&\left( P^{\b\ab} \left( \frac13\N_\b T_{abc} - \N_a T_{\b bc} - T_{ab}{}^d T_{\b dc} - R_{\b abc} \right) - \left( \N_a T_{bc}{}^{\ab}+T_{ab}{}^dT_{dc}{}^{\ab} \right) \right),
\end{align}
which is zero after using the Bianchi identities involving $\nabla_{[a}T_{bc]}\,^{\hdelta},\nabla_{[\alpha}H_{bcd]}$ and $R_{[\alpha bc]d}$. The terms with factor of $\psib^a\L^{\beta}\wb_{\hgamma}\wb_{\hdelta}$ in (\ref{GGh0}) are
\begin{align}\label{GGh37}
    \psib^a\L^\a\wb_{\bb}\wb_{\gb} P^{\r\bb} \left( \Nh_{(\a} T_{\r)a}{}^{\gb} - (\g^b)_{\a\r} T_{ba}{}^{\gb} \right),
\end{align}
which vanishes because of the Bianchi identity for $\Rh_{[\a\r a]}{}^{\gb}$. The terms with the factor $\psib^a\L^{\beta}\L^{\gamma}w_{\delta}$ in (\ref{GGh0}) are
\begin{align}\label{GGh37p}
    \psib^a\L^\a\L^\b w_\g \left( -R_{a\a\b}{}^\g +\frac12 (\g^b)_{\a\b} T_{ab}{}^\g \right),
\end{align}
which is zero because of the Bianchi identity involving $R_{[a\a\b]}{}^\g$. The terms with the factor $\psib^a w_{\beta}w_{\gamma}w_{\delta}$ in (\ref{GGh0}) are
\begin{align}\label{GGh38}
    \psib^a w_\a w_\b w_\g \left( -P^{\a\rb} \N_{\rb} Y_a{}^{\b\g} + T_{ab}{}^\a Y_c{}^{\b\g} \eta^{bc} + P^{\a\rb} T_{ab}{}^\b T_{\rb c}{}^\g \eta^{bc} \right).
\end{align}
Using the $Y_a{}^{\a\b}$ definition we obtain
\begin{align}\label{GGh39}
    \frac12\psib^a w_\a w_\b w_\g &\left( P^{\a\rb} P^{\b\overline{\sigma}} \left( \N_{(\rb} T_{\overline{\sigma})a}{}^\g +(\g^b)_{\rb\overline{\sigma}} T_{ab}{}^\g \right) \right.\cr
    &\quad +\left. P^{\a\rb}P^{\b\overline{\sigma}} P^{\g\taub} \left( 2\Oh_{\rb} (\g_a)_{\overline{\sigma}\taub} - \Th_{\rb a}{}^b (\g_b)_{\overline{\sigma}\taub} \right) \right).
\end{align}
The first line vanishes because of the Bianchi identity for $\nabla_{(\hrho}T_{\hsigma)a}\,^{\gamma}$ and the second line is also zero after noticing that the supergravity constraints imply $\Th_{\ab ab}=2(\g_{ab}\Oh)_{\ab}$ \cite{Berkovits:2001ue}. The terms with factor of $\psi^a\Lb^{\hbeta}\L^{\gamma}\L^{\delta}$ in (\ref{GGh0}) are proportional to
\begin{align}\label{GGh40}
    \psi^a \Lb^{\ab} \L^\b \L^\g \N_{\ab} T_{\b\g a} ,
\end{align}
which is zero because of the Bianchi identity for $R_{[\ab\b\g]a}$. The terms with factor of $\L^{\alpha}\L^{\beta}\psi^c\psib^d$ and the terms with $\Lb^{\halpha}\Lb^{\hbeta}\psib^c\psi^d$ (\ref{GGh0}) are trivially zero. Finally, the terms with factor of $\psi^a\L^{\beta}\L^{\gamma}\wb_{\hdelta}$ in (\ref{GGh0})  are 
\begin{align}\label{GGh41}
    \psi^a\L^\a\L^\b\wb_{\gb} \left( P^{\r\gb} \O_\r(\gamma_a)_{\alpha\beta} +\frac12(\g^b)_{\a\b} \left( T_{ab}{}^{\gb}+P^{\r\gb} T_{\r ab} \right) + (\g_a)_{\a\r} \N_\b P^{\r\gb} \right).
\end{align}
Using $T_{\a ab}=2(\g_{ab}\O)_\a$ as  implied by the supergravity constraints \cite{Berkovits:2001ue} this expression becomes proportional to 
\begin{align}\label{GGh42}
    \psi^a\L^\a\L^\b\wb_{\gb} P^{\r\gb} (\g^b)_{(\a\b} (\g_b\g_a\O)_{\r)},
\end{align}
which vanishes due to the Fierz identity $(\g^b)_{(\a\b} (\g_b)_{\r)\s}=0$. 

Above, we have covered only half of the contributions with four worldsheet fields one obtains when computing $\big\{G,\hG\big\}$. The missing half is the ``hatted'' version of the contributions we have written, that is, we detailed the vanishing of the contribution with a factor of $\psi^a\Lambda^{\beta}\Lambda^{\gamma}\hw_{\hdelta}$, but not the vanishing of $\hpsi^a\hLambda^{\hbeta}\hLambda^{\hgamma}w_{\delta}$. Unsurprisingly, the reason why these omitted contributions vanish is analogous to the reason why their ``unhatted'' counterparts shown here vanish. This concludes the proof that the constraints from $\{G,G\}=-2T$ and $\{\Gh,\Gh\}=-2\Th$ imply $\{G,\Gh\}=0$. 

\section{Holomorphicity and anti-holomorphicity of $G,\hG$} \label{HOL}
In the previous sections we have shown that requiring the conditions $\big\{G,G\big\}=-2T$ and $\big\{\hG,\hG\big\}=-2\Th$ imply $\big\{G,\hG\big\}=0$ and, more importantly, constraints for the Type II supergravity fields. In the present section we will show that these three Poisson Brackets imply  
\begin{equation}
    \B{\partial}G=0,\quad \partial\hG=0.
\end{equation}
Here we should recall the notation that Poisson brackets with omitted $\sigma, \sigma'$ imply integration over these variables. More precisely:
\begin{equation}
    \big\{G,G\big\}\equiv\oint{d\sigma}\oint{d\sigma'}\big\{G(\sigma),G(\sigma')\big\},\quad \big\{T,\alpha(\sigma')\big\}\equiv\oint{d\sigma}\big\{T(\sigma),\alpha(\sigma')\big\}
\end{equation}
and so on.

The first step in the proof is to use the Jacobi identity to write
\begin{equation} \label{Jacobi}
    \big[\hG,\big\{\hG,fG\big\}\big]+\big[fG,\big\{\hG,\hG\big\}\big]+\big[\hG,\big\{fG,\hG\big\}\big]=0,
\end{equation}
where $f(\sigma)$ is some function of $\sigma$ which does not depend on any of the fields in our action. Then, using the aforementioned anti-commutators, we find that this equation reduces to 
\begin{equation}
    \big[\Th,fG\big]=0.
\end{equation}
Here, note that $\big\{\hG,fG\big\}=0$ is true even with the inclusion of $f(\sigma)$. The analogous computation switching $G\to \hG,\ \hG\to G$ and $\hT\to T$ yields the result
\begin{equation}
    \big[T,f\hG\big]=0.
\end{equation}
It is clear that once we show
\begin{equation} \label{TgenD}
    \big[\hT,f G\big]=f\B{\partial}G\equiv\oint\frac{d\sigma}{2\pi}f(\sigma)\B{\partial}G(\sigma),\quad \big[T,f\hG\big]=f\partial\hG\equiv\oint\frac{d\sigma}{2\pi}f(\sigma)\partial\hG(\sigma),
\end{equation}
we are done with the proof, as the computation holds for any $f(\sigma)$. Indeed, here we see the importance of being able to have non-constant $f(\sigma)$: were we not, we would conclude only that $\partial_{0}G(\sigma)=0$ because the piece $\partial_{1}G(\sigma)$ would appear as a total derivative (remember that we are always dealing with an integral over $\sigma$).

In order to prove (\ref{TgenD}), we need to show that
\begin{equation} \label{commshol}
    \big[\hT,\varphi(\sigma)\big]=\B{\partial}\varphi(\sigma),\quad \big[\hT,\Phi(Z)(\sigma)\big]=\B{\partial}\Phi(Z)(\sigma),\quad \big[\hT,f d_A\big]=f\B{\partial} d_A,
\end{equation}
\begin{equation}
    \big[T,\varphi(\sigma)\big]=\partial\varphi(\sigma),\quad \big[T,\Phi(Z)(\sigma)\big]=\partial\Phi(Z)(\sigma), \quad \big[T,f d_A\big]=f\partial d_A,
\end{equation}
where $\varphi\in\big\{\psi^a,\Lambda^{\alpha},w_{\alpha},\hpsi^a,\hLambda^{\halpha},\hw_{\halpha}\big\}$, $\Phi(Z)$ is some function of the supercoordinates $Z^M$ and $d_A\in\big\{d_a,d_{\alpha},\hd_{\halpha}\big\}$. We will focus on the proof for the brackets that involve $\hT$ (the proof for the ones with $T$ is analogous).

Consider then using the equations of motion for $\hpsi^a,\hLambda^{\halpha}$ and $\hw_{\halpha}$ to rewrite $\hT$ as
\begin{equation}
\begin{split}
    \hT=&-\frac{1}{2}\B{\Pi}^a\B{\Pi}_a-\hd_{\halpha}\B{\Pi}^{\halpha}-\hpsi^a\partial_{1}\hpsi_a-\hw_{\halpha}\partial_{1}\hLambda^{\halpha}+\hLambda^{\halpha}\partial_{1}\hw_{\halpha}\\
    &+\hpsi^a\hpsi^b\partial_{1}Z^M\hOmega_{Mab}-2\hLambda^{\halpha}\hw_{\hbeta}\partial_{1}Z^M\hOmega_{M\halpha}\,^{\hbeta}-2\hpsi^a\hw_{\hbeta}\partial_{1} Z^M\hC_{Ma}\,^{\hbeta}-2\hw_{\halpha}\hw_{\hbeta}\partial_{1}Z^M\hY_M\,^{\halpha\hbeta}\\
    &+d_{\alpha}\hpsi^b\hpsi^c\HH{U}_{bc}{}^{\alpha}+d_{\alpha}\hLambda^{\hbeta}\hw_{\hgamma}D_{\hbeta}{}^{\hgamma\alpha}+d_{\alpha}\hpsi^b\hw_{\gamma}V_b{}^{\alpha\hgamma}+d_{\alpha}\hw_{\hbeta}\hw_{\hgamma}Z^{\alpha\hbeta\hgamma}\\
    &+\psi^a\psi^b\hpsi^c\hpsi^dH_{abcd}+\psi^a\psi^b\hLambda^{\hgamma}\hw_{\hdelta}\HH{E}_{ab\hgamma}{}^{\hdelta}+\Lambda^{\alpha}w_{\beta}\hpsi^c\hpsi^dE_{ cd\alpha}{}^{\beta}+\Lambda^{\alpha}w_{\beta}\hLambda^{\hgamma}\hw_{\hdelta}S_{\alpha\hgamma}{}^{\beta\hdelta}\\
    &+\psi^a\psi^b\hpsi^c\hw_{\hdelta}I_{abc}{}^{\hdelta}+\psi^a\psi^b\hw_{\hgamma}\hw_{\hdelta}J_{ab}{}^{\hgamma\hdelta}+\psi^aw_{\beta}\hpsi^c\hpsi^d\HH{I}_{acd}{}^{\beta}+w_{\alpha}w_{\beta}\hpsi^c\hpsi^d\HH{J}_{cd}{}^{\alpha\beta}\\
    &+\Lambda^{\alpha}w_{\beta}\hpsi^c\hw_{\hdelta}F_{c\alpha}{}^{\beta\hdelta}+\Lambda^{\alpha}w_{\beta}\hw_{\hgamma}\hw_{\hdelta}G_{\alpha}{}^{\beta\hgamma\hdelta}+\psi^aw_{\beta}\hLambda^{\hgamma}\hw_{\hdelta}\HH{F}_{a\hgamma}{}^{\hdelta\beta}+w_{\alpha}w_{\beta}\hLambda^{\hgamma}\hw_{\hdelta}\HH{G}_{\hgamma}{}^{\hdelta\alpha\beta}\\
    &+\psi^aw_{\beta}\hpsi^c\hw_{\hdelta}K_{ac}{}^{\beta\hdelta}+\psi^aw_{\beta}\hw_{\hgamma}\hw_{\hdelta}L_{a}{}^{\beta\hgamma\hdelta}+w_{\alpha}w_{\beta}\hpsi^c\hw_{\hdelta}\HH{L}_{c}{}^{\alpha\beta\hdelta}+w_{\alpha}w_{\beta}\hw_{\hgamma}\hw_{\hdelta}M^{\alpha\beta\hgamma\hdelta}.
\end{split}
\end{equation}
Let us start with the case for $\big[\hT,\hw_{\halpha}\big]$. Using the commutators listed in the appendix, one can show that the bracket reduces to
\begin{equation}
    \big[\hT,\hw_{\halpha}(\sigma)\big]=\Big(2\partial_{1}\hw_{\halpha}+\partial\hw_{\halpha}\Big)(\sigma)=\B{\partial}\hw_{\halpha}(\sigma),
\end{equation}
where the contribution $2\partial_{1}\hw_{\halpha}$ comes from the commutator
\begin{equation}
    \Big[-\oint d\sigma'\big(\hw_{\hrho}\partial_{1'}\hLambda^{\hrho}+\hLambda^{\rho}\partial_{1'}\hw_{\hrho}\big)(\sigma'),\hw_{\alpha}(\sigma)\Big]=2\partial_{1}\hw_{\halpha}(\sigma)
\end{equation}
and the remaining commutators amount to $\partial\hw_{\halpha}$ using the equations of motion for $\hw_{\halpha}$ again. The same logic applies to the cases for $\hpsi^a$ and $\hLambda^{\halpha}$, where we also find
\begin{equation}
    \big[\hT,\hpsi^a(\sigma)\big]=\Big(2\partial_{1}\hpsi^a+\partial\hpsi^a\Big)(\sigma)=\B{\partial}\hpsi^a(\sigma),
\end{equation}
\begin{equation}
    \big[\hT,\hLambda^{\halpha}(\sigma)\big]=\Big(2\partial_{1}\hLambda^{\halpha}+\partial\hLambda^{\halpha}\Big)(\sigma)=\B{\partial}\hLambda^{\halpha}(\sigma).
\end{equation}
For the case of $\psi^a,\Lambda^{\alpha}$ and $w_{\alpha}$, the computation is slightly simpler. The brackets with the terms
\begin{equation}
\begin{split}
\Big(&\hpsi^a\hpsi^b\partial_{1}Z^M\hOmega_{Mab}-2\hLambda^{\halpha}\hw_{\hbeta}\partial_{1}Z^M\hOmega_{M\halpha}\,^{\hbeta}-\hpsi^a\partial_{1}\hpsi_a-\hw_{\halpha}\partial_{1}\hLambda^{\halpha}+\hLambda^{\halpha}\partial_{1}\hw_{\halpha}\cr
&-2\hpsi^a\hw_{\hbeta}\partial_{1} Z^M\hC_{Ma}\,^{\hbeta}-2\hw_{\halpha}\hw_{\hbeta}\partial_{1}Z^M\hY_M\,^{\halpha\hbeta}\Big)
\end{split}
\end{equation}
are always zero, whereas the remaining brackets can be easily computed to yield $\B{\partial}\psi^a,\B{\partial}\Lambda^{\alpha}$ and $\B{\partial}w_{\alpha}$ after we use the equations of motion for these fields. This concludes the proof that
\begin{equation}
    \big[\hT,\varphi(\sigma)\big]=\B{\partial}\varphi(\sigma).
\end{equation}

The proof for $\big[\hT,\Phi(Z)(\sigma)\big]=\B{\partial}\Phi(Z)(\sigma)$ is easy. The only brackets that contribute are
\begin{equation}
\begin{split}
    \Big[-\frac{1}{2}\B{\Pi}^a\B{\Pi}_a-\hd_{\halpha}\B{\Pi}^{\halpha}-\big(\hpsi^b\hpsi^c\HH{U}_{bc}{}^{\alpha}+\hLambda^{\hbeta}\hw_{\hgamma}D_{\hbeta}{}^{\hgamma\alpha}+\hpsi^b\hw_{\gamma}V_b{}^{\alpha\hgamma}+\hw_{\hbeta}\hw_{\hgamma}Z^{\alpha\hbeta\hgamma}\big)d_{\alpha},\Phi(Z)(\sigma)\Big].
\end{split}
\end{equation}
They are easily computed and we conclude that
\begin{equation}
    \big[\hT,\Phi(Z)(\sigma)\big]=\B{\Pi}^AE_A\,^M\partial_M\Phi(Z)(\sigma)=\B{\partial}\Phi(Z)(\sigma).
\end{equation}
Note that none of the results so far are changed by the inclusion of non-constant $f(\sigma)$.

Finally, the proof that $\big[\hT, f d_A\big]=f\B{\partial} d_A$ is the one with the longer computations. Still, the only thing one has to do is to compute the relevant commutators and recombine the contributions to get the result. After some long computations using the commutators in the appendix, we conclude that
\begin{equation}
    \oint{d\sigma}\oint{d\sigma'}\big[\hT(\sigma),f(\sigma')d_{\alpha}(\sigma')\big]=\oint{d\sigma}f(\sigma)\B{\partial}d_{\alpha}(\sigma),
\end{equation}
\begin{equation}
    \oint{d\sigma}\oint{d\sigma'}\big[\hT(\sigma),f(\sigma')\hd_{\halpha}(\sigma')\big]=\oint{d\sigma}f(\sigma)\Big({\partial}\hd_{\halpha}+2\partial_{1}\hd_{\halpha}\Big)(\sigma)=\oint{d\sigma}f(\sigma)\B{\partial}\hd_{\halpha}(\sigma),
\end{equation}
\begin{equation}
    \oint{d\sigma}\oint{d\sigma'}\big[\hT(\sigma),f(\sigma')d_{a}(\sigma')\big]=\oint{d\sigma}f(\sigma)\B{\partial}d_{a}(\sigma),
\end{equation}
where we have used the equations of motion for the $d_A$ derived from Equation (\ref{eomdA}). 

We have proven all the commutators in (\ref{commshol}). Therefore, the Jacobi identity (\ref{Jacobi}) does indeed imply
\begin{equation}
    \big[\hT,f G\big]=\oint{d\sigma}f(\sigma)\B{\partial}G(\sigma)=0, \forall\ \mathrm{function}\ f(\sigma),
\end{equation}
and we finally conclude that the conditions $\big\{G,G\big\}=-2T$ and $\big\{\hG,\hG\big\}=-2\hT$ imply
\begin{equation}
    \B{\partial}G=0,\quad \partial\hG=0.
\end{equation}

\section{B-RNS-GSS formalism on $AdS_5\times S^5$}\label{AdS}
In this section, we present the B-RNS-GSS Type IIB string on an $AdS_5\times S^5$ background by specializing the results for general backgrounds that have been obtained in the preceding sections. We start with the action and equations of motion for the worldsheet fields in this theory, and then establish classical integrability of the B-RNS-GSS string in this background.

\subsection{Action and equations of motion}
In order to write the action for the B-RNS-GSS string on $AdS_5\times S^5$, we must know the values for the RR flux, the B-field, the torsions, the H-fluxes and the curvatures on this background. Using the conventions of \cite{Chandia:2017afc}, these background fields take the values 
\begin{equation}
    P^{\alpha\hbeta}=-\frac{1}{2}\eta^{\alpha\hbeta},\quad B_{\alpha\hbeta}=\eta_{\alpha\hbeta},
\end{equation}
\begin{equation}
    T_{\alpha\beta}\,^{\underline{c}}=-(\gamma^{\underline{c}})_{\alpha\beta},\quad T_{\halpha\hbeta}\,^{\underline{c}}=-(\gamma^{\underline{c}})_{\halpha\hbeta},\quad T_{\ua\beta}\,^{\hgamma}=-\frac{1}{2}(\gamma_{\ua}\eta)_{\beta}\,^{\hgamma},\quad T_{\ua\hbeta}\,^{\gamma}=\frac{1}{2}(\eta\gamma_{\ua})^{\gamma}\,_{\hbeta}.
\end{equation}
\begin{equation}
    R_{\alpha\hbeta cd}=-(\eta\gamma_{cd})_{\alpha\hbeta},\quad R_{\alpha\hbeta c'd'}=(\eta\gamma_{c'd'})_{\alpha\hbeta},\quad R_{abcd}=\eta_{a[c}\eta_{d]b},\quad R_{a'b'c'd'}=-\eta_{a'[c'}\eta_{d']b'}
\end{equation}
Furthermore, we will favor the supercoset description of $AdS_5\times S^5$
\begin{equation}\label{aalgads}
    AdS_5\times S^5=\frac{PSU(2,2|4)}{SO(4,1)\times SO(5)}
\end{equation}
and we use the conventions shown in appendix \ref{app3} for the $\mathfrak{psu}(2,2|4)$ algebra.
In this description, we introduce the left-invariant currents
\begin{equation}
    J=g^{-1}dg=\frac{1}{2}J^{[\ua\ub]}T_{\ua\ub}+J^{\alpha}T_{\alpha}+J^{\ua}T_{\ua}+J^{\halpha}T_{\halpha},\quad g\in \frac{PSU(2,2|4)}{SO(4,1)\times SO(5)},
\end{equation}
where $\big\{T_{[\ua\ub]},T_{\alpha},T_{\ua},T_{\halpha}\big\}$ denote the $\mathfrak{psu}(2,2|4)$ generators. These currents are related to our familiar background fields $\Omega^{[\ua\ub]}$ and $E^A$ as:
\begin{equation}
    J^{[\ua\ub]}=dZ^M\Omega_M\,^{[\ua\ub]},\quad J^A=dZ^ME_M\,^A=E^A
\end{equation}
Finally, note that since $T_{\ua\ub\uc}=H_{\ua\ub\uc}=0$ we only have one spin-connection $\Omega^{\ua\ub}=\hOmega^{\ua\ub}=J^{[\ua\ub]}$.

Plugging in these values and the constraints derived earlier in the general action for the Type II string, we find
\begin{equation}
\begin{split}
    S=\int d^2z\Bigg\{&\frac{1}{2}J^{\ua}\B{J}_{\ua}+\frac{1}{2}\eta_{\alpha\hbeta}\Big(J^{\alpha}\B{J}^{\hbeta}-\B{J}^{\alpha}J^{\hbeta}\Big)+d_{\alpha}\B{J}^{\alpha}+\hd_{\halpha}J^{\halpha}-\frac{1}{2}\eta^{\alpha\hbeta}d_{\alpha}\hd_{\hbeta}\\
    &+w_{\alpha}\B{\nabla}\Lambda^{\alpha}+\frac{1}{2}\psi^{\ua}\B{\nabla}\psi_{\ua}+\hw_{\halpha}\nabla\hLambda^{\halpha}+\frac{1}{2}\hpsi^{\ua}\nabla\hpsi_{\ua}\\
    &-\frac{1}{2}\psi^{\ua}w_{\beta}\B{J}^{\hgamma}(\eta\gamma_{\ua})^{\beta}\,_{\hgamma}+\frac{1}{2}\hpsi^{\ua}\hw_{\hbeta}J^{\gamma}(\gamma_{\ua}\eta)_{\gamma}\,^{\hbeta}-\frac{1}{8}w_{\alpha}w_{\beta}\B{J}^{\uc}(\gamma_{\uc})^{\alpha\beta}-\frac{1}{8}\hw_{\halpha}\hw_{\hbeta}J^{\uc}(\gamma_{\uc})^{\halpha\hbeta}\\
    &+\frac{1}{2}\big(N^{ab}\hN_{ab}-N^{a'b'}\hN_{a'b'}\big)-\frac{1}{8}\psi^{\ua}w_{\beta}\hpsi^{\uc}\hw_{\hdelta}\eta^{\gamma\hdelta}(\gamma_{\uc}\eta)_{\gamma}\,^{\hrho}(\eta\gamma_{\ua})^{\beta}\,_{\hrho}\\
    &-\frac{1}{64}w_{\alpha}w_{\beta}\hw_{\hgamma}\hw_{\hdelta}(\gamma^{\ue})^{\alpha\beta}(\gamma_{\ue})^{\hgamma\hdelta}\Bigg\}+S_{ghosts},
\end{split}
\end{equation}
where recall the notation
\begin{equation}
    N^{\ua\ub}=-\psi^{\ua}\psi^{\ub}+\frac{1}{2}\Lambda\gamma^{\ua\ub}w,\quad \hN^{\ua\ub}=-\hpsi^{\ua}\hpsi^{\ub}+\frac{1}{2}\hLambda\gamma^{\ua\ub}\hw.
\end{equation}
Note that the auxiliary fields $d_{\alpha},\hd_{\halpha}$ have the same equations of motion as they do in the pure spinor formalism, namely
\begin{equation}
    d_{\alpha}=-2\eta_{\alpha\halpha}J^{\halpha},\quad \hd_{\halpha}=2\eta_{\alpha\halpha}\B{J}^{\alpha}.
\end{equation}
This means that by integrating them out we find
\begin{equation}
\begin{split}
    S=\int d^2z\Bigg\{&\frac{1}{2}J^{\ua}\B{J}_{\ua}+\frac{1}{2}\eta_{\alpha\hbeta}J^{\alpha}\B{J}^{\hbeta}-\frac{3}{2}\eta_{\alpha\hbeta}J^{\hbeta}\B{J}^{\alpha}+w_{\alpha}\B{\nabla}\Lambda^{\alpha}+\frac{1}{2}\psi^{\ua}\B{\nabla}\psi_{\ua}+\hw_{\halpha}\nabla\hLambda^{\halpha}+\frac{1}{2}\hpsi^{\ua}\nabla\hpsi_{\ua}\\
    &-\frac{1}{2}\psi^{\ua}w_{\beta}\B{J}^{\hgamma}(\eta\gamma_{\ua})^{\beta}\,_{\hgamma}+\frac{1}{2}\hpsi^{\ua}\hw_{\hbeta}J^{\gamma}(\gamma_{\ua}\eta)_{\gamma}\,^{\hbeta}-\frac{1}{8}w_{\alpha}w_{\beta}\B{J}^{\uc}(\gamma_{\uc})^{\alpha\beta}-\frac{1}{8}\hw_{\halpha}\hw_{\hbeta}J^{\uc}(\gamma_{\uc})^{\halpha\hbeta}\\
    &+\frac{1}{2}\big(N^{ab}\hN_{ab}-N^{a'b'}\hN_{a'b'}\big)-\frac{1}{8}\psi^{\ua}w_{\beta}\hpsi^{\uc}\hw_{\hdelta}\eta^{\gamma\hdelta}(\gamma_{\uc}\eta)_{\gamma}\,^{\hrho}(\eta\gamma_{\ua})^{\beta}\,_{\hrho}\\
    &-\frac{1}{64}w_{\alpha}w_{\beta}\hw_{\hgamma}\hw_{\hdelta}(\gamma^{\ue})^{\alpha\beta}(\gamma_{\ue})^{\hgamma\hdelta}\Bigg\}+S_{ghosts}.
\end{split}
\end{equation}

Consider now the Lie-algebra valued worldsheet fields given by
\begin{equation}
    \psi\equiv \psi^{\ua}T_{\ua},\quad \Lambda\equiv \Lambda^{\alpha}T_{\alpha},\quad w\equiv w_{\alpha}\eta^{\alpha\halpha}T_{\halpha},
\end{equation}
\begin{equation}
    \hpsi\equiv \hpsi^{\ua}T_{\ua},\quad \hLambda\equiv \hLambda^{\halpha}T_{\halpha},\quad \hw\equiv \hw_{\halpha}\eta^{\alpha\halpha}T_{\alpha},
\end{equation}
and define
\begin{equation}
    N_0\equiv \big\{\psi,\psi\big\}+\big\{\Lambda,w\big\}, \quad N_1\equiv \big\{\psi,w\big\},\quad N_2\equiv \big\{w,w\big\},
\end{equation}
\begin{equation}
    \hN_0\equiv \big\{\hpsi,\hpsi\big\}-\big\{\hLambda,\hw\big\},\quad \hN_3\equiv \big\{\hpsi,\hw\big\},\quad \hN_2\equiv \big\{\hw,\hw\big\}.
\end{equation}
Using these new notations and the conventions for the traces of $\mathfrak{psu}(2,2|4)$ generators from Appendix \ref{app3}, the stress-energy tensors are written as
\begin{equation}\label{Ttr}
    T=-\mathrm{Str}\Big(\frac{1}{2}J_2{J}_2+J_1J_3+\frac{1}{4}w\nabla\Lambda+\frac{1}{4}\Lambda\nabla w+\frac{1}{2}\psi\nabla\psi+\frac{1}{2}N_1J_3-\frac{1}{8}N_2J_2\Big),
\end{equation}
\begin{equation}\label{Thattr}
    \hT=-\mathrm{Str}\Big(\frac{1}{2}\B{J}_2\B{J}_2+\B{J}_1\B{J}_3-\frac{1}{4}\hw\B{\nabla}\hLambda-\frac{1}{4}\hLambda\B{\nabla}\hw+\frac{1}{2}\hpsi\B{\nabla}\hpsi-\frac{1}{2}\hN_3\B{J}_1-\frac{1}{8}\hN_2\B{J}_2\Big)
\end{equation}
whereas the superconformal currents are written as
\begin{align}\label{Gtr}
    G=\mathrm{Str}\left(\frac12\{\L,\L\}\psi-\frac18\{w,w\}\psi-\L J_3+\psi J_2-\frac12 w J_1 \right),
\end{align}
\begin{align}\label{Ghattr}
    \hG=\mathrm{Str}\left(\frac12\{\hLambda,\hLambda\}\hpsi-\frac18\{\hw,\hw\}\hpsi-\hLambda \B{J}_1+\hpsi \B{J}_2+\frac12 \hw \B{J}_3 \right).
\end{align}
Furthermore, the action can also be rewritten as
\begin{equation} \label{AdSaction}
\begin{split}
    S=\int d^2z\mathrm{Str}\Bigg(&\frac{1}{2}J_2\B{J}_2+\frac{1}{4}J_1\B{J}_3+\frac{3}{4}J_3\B{J}_1+\frac{1}{2}w\B{\nabla}\Lambda+\frac{1}{2}\psi\B{\nabla}\psi-\frac{1}{2}\hw\nabla\hLambda+\frac{1}{2}\hpsi\nabla\hpsi\\
    &+\frac{1}{2}N_1\B{J}_3-\frac{1}{2}\hN_3J_1-\frac{1}{8}N_2\B{J}_2-\frac{1}{8}\hN_2 J_2\\
    &-\frac{1}{4}N_0\hN_0+\frac{1}{4}N_1\hN_3-\frac{1}{64}N_2\hN_2\Bigg)+S_{ghosts}.
\end{split}
\end{equation}
Note that the part of this action that involves only $J$'s, $\Lambda$'s and $w$'s has the same form as the pure spinor $AdS_5\times S^5$ action \cite{Berkovits:2004xu}. Furthermore, the new terms are the simplest possible generalization which respects the $\mathbb{Z}_4$ grading of the $\mathfrak{psu}(2,2|4)$ algebra while including all of the $N_i$'s for $i=0$ to $3$. The equations of motion for the $\big\{\Lambda, w,\psi,\hLambda,\hw,\hpsi\big\}$ are
\begin{equation}
    \B{\nabla}\Lambda=+\frac{1}{2}[\Lambda,\hN_0]-[\psi,\B{J}_3+\frac{1}{2}\hN_3]+\frac{1}{2}[w,\B{J}_2+\frac{1}{8}\hN_2],
\end{equation}
\begin{equation}
    \nabla\hLambda=+\frac{1}{2}[\hLambda,N_0]-[\hpsi,J_1-\frac{1}{2}N_1]-\frac{1}{2}[\hw,J_2+\frac{1}{8}N_2],
\end{equation}
\begin{equation}
    \B{\nabla}\psi=+\frac{1}{2}[\psi,\hN_0]-\frac{1}{2}[w,\B{J}_3+\frac{1}{2}\hN_3],\qquad \nabla\hpsi=+\frac{1}{2}[\hpsi,N_0]+\frac{1}{2}[\hw,J_1-\frac{1}{2}N_1],
\end{equation}
\begin{equation}
    \B{\nabla}w=+\frac{1}{2}[w,\hN_0],\qquad 
    \nabla\hw=+\frac{1}{2}[\hw,N_0].
\end{equation}
The graded Jacobi identity then allows us to obtain the equations of motion for the $N_i$'s and $\hN_i$'s. We find:
\begin{equation}
    \B{\nabla}N_0=+\frac{1}{2}[N_0,\hN_0]-[N_1,\B{J}_3+\frac{1}{2}\hN_3]+\frac{1}{4}[N_2,\B{J}_2+\frac{1}{8}\hN_2],
\end{equation}
\begin{equation}
    \nabla\hN_0=-\frac{1}{2}[N_0,\hN_0]-[J_1-\frac{1}{2}N_1,\hN_3]-\frac{1}{4}[J_2+\frac{1}{8}N_2,\hN_2],
\end{equation}
\begin{equation}
    \B{\nabla}N_1=+\frac{1}{2}[N_1,\hN_0]-\frac{1}{4}[N_2,\B{J}_3+\frac{1}{2}\hN_3],\quad \nabla\hN_3=-\frac{1}{2}[N_0,\hN_3]-\frac{1}{4}[J_1-\frac{1}{2}N_1,\hN_2],
\end{equation}
\begin{equation}
    \B{\nabla}N_2=+\frac{1}{2}[N_2,\hN_0],\quad 
    \nabla\hN_2=-\frac{1}{2}[N_0,\hN_2].
\end{equation}
Finally, by varying the action with respect to $\delta g=gX$ and using the Maurer-Cartan equations, we can derive the equations of motion for the $J_i$'s and $\B{J}_i$'s. These are:
\begin{equation}
    \nabla\B{J}_1=-\frac{1}{2}[N_0,\B{J}_1]+\frac{1}{2}[J_1,\hN_0]-\frac{1}{2}\B{\nabla}N_1+\frac{1}{2}[J_2,\hN_3]-\frac{1}{8}[N_2,\B{J}_3]+\frac{1}{8}[J_3,\hN_2],
\end{equation}
\begin{equation}
    \B{\nabla}J_1=[J_2,\B{J}_3]+[J_3,\B{J}_2]-\frac{1}{2}[N_0,\B{J}_1]+\frac{1}{2}[J_1,\hN_0]-\frac{1}{2}\B{\nabla}N_1+\frac{1}{2}[J_2,\hN_3]-\frac{1}{8}[N_2,\B{J}_3]+\frac{1}{8}[J_3,\hN_2],
\end{equation}
\begin{equation}
    \nabla\B{J}_2=-[J_1,\B{J}_1]-\frac{1}{2}[N_0,\B{J}_2]+\frac{1}{2}[J_2,\hN_0]+\frac{1}{8}\B{\nabla}N_2+\frac{1}{8}\nabla\hN_2+\frac{1}{2}[N_1,\B{J}_1]+\frac{1}{2}[J_3,\hN_3],
\end{equation}
\begin{equation}
    \B{\nabla}J_2=[J_3,\B{J}_3]-\frac{1}{2}[N_0,\B{J}_2]+\frac{1}{2}[J_2,\hN_0]+\frac{1}{8}\B{\nabla}N_2+\frac{1}{8}\nabla\hN_2+\frac{1}{2}[N_1,\B{J}_1]+\frac{1}{2}[J_3,\hN_3],
\end{equation}
\begin{equation}
    \nabla\B{J}_3=-[J_1,\B{J}_2]-[J_2,\B{J}_1]-\frac{1}{2}[N_0,\B{J}_3]+\frac{1}{2}[J_3,\hN_0]+\frac{1}{2}\nabla\hN_3+\frac{1}{2}[N_1,\B{J}_2]-\frac{1}{8}[N_2,\B{J}_1]+\frac{1}{8}[J_1,\hN_2]
\end{equation}
\begin{equation}
    \B{\nabla}J_3=-\frac{1}{2}[N_0,\B{J}_3]+\frac{1}{2}[J_3,\hN_0]+\frac{1}{2}\nabla\hN_3+\frac{1}{2}[N_1,\B{J}_2]-\frac{1}{8}[N_2,\B{J}_1]+\frac{1}{8}[J_1,\hN_2].
\end{equation}

\subsection{Integrability}
We will now establish classical integrability of the B-RNS-GSS string on $AdS_5\times S^5$. In order to do so, we will find a Lax connection that is flat and that transforms under the action of the BRST charge by a dressing transformation. As a further check, we will also show that the Lax connection generates the usual Noether charges associated to the isometries of $AdS_5\times S^5$.

The most standard way to tackle the problem of finding the Lax connection is to write a general expression for it using the $J_i$'s, $\B{J}_i$'s, $N_i$'s and $\hN_i$'s and then fix the coefficients associated to each of these currents by requiring the flatness condition to be satisfied. This was the method used when integrability was first established for the Green-Schwarz string in $AdS_5\times S^5$ \cite{Bena:2003wd}, and was also used to find the Lax connection for the pure spinor formalism \cite{Vallilo:2003nx}. We will, however, take a more direct route which takes advantage of the previous knowledge about the pure spinor Lax connection. Notice that the equations of motion above reduce to the ones from the pure spinor formalism when we take $N_1,N_2,\hN_2,\hN_3\to0$. Furthermore, it is well known that the flatness condition on the Lax connection implies the equations of motion. Bearing these two facts in mind is enough to easily achieve our goal, as we may start off with the pure spinor Lax connection and add corrections to compensate for the new terms in the equations of motion. It turns out that the needed corrections are really simple.

The pure spinor Lax connection is given by 
\begin{equation}
    \WT{A}=(J_0+\frac{1}{2}N_0)+\frac{1}{\mu}J_1+\frac{1}{\mu^2}J_2+\frac{1}{\mu^3}J_3-\frac{1}{2\mu^4}N_0,
\end{equation}
\begin{equation}
    \WT{\B{A}}=(\B{J}_0+\frac{1}{2}\hN_0)+{\mu}\B{J}_3+{\mu^2}\B{J}_2+{\mu^3}\B{J}_1-\frac{\mu^4}{2}\hN_0.
\end{equation}
If we compute the flatness condition with these expressions at orders $\mu^{-4}$ and $\mu^4$ using the B-RNS-GSS equations of motion, we find that we are missing two contributions:
\begin{equation}
    \frac{1}{\mu^4}\Big(\frac{1}{2}[N_1,\B{J}_3+\frac{1}{2}\hN_3]-\frac{1}{8}[N_2,\B{J}_2+\frac{1}{8}\hN_2]\Big)\quad \mathrm{and} \quad \mu^4\Big(-\frac{1}{2}[J_1-\frac{1}{2}N_1,\hN_3]-\frac{1}{8}[J_2+\frac{1}{8}N_2,\hN_2]\Big).
\end{equation}
These missing commutators are accounted for if we correct the Lax connection to
\begin{equation}
    A=(J_0+\frac{1}{2}N_0)+\frac{1}{\mu}(J_1-\frac{1}{2}N_1)+\frac{1}{\mu^2}(J_2+\frac{1}{8}N_2)+\frac{1}{\mu^3}J_3-\frac{1}{2\mu^4}N_0+\frac{1}{2\mu^5}N_1-\frac{1}{8\mu^6}N_2,
\end{equation}
\begin{equation}
    \B{A}=(\B{J}_0+\frac{1}{2}\hN_0)+\mu(\B{J}_3+\frac{1}{2}\hN_3)+\mu^2(\B{J}_2+\frac{1}{8}\hN_2)+\mu^{3}\B{J}_1-\frac{1}{2}\mu^{4}\hN_0-\frac{1}{2}\mu^5\hN_3-\frac{1}{8}\mu^6\hN_2.
\end{equation}
This means that now the flatness condition at orders $\mu^{-4},\mu^{4}$ correctly implies the equations of motion for the currents $N_0,\hN_0$. This turns out to be the sole correction needed to obtain the B-RNS-GSS Lax connection. Indeed, the fact that it satisfies the flatness equation
\begin{equation}
    \partial \B{A}-\B{\partial}A+[A,\B{A}]=0
\end{equation}
is equivalent to the equations of motion listed above for all orders from $\mu^{6}$ to $\mu^{-6}$. As claimed above, it is also amusing to check that upon taking the coefficient of the first order term in the Taylor expansion of $A,\B{A}$ around $\mu=1$ we find the Noether current associated to the $AdS_5\times S^5$ isometries
\begin{equation}
    \frac{\partial A}{\partial \mu}\Big|_{\mu=1}=-\Big(J_1+2J_2+3J_3-2N_0-\frac{1}{2}N_2+2N_1\Big),
\end{equation}
\begin{equation}
    \frac{\partial \B{A}}{\partial \mu}\Big|_{\mu=1}=3\B{J}_1+2\B{J}_2+\B{J}_3-2\hN_0-\frac{1}{2}\hN_2-2\hN_3,
\end{equation}
which is the expected behavior of a Lax connection. 

The last thing to do is to show that the eigenvalues of the monodromy matrix 
\begin{equation} \label{MonodromyM}
    \mathcal{M}(\mu)\equiv\mathrm{Pexp}\Big\{-\oint_{C}\big(A(\mu)dz+\B{A}(\mu)d\B{z}\big)\Big\},
\end{equation}
associated to this Lax connection are BRST invariant. In order to do so, we must first compute the BRST variation of the fields $\big\{\Lambda,w,\psi,\hLambda,\hw,\hpsi\big\}$ and of the currents $J$ (for the discussion on the BRST variation of the Lax connection in the context of the pure spinor formalism see \cite{Berkovits:2004jw,Adam:2007ws}). We start by noting that the action of the stress energy tensor is simply
\begin{equation}
    T(\varphi)=\partial\varphi,\quad \hT(\varphi)=\B{\partial}\varphi,\quad \varphi\in\big\{\Lambda,w,\psi,\hLambda,\hw,\hpsi,J_i,\B{J}_i\big\}
\end{equation}
from the discussion in Section \ref{HOL}. Once we recall the equations of motion $\B{\partial}c=\partial \B{c}=0$, it follows from this same discussion that the components of the Lax connection transform as
\begin{equation}
    \delta A=\partial(cA)+\B{c}\B{\partial}A,\quad \delta \B{A}=c\partial\B{A}+\B{\partial}(\B{c}\B{A}),
\end{equation}
under the action of the $(cT+\B{c}\hT)$ piece of the BRST charge. We may use the equations of motion $\B{\partial}c=\partial\B{c}=0$, the trivial relations $[A,cA]=0$ and $[\B{A},\B{c}\B{A}]=0$, as well as the flatness equation to rewrite the variations as
\begin{equation}
    \delta A=\partial(cA+\B{c}\B{A})+[A,cA+\B{c}\B{A}],\quad \delta \B{A}=\B{\partial}(cA+\B{c}\B{A})+[\B{A},cA+\B{c}\B{A}].
\end{equation}
These equations can be put together in the very suggestive expression
\begin{equation}
    (cT+\B{c}\hT)\big(A(\mu)dz+\B{A}(\mu)d\B{z}\big)=d(cA(\mu)+\B{c}\B{A}(\mu))+[A(\mu)dz+\B{A}(\mu)d\B{z},cA(\mu)+\B{c}\B{A}(\mu)].
\end{equation}
We conclude that the $(cT+\B{c}\hT)$ piece of the BRST charge transforms the Lax connection by a dressing transformation with dressing parameter $cA(\mu)+\B{c}\B{A}(\mu)$.

Next, we note that, up to $SO(4,1)\times SO(5)$ gauge transformations (see \cite{Chandia:2014sta} for their inclusion in the pure spinor case),
\begin{equation}
    (\gamma G+\hgamma\hG)(g)=g\Big(\gamma\Lambda-\gamma\psi+\frac{1}{2}\gamma w+\hgamma\hLambda-\hgamma\hpsi-\frac{1}{2}\hgamma\hw\Big)
\end{equation}
and therefore
\begin{equation}
\begin{split}
    (\gamma G+\hgamma\hG)(J_i)=~&\delta_{i+3,0}\partial\Big(\gamma\Lambda-\frac{1}{2}\hgamma\hw\Big)-\delta_{i+2,0}\partial\big(\gamma\psi+\hgamma\hpsi\big)+\delta_{i+1,0}\partial\Big(\frac{1}{2}\gamma w+\hgamma\hLambda\Big)\\
    &+\big[J_{i+3},\g\Lambda-\frac{1}{2}\gb\hw\big]-\big[J_{i+2},\g\psi+\gb\hpsi\big]+\big[J_{i+1},\frac{1}{2}\g w+\gb\hLambda\big].
\end{split}
\end{equation}
The equation for the $\B{J}_i$'s is analogous. Furthermore, we find
\begin{equation}
    (\gamma G+\hgamma\hG)(\Lambda)=\frac{1}{2}\gamma N_1-\gamma J_1+\hgamma\big\{\psi,\hLambda\big\}+\frac{1}{2}\hgamma\big\{\hpsi,w\big\},
\end{equation}
\begin{equation}
    (\gamma G+\hgamma\hG)(w)=2\gamma\big\{\psi,\Lambda\big\}-2\gamma J_3,
\end{equation}
\begin{equation}
    (\gamma G+\hgamma\hG)(\psi)=\frac{1}{2}\gamma\big\{\Lambda,\Lambda\big\}+\gamma J_2+\frac{1}{8}\gamma N_2+\frac{1}{2}\hgamma\big\{\hLambda,w\big\}
\end{equation}
\begin{equation}
    (\gamma G+\hgamma\hG)(\hLambda)=-\hgamma\B{J}_3-\frac{1}{2}\hgamma\hN_3+\gamma\big\{\hpsi,\Lambda\big\}-\frac{1}{2}\gamma\big\{\psi,\hw\big\},
\end{equation}
\begin{equation}
    (\gamma G+\hgamma\hG)(\hw)=-2\hgamma\big\{\hpsi,\hLambda\big\}+2\hgamma\B{J}_1,
\end{equation}
\begin{equation}
    (\gamma G+\hgamma\hG)(\hpsi)=\frac{1}{2}\hgamma\big\{\hLambda,\hLambda\big\}+\hgamma\B{J}_2+\hgamma\frac{1}{8}\hN_2-\frac{1}{2}\gamma\big\{\Lambda,\hw\big\}.
\end{equation}
It then follows that the variation of the $N_i$'s is
\begin{equation}
    (\gamma G+\hgamma\hG)(N_0)=2\gamma[J_2+\frac{1}{8}N_2,\psi]-\gamma[J_1-\frac{1}{2}N_1,w]-2\gamma[J_3,\Lambda]-\hgamma[N_1,\hLambda]-\frac{1}{4}\hgamma[N_2,\hpsi],
\end{equation}
\begin{equation}
    (\gamma G+\hgamma\hG)(N_1)=-\gamma[N_0,\Lambda]+\gamma[J_2+\frac{1}{8}N_2,w]-2\gamma[J_3,\psi]-\frac{1}{4}\hgamma[N_2,\hLambda],
\end{equation}
\begin{equation}
    (\gamma G+\hgamma\hG)(N_2)=-4\gamma[N_1,\Lambda]-4\gamma[N_0,\psi]-4\gamma[J_3,w].
\end{equation}
\begin{equation}
    (\gamma G+\hgamma\hG)(\hN_0)=2\hgamma[\B{J}_2+\frac{1}{8}\hN_2,\hpsi]+\hgamma[\B{J}_3+\frac{1}{2}\hN_3,\hw]-2\hgamma[\B{J}_1,\hLambda]+\gamma[\hN_3,\Lambda]-\frac{1}{4}\gamma[\hN_2,\psi],
\end{equation}
\begin{equation}
    (\gamma G+\hgamma\hG)(\hN_3)=\hgamma[\hN_0,\hLambda]+\hgamma[\B{J}_2+\frac{1}{8}\hN_2,\hw]+2\hgamma[\B{J}_1,\hpsi]+\frac{1}{4}\gamma[\hN_2,\Lambda],
\end{equation}
\begin{equation}
    (\gamma G+\hgamma\hG)(\hN_2)=4\hgamma[\hN_3,\hLambda]-4\hgamma[\hN_0,\hpsi]+4\hgamma[\B{J}_1,\hw].
\end{equation}

Now we can compute the variation of the Lax connection due to the $\gamma G+\hgamma\hG$ piece of the BRST charge. After the calculations, we verify that it is equal to the dressing transformation
\begin{equation}
    (\gamma G+\hgamma\hG)\big(A(\mu)dz+\B{A}(\mu)d\B{z}\big)=d\varphi(\mu)+[A(\mu)dz+\B{A}(\mu)d\B{z},\varphi(\mu)]
\end{equation}
with the dressing factor $\varphi(\mu)$ being given by
\begin{equation}
    \varphi(\mu)=\frac{1}{2\mu^3}\gamma w-\frac{1}{\mu^2}\gamma\psi+\frac{1}{\mu}\gamma\Lambda+\mu\hgamma\hLambda-\mu^2\hgamma\hpsi-\frac{\mu^3}{2}\hgamma\hw.
\end{equation}
Note that this is the natural generalization of the dressing factor for the pure spinor case due to the presence of the terms $\mathrm{Str}\big(\psi J_2-\frac{1}{2}wJ_1\big)$ in $G$ and $\mathrm{Str}\big(\hpsi \B{J}_2+\frac{1}{2}\hw\B{J}_3\big)$ in $\hG$. 

We thus conclude that the complete variation of the Lax connection under the BRST charge is given by a dressing transformation 
\begin{equation}
    Q\big(A(\mu)dz+\B{A}(\mu)d\B{z}\big)=d\Phi(\mu)+[A(\mu)dz+\B{A}(\mu)d\B{z},\Phi(\mu)]
\end{equation}
with dressing parameter
\begin{equation}
    \Phi(\mu)=cA(\mu)+\B{c}\B{A}(\mu)+\frac{1}{2\mu^3}\gamma w-\frac{1}{\mu^2}\gamma\psi+\frac{1}{\mu}\gamma\Lambda+\mu\hgamma\hLambda-\mu^2\hgamma\hpsi-\frac{\mu^3}{2}\hgamma\hw.
\end{equation}
Consequently, the monodromy matrix (\ref{MonodromyM}) generates an infinite set of BRST-invariant conserved charges and classical integrability has been established for the B-RNS-GSS string in $AdS_5\times S^5$.

\section{Conclusion}\label{conclusion}
In this paper, the B-RNS-GSS formalism has been extended for the description of the Type II superstring in curved supergravity backgrounds, including backgrounds with finite Ramond-Ramond flux. More precisely, nilpotence of the $\mathcal{N}=(1,1)$ worldsheet superconformal BRST charge at the classical level has been shown to imply the Type II supergravity equations of motion for the background superfields. As a direct application of this result, an action for the B-RNS-GSS string on an $AdS_5\times S^5$ background has been proposed. Classical integrability of this sigma-model has also been established.

There are several lines of investigation that can be pursued now that these results have been achieved. One possible direction is to relate the $AdS_5\times S^5$ action (\ref{AdSaction}) to the twistor superstring action proposed in \cite{Berkovits:2016xnb}. Note that this twistor superstring formalism is also manifestly invariant under $\mathcal{N}=(1,1)$ worldsheet supersymmetry and $N=2$ $D=10$ spacetime supersymmetry. Furthermore, physical states in this formalism must satisfy a U(1) charge requirement that is similar to the U(1)-charge/small Hilbert space condition discussed in Section \ref{flat}. It should also be possible to find B-RNS-GSS vertex operators in $AdS_5\times S^5$ starting from the pure spinor ones \cite{Berkovits:2000yr} by exploring the similarities between the two formalisms in this background. 

Most notably, we should stress that all of the analysis presented in this paper is classical. Consequently, quantum consistency of the B-RNS-GSS formalism in $AdS_5\times S^5$ remains to be proven. We expect that the techniques used to prove quantum consistency of the pure spinor formalism in $AdS_5\times S^5$ \cite{Berkovits:2004xu} will be useful to this goal. Moreover, given the necessity of the non-minimal variables to ensure that the formalism is consistent at the quantum level in flat space, it is likely that knowledge about their role in curved backgrounds will be relevant to understand the quantum behavior of the formalism in more general supergravity settings, in addition to $AdS_5\times S^5$.

Attempts at using the RNS formalism to describe Ramond-Ramond backgrounds have either struggled with ill-defined half-integer picture raising operators \cite{Berenstein:1999jq,Berenstein:1999ip}, or have been limited to a perturbative analysis of Ramond-Ramond deformations using String Field Theory methods \cite{Cho:2018nfn}. Since the starting point of the B-RNS-GSS formalism can be seen as the variables, action and BRST charge from the RNS formalism, perhaps a more daring application would be to use the results of this paper to better understand how to describe Ramond-Ramond backgrounds with the RNS formalism.

{\bf Acknowledgments:} The authors are especially grateful to Nathan Berkovits for useful discussions, for suggesting to investigate the integrability of the $AdS_5\times S^5$ sigma-model and for carefully reading and commenting on the manuscript. OC would like to thank fondecyt grants  1200342 and 1201550 for partial financial support. JG would also like to thank Lucas N.S. Martins and Rodrigo S. Pitombo for useful discussions, and FAPESP grants 2021/14335-0 (ICTP-SAIFR) and 2022/04105-0 for partial financial support.

\begin{appendices}

\section{Gauge Symmetries}\label{app0}

In this Appendix, we list the gauge transformations associated to background field relabelling in the action and BRST charge. The ones for the fields which appear in the action are

\begin{equation} \label{g1}
    \delta d_{\alpha}=-\frac{1}{2}\omega_{\alpha bc}\psi^b\psi^c-\omega_{\alpha\beta}\,^{\gamma}\Lambda^{\beta}w_{\gamma}+\psi^bw_{\gamma}\rho_{\alpha b}\,^{\gamma}-w_{\beta}w_{\gamma}\WT{\rho}_{\alpha}\,^{\beta\gamma},
 \end{equation}
\begin{equation}
    \delta\hd_{\halpha}=-\frac{1}{2}\homega_{\halpha bc}\hpsi^b\hpsi^c-\homega_{\halpha\hbeta}\,^{\hgamma}\hLambda^{\hbeta}\hw_{\hgamma}+\hpsi^b\hw_{\hgamma}\hrho_{\halpha b}\,^{\hgamma}-\hw_{\hbeta}\hw_{\hgamma}\HH{\WT{\rho}}_{\halpha}\,^{\hbeta\hgamma}
\end{equation}
\begin{equation}
    \delta\Omega_{\alpha bc}=\omega_{\alpha bc},\quad \delta\Omega_{\alpha\beta}\,^{\gamma}=-\omega_{\alpha\beta}\,^{\gamma},\quad \delta C_{\alpha b}\,^{\gamma}=-\rho_{\alpha b}\,^{\gamma},\quad \delta Y_{\alpha}\,^{\beta\gamma}=-\WT{\rho}_{\alpha}\,^{\beta\gamma},
\end{equation}
\begin{equation}
    \delta \hOmega_{\halpha bc}=\homega_{\halpha bc},\quad \delta\hOmega_{\halpha\hbeta}\,^{\hgamma}=-\homega_{\halpha\hbeta}\,^{\hgamma},\quad \delta\hC_{\halpha b}\,^{\hgamma}=-\hrho_{\halpha b}\,^{\hgamma},\quad \delta\hY_{\halpha}\,^{\hbeta\hgamma}=-\HH{\WT{\rho}}_{\halpha}\,^{\hbeta\hgamma}
\end{equation}
\begin{equation}
    \delta U^{\hgamma}_{bc}=-\frac{1}{2}P^{\gamma\hgamma}\omega_{\gamma bc},\quad\delta \hD_{\alpha}^{\beta\hgamma}=-P^{\gamma\hgamma}\omega_{\gamma\alpha}\,^{\beta},\quad \delta \hV_a^{\beta\hgamma}=P^{\gamma\hgamma}\rho_{\gamma a}\,^{\beta},\quad \delta \HH{Z}^{\alpha\beta\hgamma}=-P^{\gamma\hgamma}\WT{\rho}_{\gamma}\,^{\alpha\beta},
\end{equation}
\begin{equation}
    \delta \HH{U}^{\gamma}_{bc}=\frac{1}{2}P^{\gamma\hgamma}\homega_{\hgamma bc},\quad \delta D_{\halpha}^{\gamma\hbeta}=P^{\gamma\hgamma}\homega_{\hgamma\halpha}\,^{\hbeta},\quad \delta V_a^{\gamma\hbeta}=-P^{\gamma\hgamma}\hrho_{\hgamma a}\,^{\hbeta},\quad \delta Z^{\gamma\halpha\hbeta}=P^{\gamma\hgamma}\HH{\WT{\rho}}_{\hgamma}\,^{\halpha\hbeta}
\end{equation}
\begin{equation}
    \delta H_{abcd}=-\frac{1}{2}\HH{U}^{\lambda}_{cd}\omega_{\lambda ab}-\frac{1}{2}U^{\hlambda}_{ab}\homega_{\hlambda cd},\quad \delta S_{\alpha\hgamma}^{\beta\hdelta}=-D_{\hgamma}^{\lambda\hdelta}\omega_{\lambda\alpha}\,^{\beta}-\hD_{\alpha}^{\beta\hlambda}\homega_{\hlambda\hgamma}\,^{\hdelta},
\end{equation}
\begin{equation}
    \delta \HH{E}_{ab\hgamma}\,^{\hdelta}=-\frac{1}{2}D_{\hgamma}^{\lambda\hdelta}\omega_{\lambda ab}-U_{ab}^{\hlambda}\homega_{\hlambda\hgamma}\,^{\hdelta},\quad \delta E_{\alpha cd}^{\beta}=-\HH{U}_{cd}^{\lambda}\omega_{\lambda\alpha}\,^{\beta}-\frac{1}{2}\hD_{\alpha}^{\beta\hlambda}\homega_{\hlambda cd},
\end{equation}
\begin{equation}
    \delta I_{abc}\,^{\hdelta}=-\frac{1}{2}V_c^{\lambda\hdelta}\omega_{\lambda ab}-U_{ab}^{\hlambda}\hrho_{\hlambda c}\,^{\hdelta},\quad \delta \HH{I}_{acd}^{\beta}=-\frac{1}{2}\hV_a^{\beta\hlambda}\homega_{\hlambda cd}-\HH{U}_{cd}^{\lambda}\rho_{\lambda a}\,^{\beta},
\end{equation}
\begin{equation}
    \delta F_{\alpha c}^{\beta\hdelta}=-V_c^{\lambda\hdelta}\omega_{\lambda\alpha}\,^{\beta}-\hD_{\alpha}^{\beta\hlambda}\hrho_{\hlambda c}\,^{\hdelta},\quad \delta \HH{F}_{a\hgamma}^{\beta\hdelta}=-\hV_a^{\beta\hlambda}\homega_{\hlambda\hgamma}\,^{\hdelta}-D_{\hgamma}^{\lambda\hdelta}\rho_{\lambda a}\,^{\beta},
\end{equation}
\begin{equation}
    \delta \HH{G}_{\hgamma}^{\alpha\beta\hdelta}=-D_{\hgamma}^{\lambda\hdelta}\WT{\rho}_{\lambda}\,^{\alpha\beta}-\HH{Z}^{\alpha\beta\hlambda}\homega_{\hlambda\hgamma}\,^{\hdelta},\quad \delta G_{\alpha}^{\beta\hgamma\hdelta}=-\hD_{\alpha}^{\beta\hlambda}\HH{\WT{\rho}}_{\hlambda}\,^{\hgamma\hdelta}-Z^{\lambda\hgamma\hdelta}\omega_{\lambda\alpha}\,^{\beta},
\end{equation}
\begin{equation}
    \delta \HH{J}_{cd}^{\alpha\beta}=-\HH{U}_{cd}^{\lambda}\WT{\rho}_{\lambda}\,^{\alpha\beta}-\frac{1}{2}\HH{Z}^{\alpha\beta\hlambda}\homega_{\hlambda cd},\quad \delta J_{ab}^{\hgamma\hdelta}=-U^{\hlambda}_{ab}\HH{\WT{\rho}}_{\hlambda}\,^{\hgamma\hdelta}-\frac{1}{2}Z^{\lambda\hgamma\hdelta}\omega_{\lambda ab},
\end{equation}
\begin{equation}
    \delta L_a^{\beta\hgamma\hdelta}=-Z^{\lambda\hgamma\hdelta}\rho_{\lambda a}\,^{\beta}-\hV_a^{\beta\hlambda}\HH{\WT{\rho}}_{\hlambda}\,^{\hgamma\hdelta},\quad \delta \HH{L}_c^{\alpha\beta\hdelta}=-\HH{Z}^{\alpha\beta\hlambda}\hrho_{\hlambda c}\,^{\hdelta}-V_c^{\lambda\hdelta}\WT{\rho}_{\lambda}\,^{\alpha\beta},
\end{equation}
\begin{equation} \label{gf}
    \delta K_{ac}^{\beta\hdelta}=-V_c^{\lambda\hdelta}\rho_{\lambda a}\,^{\beta}+\hV_a^{\beta\hlambda}\hrho_{\hlambda c}\,^{\hdelta},\quad \delta M^{\alpha\beta\hgamma\hdelta}=-Z^{\lambda\hgamma\hdelta}\WT{\rho}_{\lambda}\,^{\alpha\beta}-\HH{Z}^{\alpha\beta\hlambda}\HH{\WT{\rho}}_{\hlambda}\,^{\hgamma\hdelta}.
\end{equation}
The gauge transformations for the superfields in the supercurrents $G,\hG$ are
\begin{equation}
    \delta G_{\alpha\beta}\,^{\gamma}=\omega_{(\alpha\beta)}\,^{\gamma},\quad \delta G_{\alpha bc}=\omega_{\alpha bc},\quad \delta G_{\alpha b}\,^{\gamma}=-\rho_{\alpha b}\,^{\gamma},\quad \delta G_{\alpha}\,^{\beta\gamma}=2\WT{\rho}_{\alpha}\,^{\beta\gamma},
\end{equation}
\begin{equation}
    \delta \hG_{\halpha\hbeta}\,^{\hgamma}=\homega_{(\halpha\hbeta)}\,^{\hgamma},\quad \delta \hG_{\halpha bc}=\homega_{\halpha bc},\quad \delta \hG_{\halpha b}\,^{\hgamma}=-\hrho_{\halpha b}\,^{\hgamma},\quad \delta \hG_{\halpha}\,^{\hbeta\hgamma}=2\HH{\WT{\rho}}_{\halpha}\,^{\hbeta\hgamma}.
\end{equation}

\section{Canonical commutators}\label{app1}
In this appendix the canonical commutators relevant for the computations of this paper are listed. The canonical variables are the Type II superspace variables $(Z^M, P_M)$ together with $(\psi^a, \psib^a, \L^\a, w_\a, \Lb^{\ab}, \wb_{\ab})$. The canonical commutators for these variables are
\begin{align} \label{brack1}
    &\big[Z^M(\s), P_N(\s')\big\} = \d_N^M \d(\s-\s') ,\cr  
    &\big\{\psi^a(\s),\psi^b(\s')\big\} = \eta^{ab} \d(\s-\s'),\quad \big\{\psib^a(\s),\psib^b(\s')\big\} = \eta^{ab} \d(\s-\s') ,\cr
    &\big[\L^\a(\s),w_\b(\s')\big] = \d_\b^\a \d(\s-\s'),\quad \big[\Lb^{\ab}(\sigma),\wb_{\bb}(\sigma')\big] = \d_{\bb}^{\ab} \d(\s-\s') .
\end{align}
We need the commutators for the supersymmetric combinations $d_\a, \dd_{\ab}, \Pi^A=\p Z^M E_M{}^A, \Pib^A=\pb Z^M E_M{}^A$. They are
\begin{align}
&\big[d_\a(\sigma) , \Phi(Z)(\sigma')\big\} = \p_\a \Phi(Z)\delta(\sigma-\sigma'),\quad \big[\Pi^{\ab}(\sigma),\Phi(Z)(\sigma')\big\}=P^{\b\ab} \p_\b \Phi(Z)\delta(\sigma-\sigma'),\cr 
&\big[\dd_{\ab}(\sigma),\Phi(Z)(\sigma')\big\}=\p_{\ab} \Phi(Z)\delta(\sigma-\sigma'),\quad \big[\Pi^\a(\sigma),\Phi(Z)(\sigma')\big\}=-P^{\a\bb}\p_{\bb} \Phi(Z)\delta(\sigma-\sigma'),\cr
&\big[d_a(\sigma),\Phi(Z)(\sigma')\big]=-\p_a \Phi(Z)\delta(\sigma-\sigma'),
\label{}
\end{align}

\begin{align}
&\big\{d_\a(\s) , \psi_a(\s')\big\} = \left( -\psi^b \O_{\a ab} - w_\b C_{\a a}{}^\b \right) \d(\s-\s') ,\cr 
&\big\{d_{\ab}(\s) , \psi_a(\s')\big\} = \left( -\psi^b \O_{\ab ab} - w_\b C_{\ab a}{}^\b \right) \d(\s-\s') ,\cr 
&\big[d_a(\s) , \psi_b(\s')\big] = \left( -\psi^c \O_{abc} + w_\a C_{ab}{}^\a \right) \d(\s-\s') ,
\label{}
\end{align}

\begin{align}
&\big\{d_\a(\s) , \psib_a(\s')\big\} = \left( - \psib^b \Oh_{\a ab} - \wb_{\bb} \Ch_{\a a}{}^{\bb} \right) \d(\s-\s'), \cr 
&\big\{\dd_{\ab}(\s) , \psib_a(\s')\big\} = \left( -\psib^b \Oh_{\ab ab} - \wb_{\bb} \Ch_{\ab a}{}^{\bb} \right) \d(\s-\s'), \cr 
&\big[d_a(\s) , \psib_b(\s') \big] = \left( -\psib^c \Oh_{abc} + \wb_{\ab} \Ch_{ab}{}^{\ab} \right) \d(\s-\s') ,
\label{}
\end{align}

\begin{align}
&\big[d_\a(\s) , \L^\b(\s')\big] = \left( -\L^\g \O_{\a\g}{}^\b + \psi^a C_{\a a}{}^\b - 2 w_\g Y_\a{}^{\g\b} \right) \d(\s-\s'), \cr 
&\big[\dd_{\ab}(\s) , \L^\b(\s')\big] = \left( -\L^\g \O_{\ab\g}{}^\b + \psi^a C_{\ab a}{}^\b - 2 w_\g Y_{\ab}{}^{\g\b} \right) \d(\s-\s'), \cr 
&\big[d_a(\s) , \L^\a(\s') 
\big] = \left( \L^\b \O_{a\b}{}^\a + \psi^b C_{ab}{}^\a + 2w_\b Y_a{}^{\b\a} \right) \d(\s-\s') ,
\label{}
\end{align}

\begin{align}
&\big[d_\a(\s) , \Lb^{\bb}(\s')\big] = \left( -\Lb^{\gb}\Oh_{\a\gb}{}^{\bb}+ \psib^a \Ch_{\a a}{}^{\bb} - 2 \wb_{\gb} \Yh_\a{}^{\gb\bb} \right) \d(\s-\s'), \cr 
&\big[\dd_{\ab}(\s) , \Lb^{\bb}(\s')\big] = \left( -\Lb^{\gb} \Oh_{\ab\gb}{}^{\bb} + \psib^a \Ch_{\ab a}{}^{\bb} - 2\wb_{\gb} \Yh_{\ab}{}^{\gb\bb} \right) \d(\s-\s'), \cr 
&\big[d_a(\s),\Lb^{\ab}(\s')\big] = \left( \Lb^{\bb} \Oh_{a\bb}{}^{\ab} + \psib^b \Ch_{ab}{}^{\ab} + 2\wb_{\bb} \Yh_a{}^{\bb\ab} \right) \d(\s-\s') ,
\label{}
\end{align}

\begin{align}
&\big[d_\a(\s),w_\b(\s')\big] = w_\g \O_{\a\b}{}^\g \d(\s-\s'), \cr 
&\big[\dd_{\ab}(\s),w_\b(\s')\big] = w_\g \O_{\ab\b}{}^\g \d(\s-\s'), \cr 
&\big[d_a(\s),w_\a(\s')\big] = -w_\b \O_{a\a}{}^\b \d(\s-\s') ,
\label{}
\end{align}

\begin{align}
&\big[d_\a(\s),\wb_{\bb}(\s')\big] = \wb_{\gb} \Oh_{\a\bb}{}^{\gb} \d(\s-\s'), \cr 
&\big[\dd_{\ab}(\s),\wb_{\bb}(\s')\big] = \wb_{\gb} \Oh_{\ab\bb}{}^{\gb} \d(\s-\s') ,\cr 
&\big[d_a(\s),\wb_{\ab}(\s')\big] = -\wb_{\bb} \Oh_{a\ab}{}^{\bb} \d(\s-\s') ,
\label{}
\end{align}

\begin{align}
\big\{\Pi^\a(\s),\psi_a(\s')\big\} &= P^{\a\bb}  \left( \psi^b \O_{\bb ab} + w_\g C_{\bb a}{}^{\g} \right)  \d(\s-\s'), \cr
\big\{\Pi^{\ab}(\s),\psi_a(\s')\big\} &= \left( -P^{\b\ab} \left( \psi^b \O_{\b ab} + w_\g C_{\b a}{}^\g \right) - 2\psi^b U_{ab}{}^{\ab} - w_\b \Vh_a{}^{\ab\b}  \right)\d(\s-\s') ,\cr 
&
\label{}
\end{align}

\begin{align}
    &\big\{\Pi^{\ab}(\s),\psib_a(\s')\big\} = P^{\b\ab}  \left( -\psib^b \Oh_{\b ab} - \wb_{\gb} \Ch_{\b a}{}^{\gb} \right) \d(\s-\s'), \cr 
    &\big\{\Pib^\a(\s),\psib_a(\s')\big\} = \left( P^{\a\bb} \left( \psib^b \Oh_{\bb ab} + \wb_{\gb} \Ch_{\bb a}{}^{\gb} \right) - 2\psib^b \Uh_{ab}{}^\a - \wb_{\bb} V_a{}^{\bb\a} \right) \d(\s-\s')   ,
\end{align}

\begin{align}
    &\big[\Pib^\a(\s),\L^\b(\s')\big] = P^{\a\gb} \left( \L^\r \O_{\gb\r}{}^\b - \psi^a C_{\gb a}{}^\b + 2 w_\r Y_{\gb}{}^{\r\b} \right) \d(\s-\s'), \cr 
    &\big[\Pi^{\bb}(\s),\L^\a(\s')\big] = \left( P^{\g\bb} \left( -\L^\r \O_{\g\r}{}^\a + \psi^a C_{\g a}{}^\a - 2w_\r Y_\g{}^{\r\a} \right)\right.\cr 
    &\qquad\qquad\qquad\qquad~+\left. 
    \L^\g \Dh_\g{}^{\a\bb} + \psi^a \Vh_a{}^{\bb\a} + 2 w_\g \Zh^{\bb\g\a} \right) \d(\s-\s') 
\end{align}

\begin{align}
    &\big[\Pi^{\ab}(\s),\Lb^{\bb}(\s')\big] = P^{\g\ab} \left( -\Lb^{\rb} \Oh_{\g\rb}{}^{\bb} + \psib^a \Ch_{\g a}{}^{\bb} - 2 \wb_{\rb} \Yh_\g{}^{\rb\bb} \right) \d(\s-\s') ,\cr 
    &\big[\Pib^\b(\s),\Lb^{\ab}(\s)\big] = \Big( -P^{\b\gb} \left( -\Lb^{\rb} \Oh_{\gb\rb}{}^{\ab} + \psib^a \Ch_{\gb a}{}^{\ab} - 2 \wb_{\rb} \Yh_{\gb}{}^{\rb\ab} \right) \cr
    &\qquad\qquad\qquad\qquad~    + \Lb^{\gb} D_{\gb}{}^{\ab\b} + \psib^a V_a{}^{\b\ab} + 2 \wb_{\gb} Z^{\b\gb\ab} \Big) \d(\s-\s') \cr 
    &
\end{align}

\begin{align}
    &\big[\Pib^\a(\s),w_\b(\s')\big] = - P^{\a\gb} w_\r \O_{\gb\b}{}^\r  \d(\s-\s'), \cr 
    &\big[\Pi^{\ab}(\s),w_\b(\s')\big] = (-w_\g \Dh_\b{}^{\g\ab} + P^{\g\ab} w_\r \O_{\g\b}{}^\r ) \d(\s-\s') ,
\end{align}

\begin{align}
&\big[\Pib^\a(\s),\wb_{\bb}(\s')\big] = (-\wb_{\gb} D_{\bb}{}^{\gb\a} - P^{\a\gb} \wb_{\rb} \Oh_{\gb\bb}{}^{\rb} ) \d(\s-\s'), \cr
&\big[\Pi^{\ab}(\s),\wb_{\bb}(\s')\big] = P^{\g\ab} \wb_{\rb} \Oh_{\g\bb}{}^{\rb} \d(\s-\s') .
\label{}
\end{align}

\begin{align}
&\big[d_a(\s),d_b(\s')\big] = E_a{}^M(\s) B_{Mb}(\s') \frac{\p}{\p\s'} \d(\s-\s') -  E_b{}^M(\s') B_{Ma}(\s) \frac{\p}{\p\s} \d(\s-\s') \cr
&+\left[ -\frac12 \psi^c \psi^d R_{abcd} - \frac12 \psib^c \psib^d \Rh_{abcd} + \L^\a w_\b R_{ab\a}{}^\b + \Lb^{\ab} \wb_{\bb} \Rh_{ab\ab}{}^{\bb} + \p_1 Z^N E_{[a}{}^M \p_{b]} B_{MN} \right. \cr
&+\left. \p_0 Z^M E_M{}^c T_{abc} - \p_0 Z^M  E_M{}^c \O_{[ab]c} - T_{ab}{}^\a d_\a - T_{ab}{}^{\ab} \dd_{\ab} - w_\a w_\b C_{ac}{}^\a C_{bd}{}^\b \eta^{cd} \right. \cr
&-\left. \wb_{\ab} \wb_{\bb} \Ch_{ac}{}^{\ab} \Ch_{bd}{}^{\bb} \eta^{cd} + \psi^c w_\a \left( \N_{[a} C_{b]c}{}^\a + T_{ab}{}^A C_{Ac}{}^\a \right) + \psib^c \wb_{\ab} \left( \Nh_{[a} \Ch_{b]c}{}^{\ab} + \Th_{ab}{}^A \Ch_{Ac}{}^{\ab} \right) \right. \cr
&+\left. w_\a w_\b \left( \N_{[a} Y_{b]}{}^{\a\b} + T_{ab}{}^A Y_A{}^{\a\b} \right) + \wb_{\ab} \wb_{\bb} \left( \Nh_{[a} \Yh_{b]}{}^{\ab\bb} + \Th_{ab}{}^A \Yh_A{}^{\ab\bb} \right)  \right] \d(\s-\s').
\label{}
\end{align}

\begin{align}
&\big\{d_\a(\s),d_\b(\s')\big\} = E_\a{}^M(\s) B_{M\b}(\s') \frac{\p}{\p\s'} \d(\s-\s') + E_\b{}^M(\s') B_{M\a}(\s) \frac{\p}{\p\s} \d(\s-\s') \cr
&+\left[ w_\g w_\r C_{\a a}{}^\g C_{\b b}{}^\r \eta^{ab} + \wb_{\gb} \wb_{\rb} \Ch_{\a a}{}^{\gb} \Ch_{\b b}{}^{\rb} \eta^{ab} + (-1)^M \p_1 Z^N E_{(\a}{}^M \p_{\b)} B_{MN} \right. \cr
&-\left. \frac12 \psi^a \psi^b R_{\a\b ab} - \frac12 \psib^a \psib^b \Rh_{\a\b ab} + \L^\g w_\r R_{\a\b\g}{}^\r + \Lb^{\gb} \wb_{\rb} \Rh_{\a\b\gb}{}^{\rb}  + \p_0 Z^M E_M{}^a T_{\a\b a} \right. \cr
&-\left. T_{\a\b}{}^\g d_\g + \O_{(\a\b)}{}^\g d_\g - T_{\a\b}{}^{\gb} \dd_{\gb}  \right. \cr
&+\left. \psi^a w_\g \left( \N_{(\a} C_{\b)a}{}^\g + T_{\a\b}{}^A C_{Aa}{}^\g \right) + \psib^a \wb_{\gb} \left(\Nh_{(\a} \Ch_{\b)a}{}^{\gb} + T_{\a\b}{}^A \Ch_{A a}{}^{\gb} \right) \right. \cr
&+\left. w_\g w_\r \left( \N_{(\a} Y_{\b)}{}^{\g\r} + T_{\a\b}{}^A Y_A{}^{\g\r} \right) + \wb_{\gb} \wb_{\rb} \left( \N_{(\a} \Yh_{\b)}{}^{\gb\rb} + T_{\a\b}{}^A \Yh_A{}^{\gb\bb} \right) \right] \d(\s-\s') .
\label{}
\end{align}

\begin{align}
&\big\{\dd_{\ab}(\s),\dd_{\bb}(\s')\big\} = E_{\ab}{}^M(\s) B_{M\bb}(\s') \frac{\p}{\p\s'} \d(\s-\s') + E_{\bb}{}^M(\s') B_{M\ab}(\s) \frac{\p}{\p\s} \d(\s-\s') \cr
&+\left[ w_\g w_\r C_{\ab a}{}^\g C_{\bb b}{}^\r \eta^{ab} + \wb_{\gb} \wb_{\rb} \Ch_{\ab a}{}^{\gb} \Ch_{\bb b}{}^{\rb} \eta^{ab} + (-1)^M \p_1 Z^N E_{(\ab}{}^M \p_{\bb)} B_{MN} \right. \cr
&-\left. \frac12 \psi^a \psi^b R_{\ab\bb ab} - \frac12 \psib^a \psib^b \Rh_{\ab\bb ab} + \L^\g w_\r R_{\ab\bb\g}{}^\r + \Lb^{\gb} \wb_{\rb} \Rh_{\ab\bb\gb}{}^{\rb} + \p_0 Z^M E_M{}^a T_{\ab\bb a} \right. \cr
&-\left. T_{\ab\bb}{}^\g d_\g - T_{\ab\bb}{}^{\gb} \dd_{\gb} + \Oh_{(\ab\bb)}{}^{\gb} \dd_{\gb} \right. \cr
&+\left. \psi^a w_\g \left( \N_{(\ab} C_{\bb)a}{}^\g + T_{\ab\bb}{}^A C_{Aa}{}^\g \right) + \psib^a \wb_{\gb} \left( \Nh_{(\ab} \Ch_{\bb)a}{}^{\gb} + T_{\ab\bb}{}^A \Ch_{Aa}{}^{\gb} \right) \right. \cr
&+\left. w_\g w_\r \left( \N_{(\ab} Y_{\bb)}{}^{\g\r} + T_{\ab\bb}{}^A Y_A{}^{\g\r} \right) + \wb_{\gb} \wb_{\rb} \left( \N_{(\ab} \Yh_{\bb)}{}^{\gb\rb} + T_{\ab\bb}{}^A \Yh_A{}^{\gb\rb} \right) \right] \d(\s-\s') .
\label{}
\end{align}

\begin{align}
&\big[d_a(\s),d_\a(\s')\big] = -E_a{}^M(\s) B_{M\a}(\s') \frac{\p}{\p\s'} \d(\s-\s') + E_\a{}^M(\s') B_{Ma}(\s) \frac{\p}{\p\s} \d(\s-\s') \cr
&+\left[ -w_\b w_\g C_{ab}{}^\b C_{\a c}{}^\g \eta^{bc} - \wb_{\bb} \wb_{\gb} \Ch_{ab}{}^{\bb} \Ch_{\a c}{}^{\gb} \eta^{bc} + (-1)^N \p_1 Z^N E_\a{}^M \p_a B_{MN} \right. \cr
&-\left. (-1)^{M+N} \p_1 Z^N E_a{}^M \p_\a B_{MN} - \frac12 \psi^b \psi^c R_{\a abc} - \frac12 \psib^b \psib^c \Rh_{\a abc} - \L^\b w_\g R_{a\a\b}{}^\g - \Lb^{\bb} \wb_{\gb} \Rh_{a\a\bb}{}^{\gb} \right. \cr
&+\left. \p_0 Z^M E_M{}^b T_{\a ab} - \p_0 Z^M E_M{}^b \O_{\a ab}  + T_{a\a}{}^\b d_\b - \O_{a\a}{}^\b d_\b + T_{a\a}{}^{\bb} \dd_{\bb} \right. \cr
&+\left. \psi^b w_\b \left( \N_{[a} C_{\a]b}{}^\b + T_{a\a}{}^A C_{Ab}{}^\b \right) + \psib^b \wb_{\bb} \left( \Nh_{[a} \Ch_{\a]b}{}^{\bb} + \Th_{a\a}{}^A \Ch_{Ab}{}^{\bb} \right) \right. \cr
&+\left. w_\b w_\g \left( \N_{[\a} Y_{a]}{}^{\b\g} + T_{\a a}{}^A Y_A{}^{\b\g} \right) + \wb_{\bb} \wb_{\gb} \left( \Nh_{[\a} \Yh_{a]}{}^{\bb\gb} + \Th_{\a a}{}^A \Yh_A{}^{\bb\gb} \right) \right] \d(\s-\s') .
\label{}
\end{align}

\begin{align}
&\big[d_a(\s),\dd_{\ab}(\s')\big] = -E_a{}^M(\s) B_{M\ab}(\s') \frac{\p}{\p\s'} \d(\s-\s') + E_{\ab}{}^M(\s') B_{Ma}(\s) \frac{\p}{\p\s} \d(\s-\s') \cr
&+\left[ -w_\b w_\g C_{ab}{}^\b C_{\ab c}{}^\g \eta^{bc} - \wb_{\bb} \wb_{\gb} \Ch_{ab}{}^{\bb} \Ch_{\ab c}{}^{\gb} \eta^{bc} + (-1)^N \p_1 Z^N E_{\ab}{}^M \p_a B_{MN} \right. \cr
&-\left. (-1)^{M+N} \p_1 Z^N E_a{}^M \p_{\ab} B_{MN} - \frac12 \psi^b \psi^c R_{\ab abc} - \frac12 \psib^b \psib^c \Rh_{\ab abc} - \L^\b w_\g R_{a\ab\b}{}^\g - \Lb^{\bb} \wb_{\gb} \Rh_{a\ab\bb}{}^{\gb} \right. \cr
&+\left. \p_0 Z^M E_M{}^b \Th_{\ab ab} - \p_0 Z^M E_M{}^b \Oh_{\ab ab}  + T_{a\ab}{}^\b d_\b - \Oh_{a\ab}{}^{\bb} \dd_{\bb} + T_{a\ab}{}^{\bb} \dd_{\bb} \right. \cr
&+\left. \psi^b w_\b \left( \N_{[a} C_{\ab]b}{}^\b + T_{a\ab}{}^A C_{Ab}{}^\b \right) + \psib^b \wb_{\bb} \left( \Nh_{[a} \Ch_{\ab]b}{}^{\bb} + \Th_{a\ab}{}^A \Ch_{Ab}{}^{\bb} \right) \right. \cr
&+\left. w_\b w_\g \left( \N_{[\ab} Y_{a]}{}^{\b\g} + T_{\ab a}{}^A Y_A{}^{\b\g} \right) + \wb_{\bb} \wb_{\gb} \left( \Nh_{[\ab} \Yh_{a]}{}^{\bb\gb} + \Th_{\ab a}{}^A \Yh_A{}^{\bb\gb} \right) \right] \d(\s-\s') .
\label{}
\end{align}

\begin{align}
&\big\{d_\a(\s),\dd_{\ab}(\s')\big\} = E_\a{}^M(\s) B_{M\ab}(\s') \frac{\p}{\p\s'} \d(\s-\s') + E_{\ab}{}^M(\s') B_{M\a}(\s) \frac{\p}{\p\s} \d(\s-\s')  \cr
&+\left[ w_\b w_\g C_{\a a}{}^\b C_{\ab b}{}^\g \eta^{ab} + \wb_{\bb}\wb_{\gb} \Ch_{\a a}{}^{\bb} \Ch_{\ab b}{}^{\gb}\eta^{ab} + (-1)^M \p_1 Z^N E_{(\a}{}^M \p_{\ab)} B_{MN} \right. \cr
&-\left. \frac12 \psi^a \psi^b R_{\a\ab ab} - \frac12 \psib^a \psib^b \Rh_{\a\ab ab} + \L^\b w_\g R_{\a\ab\b}{}^\g + \Lb^{\bb} w_{\gb} \Rh_{\a\ab\bb}{}^{\gb} \right. \cr
&+\left. \p_0 Z^M E_M{}^a T_{\a\ab a} - T_{\a\ab}{}^\b d_\b + \O_{\ab\a}{}^\b d_\b - T_{\a\ab}{}^{\bb} \dd_{\bb} + \Oh_{\a\ab}{}^{\bb} \dd_{\bb} \right. \cr
&+\left. \psi^a w_\b\left( \N_{(\a} C_{\ab)a}{}^\b + T_{\a\ab}{}^A C_{Aa}{}^\b\right) + \psib^a \wb_{\bb}\left( \Nh_{(\a} \Ch_{\ab)a}{}^{\bb} + T_{\a\ab}{}^A \Ch_{Aa}{}^{\bb} \right) \right. \cr
&+\left. w_\b w_\g \left( \N_{(\a} Y_{\ab)}{}^{\b\g} + T_{\a\ab}{}^A Y_A{}^{\b\g} \right) + \wb_{\bb} \wb_{\gb} \left( \N_{(\a} \Yh_{\ab)}{}^{\bb\gb} + T_{\a\ab}{}^A \Yh_A{}^{\bb\gb} \right) \right] \d(\s-\s') . 
\label{}
\end{align}

\begin{align}
&\big\{d_\a(\s),\Pi^\b(\s')\big\} = -2E_\a{}^M(\s) E_M{}^\b(\s') \frac{\p}{\p\s'} \d(\s-\s') - 2(-1)^M \p_1 Z^M \p_\a E_M{}^\b \d(\s-\s') \cr
&-\left( E_\a{}^M(\s) B_{M\gb}(\s') \frac{\p}{\p\s'} \d(\s-\s') + E_{\gb}{}^M(\s') B_{M\a}(\s) \frac{\p}{\p\s} \d(\s-\s') \right) P^{\b\gb}(\s')   \cr
&+\left[ \dd_{\gb} \N_\a P^{\b\gb}-\Lb^{\gb} \wb_{\rb}\N_\a D_{\gb}{}^{\rb\b}  - \psib^a \psib^b\Nh_\a \Uh_{ab}{}^\b + \psib^a \ob_{\gb}\Nh_\a V_a{}^{\b\gb} \right. \cr
&-\left.\wb_{\gb} \wb_{\rb} \N_\a Z^{\b\gb\rb} - \O_{\a\g}{}^\b\Pib^\g
\right] \d(\s-\s') \cr
&+\left[ -w_\r w_\s C_{\a a}{}^\r C_{\gb b}{}^\s \eta^{ab} - \wb_{\rb} \wb_{\overline\sigma} \Ch_{\a a}{}^{\rb} \Ch_{\gb b}{}^{\overline\sigma} \eta^{ab} - (-1)^M \p_1 Z^N E_{(\a}{}^M  
\p_{\gb)} B_{MN} + \frac12 \psi^a \psi^b R_{\a\gb ab} \right. \cr
&+\left. \frac12 \psib^a \psib^b \Rh_{\a\gb ab} - \L^\r w_\s R_{\a\gb\r}{}^\s - \Lb^{\rb} \wb_{\overline\sigma} \Rh_{\a\gb\rb}{}^{\overline\sigma} - \p_0 Z^M E_M{}^a T_{\a\gb a} + T_{\a\gb}{}^\r d_\r -  \O_{\gb\a}{}^\r d_\r + T_{\a\gb}{}^{\rb} \dd_{\rb}  \right. \cr
&-\left. \psi^a w_\r \left( \N_{(\a} C_{\gb)a}{}^\r + T_{\a\gb}{}^A C_{Aa}{}^\r \right) - \psib^a \wb_{\rb} \left( \Nh_{(\a} \Ch_{\gb)a}{}^{\rb} + T_{\a\gb}{}^A \Ch_{Aa}{}^{\rb} \right) \right. \cr
&-\left. w_\r w_\s \left( \N_{(\a} Y_{\gb)}{}^{\r\s} + T_{\a\gb}{}^A Y_A{}^{\r\s}\right) - \wb_{\rb} \wb_{\overline\sigma} \left( \N_{(\a} \Yh_{\gb)}{}^{\rb\overline\sigma} + T_{\a\gb}{}^A \Yh_A{}^{\rb\overline\sigma} \right) \right] P^{\b\gb} \d(\s-\s') \cr
&+\left[ \psib^a \wb_{\gb} \left( -\Ch_{\a a}{}^{\rb} D_{\rb}{}^{\gb\b} + 2\Ch_{\a b}{}^{\gb} \Uh_{ca}{}^\b \eta^{bc} \right) + \wb_{\gb} \wb_{\rb} \left( 2\Yh_\a{}^{\gb\overline\sigma} D_{\overline\sigma}{}^{\rb\b} + \Ch_{\a a}{}^{\gb} V_b{}^{\b\rb} \eta^{ab} \right) \right] \d(\s-\s') .
\label{}
\end{align}

\begin{align}
&\big[d_a(\s),\Pi^\a(\s')\big] = 2E_a{}^M(\s) E_M{}^\a(\s') \frac{\p}{\p\s'} \d(\s-\s') + 2 \p_1 Z^M \p_a E_M{}^\a \d(\s-\s') \cr
&+\left( E_a{}^M(\s) B_{M\bb}(\s') \frac{\p}{\p\s'} \d(\s-\s') - E_{\bb}{}^M(\s') B_{Ma}(\s) \frac{\p}{\p\s} \d(\s-\s') \right) P^{\a\bb}(\s')   \cr
&+\left[ \dd_{\bb}\N_a P^{\a\bb}+ \Lb^{\b} \wb_{\gb}\N_a D_{\bb}{}^{\gb\a}+ \psib^b \psib^c\Nh_a \Uh_{bc}{}^\a  \right. \cr
&+\left. \psib^b \ob_{\bb}\Nh_a V_b{}^{\a\bb}+  \wb_{\bb} \wb_{\gb}\N_a Z^{\a\bb\gb}+\O_{a\b}{}^\a\Pib^\b\right] \d(\s-\s') \cr
&+\left[ - (-1)^N \p_1 Z^N E_{\bb}{}^M  \p_a B_{MN} + (-1)^{M+N} \p_1 Z^N E_a{}^M \p_{\bb} B_{MN} \right. \cr
&+\left. \frac12 \psi^b \psi^c R_{\bb abc}  + \frac12 \psib^b \psib^c \Rh_{\bb abc} + \L^\g w_\r R_{a\bb\g}{}^\r + \Lb^{\gb} \wb_{\rb} \Rh_{a\bb\gb}{}^{\rb} - \p_0 Z^M E_M{}^b \Th_{\bb ab} \right. \cr
&+ w_\g w_\r C_{ab}{}^\g C_{\bb c}{}^\r \eta^{bc} + \wb_{\gb} \wb_{\rb} \Ch_{ab}{}^{\gb} \Ch_{\bb c}{}^{\rb} \eta^{bc}-\left. T_{a\bb}{}^\g d_\g  - T_{a\bb}{}^{\gb} \dd_{\gb} +\p_0 Z^M E_M{}^b \Oh_{\bb ab} \right. \cr
&-\left. \psi^b w_\g \left( \N_{[a} C_{\bb]b}{}^\g + T_{a\bb}{}^A C_{Ab}{}^\g \right) - \psib^b \wb_{\gb} \left( \Nh_{[a} \Ch_{\bb]b}{}^{\gb} + \Th_{a\bb}{}^A \Ch_{Ab}{}^{\gb} \right) \right. \cr
&-\left. w_\g w_\r \left( \N_{[\bb} Y_{a]}{}^{\g\r} + T_{\bb a}{}^A Y_A{}^{\g\r} \right) - \wb_{\gb} \wb_{\rb} \left( \Nh_{[\bb} \Yh_{a]}{}^{\gb\rb} + \Th_{\bb a}{}^A \Yh_A{}^{\gb\rb} \right) \right] P^{\a\bb} \d(\s-\s') \cr
&+\left[ \psib^b \wb_{\bb} \left( -\Ch_{ab}{}^{\gb} D_{\gb}{}^{\bb\a} + 2\Ch_{a c}{}^{\bb} \Uh_{db}{}^\a \eta^{cd} \right) + \wb_{\bb} \wb_{\gb} \left( -2\Yh_a{}^{\bb\rb} D_{\rb}{}^{\gb\a} - \Ch_{ab}{}^{\bb} V_c{}^{\a\gb} \eta^{bc} \right) \right] \d(\s-\s') \cr 
&
\label{}
\end{align}

\begin{align}
&\big\{\Pi^\a(\s),\Pi^\b(\s')\big\} = 2 P^{\a\gb} E_{\gb}{}^M(\s) E_M{}^\b(\s') \frac{\p}{\p\s'} \d(\s-\s') + 2 P^{\b\gb} E_{\gb}{}^M(\s') E_M{}^\a(\s) \frac{\p}{\p\s} \d(\s-\s') \cr
&+\left( 2(-1)^M \p_1 Z^M \p_{\gb} E_M{}^{(\a} P^{\b)\gb} - \dd_{\gb} P^{(\a \rb} \p_{\rb} P^{\b)\gb} \right)  \d(\s-\s') \cr
&+ P^{\a\gb} E_{\gb}{}^M(\s) B_{M\rb} P^{\b\rb}(\s') \frac{\p}{\p\s'} \d(\s-\s') + P^{\b\rb} E_{\rb}{}^M(\s') B_{M\gb} P^{\a\gb}(\s) \frac{\p}{\p\s} \d(\s-\s') \cr
&+\left[ w_\s w_\tau C_{\gb a}{}^\s C_{\rb b}{}^\tau \eta^{ab} + \wb_{\overline\sigma} \wb_{\taub} \Ch_{\gb a}{}^{\overline\sigma} \Ch_{\rb b}{}^{\taub} \eta^{ab} + (-1)^M \p_1 Z^N E_{(\gb}{}^M \p_{\rb)} B_{MN} \right.\cr
&-\left. \frac12 \psi^a \psi^b R_{\gb\rb ab} - \frac12 \psib^a \psib^b \Rh_{\gb\rb ab} + \L^\s w_\tau R_{\gb\rb\s}{}^\tau + \Lb^{\overline\sigma} \wb_{\taub} \Rh_{\gb\rb\overline\sigma}{}^{\taub} \right.\cr
&+\left. \p_0 Z^M E_M{}^a T_{\gb\rb a} - T_{\gb\rb}{}^\s d_\s - T_{\gb\rb}{}^{\overline\sigma} \dd_{\overline\sigma} + \Oh_{(\gb\rb)}{}^{\overline\sigma} \dd_{\overline\sigma}  \right. \cr
&+\left. \psi^a w_\s \left( \N_{(\gb} C_{\rb)a}{}^\s + T_{\gb\rb}{}^A C_{Aa}{}^\s \right) + \psib^a \wb_{\overline\sigma} \left( \Nh_{(\gb} \Ch_{\rb)a}{}^{\overline\sigma} + T_{\gb\rb}{}^A \Ch_{Aa}{}^{\overline\sigma} \right) \right. \cr
&+\left. w_\s w_\tau \left( \N_{(\gb} Y_{\rb)}{}^{\s\tau} + T_{\gb\rb}{}^A Y_A{}^{\s\tau} \right) + \wb_{\overline\sigma} \wb_{\taub} \left( \N_{(\gb} \Yh_{\rb)}{}^{\overline\sigma\taub} + T_{\gb\rb}{}^A \Yh_A{}^{\overline\sigma\taub} \right) \right] P^{\a\gb} P^{\b\rb} \d(\s-\s')  \cr
&+\left[ -\Lb^{\gb} \wb_{\rb} D_{\gb}{}^{\overline\sigma(\a} D_{\overline\sigma}{}^{\rb\b)} - 4 \psib^a \psib^b \Uh_{ac}{}^\a \Uh_{bd}{}^\b \eta^{cd} + \wb_{\gb} \wb_{\rb} V_a{}^{\gb\a} V_b{}^{\rb\b} \eta^{ab} \right. \cr
&+\left. \Lb^{\gb} \wb_{\rb} P^{(\a\overline\sigma} \N_{\overline\sigma} D_{\gb}{}^{\rb\b)} + \psib^a \wb_{\gb} P^{(\a\rb} \Ch_{\rb a}{}^{\taub} D_{\taub}{}^{\gb\b)} - 2\wb_{\gb} \wb_{\rb} P^{(\a\overline\sigma} \Yh_{\overline\sigma}{}^{\gb\taub} D_{\taub}{}^{\rb\b)} \right. \cr
&+\left. \psib^a \psib^b P^{(\a\gb} \Nh_{\gb} \Uh_{ab}{}^{\b)}- 2 \psib^a \wb_{\gb} P^{(\a\rb} \Ch_{\rb b}{}^{\gb} \Uh_{ca}{}^{\b)} \eta^{bc} \right. \cr
&-\left. \psib^a \wb_{\gb} P^{(\a\rb} \Nh_{\rb}V_a{}^{\gb\b)} - \wb_{\gb} \wb_{\rb} P^{(\a\overline\sigma} \Ch_{\overline\sigma a}{}^{\gb} V_b{}^{\rb\b)} \eta^{ab} \right. \cr
&+\left. \wb_{\gb} \wb_{\rb} P^{(\a\overline\sigma} \N_{\overline\sigma} Z^{\b)\gb\rb} - \psib^a \wb_{\gb} D_{\rb}{}^{\gb(\a} V_a{}^{\rb\b)} \right. \cr
&+\left. 2 \wb_{\gb}\wb_{\rb}D_{\overline\sigma}{}^{\gb(\a} Z^{\b)\overline\sigma\rb} -2 \psib^a \wb_{\gb} \Uh_{ab}{}^{(\a} V_c{}^{\gb\b)} \eta^{bc}+P^{(\a\gb}\O_{\gb\r}{}^{\b)} \left(\Pib^\r+\dd_{\overline\sigma}P^{\r\overline\sigma}\right)\right] \d(\s-\s') .
\label{}
\end{align}

\begin{align}
&\big\{\dd_{\ab}(\s),\Pib^{\bb}(\s')\big\} = 2E_{\ab}{}^M(\s) E_M{}^{\bb}(\s') \frac{\p}{\p\s'} \d(\s-\s') + 2(-1)^M \p_1 Z^M \p_{\ab} E_M{}^{\bb} \d(\s-\s') \cr
&+\left( E_\g{}^M(\s') B_{M\ab}(\s) \frac{\p}{\p\s} \d(\s-\s') + E_{\ab}{}^M(\s) B_{M\g}(\s') \frac{\p}{\p\s'} \d(\s-\s') \right) P^{\g\bb}(\s')   \cr
&+\left[ -d_{\g} \N_{\ab} P^{\g\bb}  - \L^{\g} w_{\r} \N_{\ab} \Dh_{\g}{}^{\r\bb}- \psi^a \psi^b\N_{\ab} U_{ab}{}^{\bb}  \right. \cr
&+\left. \psi^a w_{\g} \N_{\ab} \Vh_a{}^{\bb\g} - w_{\g} w_{\r}\N_{\ab} \Zh^{\bb\g\r}-\Oh_{\ab\gb}{}^{\bb} \Pi^{\gb}\right] \d(\s-\s') \cr
&+\left[ w_\r w_\s C_{\g a}{}^\r C_{\ab b}{}^\s \eta^{ab} + \wb_{\rb} \wb_{\overline\sigma} \Ch_{\g a}{}^{\rb} \Ch_{\ab b}{}^{\overline\sigma} \eta^{ab} + (-1)^M \p_1 Z^N E_{(\g}{}^M  
\p_{\ab)} B_{MN} - \frac12 \psi^a \psi^b R_{\g\ab ab} \right. \cr
&-\left. \frac12 \psib^a \psib^b \Rh_{\g\ab ab} + \L^\r w_\s R_{\g\ab\r}{}^\s + \Lb^{\rb} \wb_{\overline\sigma} \Rh_{\g\ab\rb}{}^{\overline\sigma} + \p_0 Z^M E_M{}^a T_{\g\ab a} - T_{\g\ab}{}^\r d_\r  - T_{\g\ab}{}^{\rb} \dd_{\rb} + \Oh_{\g\ab}{}^{\rb} \dd_{\rb} \right. \cr
&+\left. \psi^a w_\r \left( \N_{(\g} C_{\ab)a}{}^\r + T_{\g\ab}{}^A C_{Aa}{}^\r \right) + \psib^a \wb_{\rb} \left( \Nh_{(\g} \Ch_{\ab)a}{}^{\rb} + T_{\g\ab}{}^A \Ch_{Aa}{}^{\rb} \right) \right. \cr
&+\left. w_\r w_\s \left( \N_{(\g} Y_{\ab)}{}^{\r\s} + T_{\g\ab}{}^A Y_A{}^{\r\s}\right) + \wb_{\rb} \wb_{\overline\sigma} \left( \N_{(\g} \Yh_{\ab)}{}^{\rb\overline\sigma} + T_{\g\ab}{}^A \Yh_A{}^{\rb\overline\sigma} \right) \right] P^{\g\bb} \d(\s-\s') \cr
&+\Big[ \psi^a w_{\g} \left( -C_{\ab a}{}^{\r} \Dh_{\r}{}^{\g\bb} + 2 C_{\ab b}{}^{\g} U_{ca}{}^{\bb} \eta^{bc} \right)\cr
&+ w_{\g} w_{\r} \left( 2 Y_{\ab}{}^{\g\s} \Dh_{\s}{}^{\r\bb} + C_{\ab a}{}^{\g} \Vh_b{}^{\bb\r} \eta^{ab} \right) \Big] \d(\s-\s') .
\end{align}

\begin{align}
&\big[d_a(\s),\Pib^{\ab}(\s')\big] = -2E_a{}^M(\s) E_M{}^{\ab}(\s') \frac{\p}{\p\s'} \d(\s-\s') - 2 \p_1 Z^M \p_a E_M{}^{\ab} \d(\s-\s') \cr
&+\left( -E_a{}^M(\s) B_{M\b}(\s') \frac{\p}{\p\s'} \d(\s-\s') + E_{\b}{}^M(\s') B_{Ma}(\s) \frac{\p}{\p\s} \d(\s-\s') \right) P^{\b\ab}(\s')   \cr
&+\left[ -d_{\b}\N_a P^{\b\ab} + \L^{\b} w_{\g}  \N_a \Dh_{\b}{}^{\g\ab}  + \psi^b \psi^c \N_a U_{bc}{}^{\ab}  \right. \cr
&+\left. \psi^b w_{\b} \N_a \Vh_b{}^{\ab\b} + w_{\b} w_{\g}  \N_a \Zh^{\ab\b\g}+\Oh_{a\bb}{}^{\ab}\Pi^{\bb}  \right] \d(\s-\s') \cr
&+\left[ + (-1)^N \p_1 Z^N E_{\b}{}^M  \p_a B_{MN} - (-1)^{M+N} \p_1 Z^N E_a{}^M \p_{\b} B_{MN} \right. \cr
&-\left. \frac12 \psi^b \psi^c R_{\b abc}  - \frac12 \psib^b \psib^c \Rh_{\b abc} - \L^\g w_\r R_{a\b\g}{}^\r - \Lb^{\gb} \wb_{\rb} \Rh_{a\b\gb}{}^{\rb} + \p_0 Z^M E_M{}^b T_{\b ab} \right. \cr
&-\left.w_\g w_\r C_{ab}{}^\g C_{\b c}{}^\r \eta^{bc} - \wb_{\gb} \wb_{\rb} \Ch_{ab}{}^{\gb} \Ch_{\b c}{}^{\rb} \eta^{bc}+ T_{a\b}{}^\g d_\g  + T_{a\b}{}^{\gb} \dd_{\gb} -\p_0 Z^M E_M{}^b \O_{\b ab} \right. \cr
&+\left. \psi^b w_\g \left( \N_{[a} C_{\b]b}{}^\g + T_{a\b}{}^A C_{Ab}{}^\g \right) + \psib^b \wb_{\gb} \left( \Nh_{[a} \Ch_{\b]b}{}^{\gb} + \Th_{a\b}{}^A \Ch_{Ab}{}^{\gb} \right) \right. \cr
&+\left. w_\g w_\r \left( \N_{[\b} Y_{a]}{}^{\g\r} + T_{\b a}{}^A Y_A{}^{\g\r} \right) + \wb_{\gb} \wb_{\rb} \left( \Nh_{[\b} \Yh_{a]}{}^{\gb\rb} + \Th_{\b a}{}^A \Yh_A{}^{\gb\rb} \right) \right] P^{\b\ab} \d(\s-\s') \cr
&+\Big[ \psi^b w_{\b} \left( -C_{ab}{}^{\g} \Dh_{\g}{}^{\b\ab} + 2 C_{ac}{}^{\b} U_{db}{}^{\ab} \eta^{cd} \right)\cr 
&+ w_{\b} w_{\g} \left( -2 Y_a{}^{\b\r} \Dh_{\r}{}^{\g\ab} - C_{ab}{}^{\b} \Vh_c{}^{\ab\g} \eta^{bc} \right) \Big] \d(\s-\s')
\end{align}

\begin{align}
&\big\{\Pib^{\ab}(\s),\Pib^{\bb}(\s')\big\} = 2 P^{\g\ab} E_{\g}{}^M(\s) E_M{}^{\bb}(\s') \frac{\p}{\p\s'} \d(\s-\s') + 2 P^{\g\bb} E_{\g}{}^M(\s') E_M{}^{\ab}(\s) \frac{\p}{\p\s} \d(\s-\s') \cr
&+\left( 2(-1)^M \p_1 Z^M \p_{\g} E_M{}^{(\ab} P^{\g\bb)} - d_{\g} P^{\r (\ab} \p_{\r} P^{\g\bb)} \right)  \d(\s-\s') \cr
&+ P^{\g\ab} E_{\g}{}^M(\s) B_{M\r} P^{\r\bb}(\s') \frac{\p}{\p\s'} \d(\s-\s') + P^{\r\bb} E_{\r}{}^M(\s') B_{M\g} P^{\g\ab}(\s) \frac{\p}{\p\s} \d(\s-\s') \cr
&+\left[ w_\s w_\tau C_{\g a}{}^\s C_{\r b}{}^\tau \eta^{ab} + \wb_{\overline\sigma} \wb_{\taub} \Ch_{\g a}{}^{\overline\sigma} \Ch_{\r b}{}^{\taub} \eta^{ab} + (-1)^M \p_1 Z^N E_{(\g}{}^M \p_{\r)} B_{MN} \right.\cr
&-\left. \frac12 \psi^a \psi^b R_{\g\r ab} - \frac12 \psib^a \psib^b \Rh_{\g\r ab} + \L^\s w_\tau R_{\g\r\s}{}^\tau + \Lb^{\overline\sigma} \wb_{\taub} \Rh_{\g\r\overline\sigma}{}^{\taub} \right.\cr
&+\left. \p_0 Z^M E_M{}^a T_{\g\r a} - T_{\g\r}{}^\s d_\s - T_{\g\r}{}^{\overline\sigma} \dd_{\overline\sigma} + \O_{(\g\r)}{}^{\s} d_{\s}  \right. \cr
&+\left. \psi^a w_\s \left( \N_{(\g} C_{\r)a}{}^\s + T_{\g\r}{}^A C_{Aa}{}^\s \right) + \psib^a \wb_{\overline\sigma} \left( \Nh_{(\g} \Ch_{\r)a}{}^{\overline\sigma} + T_{\g\r}{}^A \Ch_{Aa}{}^{\overline\sigma} \right) \right. \cr
&+\left. w_\s w_\tau \left( \N_{(\g} Y_{\r)}{}^{\s\tau} + T_{\g\r}{}^A Y_A{}^{\s\tau} \right) + \wb_{\overline\sigma} \wb_{\taub} \left( \N_{(\g} \Yh_{\r)}{}^{\overline\sigma\taub} + T_{\g\r}{}^A \Yh_A{}^{\overline\sigma\taub} \right) \right] P^{\g\ab} P^{\r\bb} \d(\s-\s')  \cr
&+\left[ -\L^{\g} w_{\r} \Dh_{\g}{}^{\s(\ab} \Dh_{\s}{}^{\r\bb)} - 4 \psi^a \psi^b U_{ac}{}^{\ab} U_{bd}{}^{\bb} \eta^{cd} + w_{\g} w_{\r} \Vh_a{}^{\ab\g} \Vh_b{}^{\bb\r} \eta^{ab} \right. \cr
&-\left. \L^{\g} w_{\r} P^{\s(\ab} \N_{\s} \Dh_{\g}{}^{\r\bb)}  - \psi^a w_{\g} P^{\r(\ab} C_{\r a}{}^{\s} \Dh_{\s}{}^{\g\bb)} + 2 w_{\g} w_{\r} P^{\s(\ab} Y_{\s}{}^{\g\tau} \Dh_{\tau}{}^{\r\bb)} \right. \cr
&-\left. \psi^a \psi^b P^{\g(\ab} \N_{\g} U_{ab}{}^{\bb)}+ 2 \psi^a w_{\g} P^{\r(\ab} C_{\r b}{}^{\g} U_{ca}{}^{\bb)} \eta^{bc} \right. \cr
&+\left. \psi^a w_{\g} P^{\r(\ab}\N_{\r} \Vh_a{}^{\bb)\g} + w_{\g} w_{\r} P^{\s(\ab} C_{\s a}{}^{\g} \Vh_b{}^{\bb)\r} \eta^{ab} \right. \cr
&-\left. w_{\g} w_{\r} P^{\s(\ab}\N_{\s} \Zh^{\bb)\g\r}- \psi^a w_{\g} \Dh_{\r}{}^{\g(\ab} \Vh_a{}^{\bb)\r} \right. \cr
&+\left. 2 w_{\g}w_{\r}\Dh_{\s}{}^{\g(\ab} \Zh^{\bb)\s\r} -2 \psi^a w_{\g} U_{ab}{}^{(\ab} \Vh_c{}^{\bb)\g} \eta^{bc} +P^{\g(\ab}\Oh_{\g\rb}{}^{\bb)} \left(-\Pi^{\rb}+d_\s P^{\s\rb} \right) \right] \d(\s-\s') .
\end{align}

\begin{align}
    &\big\{\Pi^\a(\s),\Pib^{\bb}(\s')\big\} = -2 P^{\a\gb} E_{\gb}{}^M (\s) E_M{}^{\bb}(\s') \frac{\p}{\p\s'} \d(\s-\s') - 2 P^{\g\bb} E_\g{}^M(\s') E_M{}^\a(\s) \frac{\p}{\p\s} \d(\s-\s') \cr 
    &+\left( -2(-1)^M P^{\a\gb} \p_1 Z^M \p_{\gb} E_M{}^{\bb} - 2 (-1)^M P^{\g\bb} \p_1 Z^M \p_\g E_M{}^\a \right)\d(\s-\s') \cr 
    &+\left( P^{\r\bb} \dd_{\gb} \p_\r P^{\a\gb} + P^{\a\rb} d_\g \p_{\rb} P^{\g\bb} \right) \d(\s-\s') \cr 
    &-P^{\r\bb} E_\r{}^M(\s') B_{M\gb} P^{\a\gb}(\s) \frac{\p}{\p\s} \d(\s-\s') - P^{\a\gb} E_{\gb}{}^M(\s) B_{M\r} P^{\r\bb} (\s') \frac{\p}{\p\s'} \d(\s-\s') \cr 
    &-P^{\a\gb} P^{\r\bb} \left[ w_\s w_\tau C_{\r a}{}^\s C_{\gb b}{}^\tau \eta^{ab} + \wb_{\overline\sigma} \wb_{\taub}\Ch_{\r a}{}^{\overline\sigma} \Ch_{\gb b}{}^{\taub} \eta^{ab} + (-1)^M \p_1 Z^N E_{(\r}{}^M \p_{\gb)} B_{MN} \right. \cr 
    &-\left. \frac12 \psi^a \psi^b R_{\r\gb ab} -\frac12 \psib^a \psib^b \Rh_{\r\gb ab} + \L^\s w_\tau R_{\r\gb\s}{}^\tau + \Lb^{\overline\sigma} \wb_{\taub} \Rh_{\r\gb\overline\sigma}{}^{\taub} + \p_0 Z^M E_M{}^a T_{\r\gb a}  \right. \cr  
    &-\left. T_{\r\gb}{}^\s d_\s + \O_{\gb\r}{}^\s d_\s - T_{\r\gb}{}^{\overline\sigma} \dd_{\overline\sigma} + \Oh_{\r\gb}{}^{\overline\sigma} \dd_{\overline\sigma} \right. \cr 
    &+\left. \psi^a w_\s \left( \N_{(\r} C_{\gb)a}{}^\s + T_{\r\overline\sigma}{}^A C_{Aa}{}^\s \right) + \psib^a \wb_{\overline\sigma} \left( \Nh_{(\r} \Ch_{\gb)a}{}^{\overline\sigma} + T_{\r\gb}{}^A \Ch_{Aa}{}^{\overline\sigma} \right) \right. \cr 
    &+\left. w_\s w_\tau \left( \N_{(\r} Y_{\gb)}{}^{\s\tau} + T_{\r\gb}{}^A Y_A{}^{\s\tau} \right) + \wb_{\overline\sigma} \wb_{\taub} \left( \N_{(\r} \Yh_{\gb)}{}^{\overline\sigma\taub} + T_{\r\gb}{}^A \Yh_A{}^{\overline\sigma\taub} \right) \right] \d(\s-\s') \cr 
    &+\left[ \L^\g w_\r P^{\a\overline\sigma} \N_{\overline\sigma}\Dh_\g{}^{\r\bb} + \psi^a w_\g P^{\a\rb} C_{\rb a}{}^\s \Dh_\s{}^{\g\bb} - 2 w_\g w_\r P^{\a\overline\sigma} Y_{\overline\sigma}{}^{\g\tau} \Dh_\tau{}^{\r\bb} \right. \cr 
    &-\left. \Lb^{\gb} \wb_{\rb} P^{\s\bb}\N_\s D_{\gb}{}^{\rb\a} - \psib^a \wb_{\gb} P^{\r\bb} \Ch_{\r a}{}^{\overline\sigma} D_{\overline\sigma}{}^{\gb\a} + 2 \wb_{\gb}\wb_{\rb} P^{\s\bb} \Yh_\s{}^{\gb\taub} D_{\taub}{}^{\rb\a} \right.\cr 
    &+\left. \psi^a \psi^b P^{\a\gb} \N_{\gb} U_{ab}{}^{\bb} + 2 \psi^a w_\g P^{\a\rb} C_{\rb b}{}^\g U_{ac}{}^{\bb}\eta^{bc} \right.\cr 
    &-\left. \psib^a \psib^b P^{\g\bb} \Nh_\g \Uh_{ab}{}^\a- 2 \psib^a \wb_{\gb} P^{\r\bb} \Ch_{\r b}{}^{\gb} \Uh_{ac}{}^\a \eta^{bc} \right.\cr 
    &-\left. \psi^a w_\g P^{\a\rb}\N_{\rb} \Vh_a{}^{\bb\g}- w_\g w_\r P^{\a\overline\sigma} C_{\overline\sigma a}{}^\g \Vh_b{}^{\bb\r} \eta^{ab} \right. \cr 
    &+\left. \psib^a \wb_{\gb} P^{\r\bb}\Nh_\r V_a{}^{\gb\a}+ \wb_{\gb} \wb_{\rb} P^{\s\bb} \Ch_{\s a}{}^{\gb} V_b{}^{\rb\a} \eta^{ab} \right. \cr 
    &+\left. w_\g w_\r P^{\a\overline\sigma}\N_{\overline\sigma} \Zh^{\bb\g\r}- \wb_{\gb}\wb_{\rb} P^{\s\bb}\N_\s Z^{\a\gb\rb} \right. \cr 
    &+\left. P^{\a\gb} \Oh_{\gb\rb}{}^{\bb} \left( \Pi^{\rb}-d_\s P^{\s\rb} \right) - P^{\g\bb} \O_{\g\r}{}^\a \left(\Pib^\r+\dd_{\overline\sigma} P^{\r\overline\sigma}\right) \right] \d(\s-\s') .
 \end{align}

\begin{align}
    &\big\{d_\a(\s) , \Pib^{\bb}(\s')\big\} = 2 E_\a{}^M (\s) E_M{}^{\bb}(\s') \frac{\p}{\p\s'} \d(\s-\s') + 2 (-1)^M \p_1 Z^M \p_\a E_M{}^{\bb} \d(\s-\s') \cr 
    &+E_\a{}^M(\s) B_{M\g} P^{\g\bb}(\s') \frac{\p}{\p\s'} \d(\s-\s') + P^{\g\bb} E_\g{}^M(\s') B_{M\a}(\s) \frac{\p}{\p\s} \d(\s-\s') \cr 
    &+\left[ -d_\g \p_\a P^{\g\bb} - \L^\g w_\r \N_\a \Dh_\g{}^{\r\bb} - \psi^a \psi^b \N_\a U_{ab}{}^{\bb} \right.\cr 
    &+\left. \psi^a w_\g \N_\a \Vh_a{}^{\bb\g} - w_\g w_\r\N_\a \Zh^{\bb\g\r} - \Oh_{\a\gb}{}^{\bb} \left(\Pi^{\gb}-d_\r P^{\r\gb} \right) \right. \cr 
    &-\left. \psi^a w_\g C_{\a a}{}^\r \Dh_\r{}^{\g\bb} + 2 w_\g w_\r Y_\a{}^{\g\s}\Dh_\s{}^{\r\bb} -2 \psi^a w_\g C_{\a b}{}^\g U_{ac}{}^{\bb} \eta^{bc} + w_\g w_\r C_{\a a}{}^\g \Vh_b{}^{\bb\r} \eta^{ab} \right] \d(\s-\s') \cr 
    &+P^{\g\bb} \left[ w_\r w_\s C_{\a a}{}^\r C_{\g b}{}^\s \eta^{ab} + \wb_{\rb} \wb_{\overline\sigma} \Ch_{\a a}{}^{\rb} \Ch_{\g b}{}^{\overline\sigma} \eta^{ab} + (-1)^M \p_1 Z^N E_{(\a}{}^M \p_{\g)} B_{MN} \right. \cr 
    &-\left. \frac12 \psi^a \psi^b R_{\a\g ab} - \frac12 \psib^a \psib^b \Rh_{\a\g ab} + \L^\r w_\s R_{\a\g\r}{}^\s +\Lb^{\rb}\wb_{\overline\sigma} \Rh_{\a\g\rb}{}^{\overline\sigma}  \right. \cr 
    &+\left. \p_0 Z^M E_M{}^a T_{\a\g a} - T_{\a\g}{}^\r d_\r + \O_{(\a\g)}{}^\r d_\r - T_{\a\g}{}^{\rb} \dd_{\rb} \right. \cr 
    &+\left. \psi^a w_\r \left( \N_{(\a} C_{\g)a}{}^\r + T_{\a\g}{}^A C_{Aa}{}^\r \right) + \psib^a \wb_{\rb} \left( \Nh_{(\a} \Ch_{\g)a}{}^{\rb} + T_{\a\g}{}^A \Ch_{Aa}{}^{\rb} \right) \right. \cr 
    &+\left. w_\r w_\s \left( \N_{(\a} Y_{\g)}{}^{\r\s} + T_{\a\g}{}^A Y_A{}^{\r\s} \right) + \wb_{\rb} \wb_{\overline\sigma} \left( \N_{(\a} \Yh_{\g)}{}^{\rb\overline\sigma} + T_{\a\g}{}^A \Yh_A{}^{\rb\overline\sigma} \right) \right] \d(\s-\s') .
 \end{align}

\begin{align}
    &\big\{\dd_{\ab}(\s),\Pi^\b(\s')\big\}  = -2 E_{\ab}{}^M (\s) E_M{}^{\b}(\s') \frac{\p}{\p\s'} \d(\s-\s') - 2 (-1)^M \p_1 Z^M \p_{\ab} E_M{}^{\b} \d(\s-\s') \cr
    &-E_{\ab}{}^M(\s) B_{M\gb} P^{\b\gb}(\s') \frac{\p}{\p\s'} \d(\s-\s') - P^{\b\gb} E_{\gb}{}^M(\s') B_{M\ab}(\s) \frac{\p}{\p\s} \d(\s-\s') \cr 
    &+\left[ \dd_{\gb} \p_{\ab} P^{\b\gb} - \Lb^{\gb} \wb_{\rb} \N_{\ab} D_{\gb}{}^{\rb\b}- \psib^a \psib^b \Nh_{\ab} \Uh_{ab}{}^{\b}  \right.\cr 
    &+\left. \psib^a \wb_{\gb} \Nh_{\ab} V_a{}^{\gb\b} - \wb_{\gb} \wb_{\rb} \N_{\ab} Z^{\b\gb\rb} - \O_{\ab\g}{}^\b \left( \Pib^\g + \dd_{\rb} P^{\g\rb} \right) \right. \cr 
    &-\left. \psib^a \wb_{\gb} \Ch_{\ab a}{}^{\rb} D_{\rb}{}^{\gb\b} + 2 \wb_{\gb} \wb_{\rb} \Yh_{\ab}{}^{\gb\overline\sigma} D_{\overline\sigma}{}^{\rb\b} -2 \psib^a \wb_{\gb} \Ch_{\ab b}{}^{\gb} \Uh_{ac}{}^{\b} \eta^{bc} + \wb_{\gb} \wb_{\rb} \Ch_{\ab a}{}^{\gb} V_b{}^{\rb\b} \eta^{ab}  \right] \d(\s-\s') \cr 
    &-P^{\b\gb} \left[ w_\r w_\s C_{\ab a}{}^\r C_{\gb b}{}^\s \eta^{ab} + \wb_{\rb} \wb_{\overline\sigma} \Ch_{\ab a}{}^{\rb} \Ch_{\gb b}{}^{\overline\sigma} \eta^{ab} + (-1)^M \p_1 Z^N E_{(\ab}{}^M \p_{\gb)} B_{MN} \right. \cr 
    &-\left. \frac12 \psi^a \psi^b R_{\ab\gb ab} - \frac12 \psib^a \psib^b \Rh_{\ab\gb ab} + \L^\r w_\s R_{\ab\gb\r}{}^\s +\Lb^{\rb}\wb_{\overline\sigma} \Rh_{\ab\gb\rb}{}^{\overline\sigma}  \right. \cr 
    &+\left. \p_0 Z^M E_M{}^a T_{\ab\gb a} - T_{\ab\gb}{}^\r d_\r + \Oh_{(\ab\gb)}{}^{\rb} \dd_{\rb} - T_{\ab\gb}{}^{\rb} \dd_{\rb} \right. \cr 
    &+\left. \psi^a w_\r \left( \N_{(\ab} C_{\gb)a}{}^\r + T_{\ab\gb}{}^A C_{Aa}{}^\r \right) + \psib^a \wb_{\rb} \left( \Nh_{(\ab} \Ch_{\gb)a}{}^{\rb} + T_{\ab\gb}{}^A \Ch_{Aa}{}^{\rb} \right) \right. \cr 
    &+\left. w_\r w_\s \left( \N_{(\ab} Y_{\gb)}{}^{\r\s} + T_{\ab\gb}{}^A Y_A{}^{\r\s} \right) + \wb_{\rb} \wb_{\overline\sigma} \left( \N_{(\ab} \Yh_{\gb)}{}^{\rb\overline\sigma} + T_{\ab\gb}{}^A \Yh_A{}^{\rb\overline\sigma} \right) \right] \d(\s-\s') .
\end{align}

\section{Equations of motion}\label{app2}
In this section we list the equations of motion derived from the action (\ref{action}). We start by listing the equations of motion for the worldsheet fields $\big\{\psi^a,\Lambda^{\alpha},w_{\alpha},\hpsi^a\,\hLambda^{\halpha},\hw_{\halpha}\big\}$ which are easy to compute. These are
\begin{align}
    &\Nb\psi_a + w_\a \Pib^A C_{Aa}{}^\a - 2d_{\ab} \psi^b U_{ab}{}^{\ab} - \dd_{\bb} w_\a \Vh_a{}^{\bb\a} + 2 \Lb^{\ab} \wb_{\bb} \psi^b \Eh_{ab\ab}{}^{\bb} + \Lb^{\ab} \wb_{\bb} w_\g \Fh_{a\ab}{}^{\bb\g} \cr 
    &+ 2 \psi^b \psib^c \psib^d H_{abcd} + 2\psi^b \psib^c \wb_{\ab} I_{abc}{}^{\ab} + \psib^b \psib^c w_\a \Ih_{bca}{}^\a + 2 \psi^b \wb_{\ab} \wb_{\bb} J_{ab}{}^{\ab\bb} + \psib^b  w_\a
    \wb_{\bb} K_{ab}{}^{\a\bb} \cr 
    &+ w_\a \wb_{\bb} \wb_{\gb} L_a{}^{\a\bb\gb} = 0 . 
\end{align}

\begin{align}
    &-\Nb w_\a + w_\b \dd_{\gb} \Dh_\a{}^{\b\gb} + w_\b \Lb^{\gb} \wb_{\rb} S_{\a\gb}{}^{\b\rb} + w_\b \psib^a \psib^b E_{ab\a}{}^\b \cr 
    &+ w_\b \psib^a \wb_{\gb} F_{a\a}{}^{\b\gb} + w_\b \wb_{\gb} \wb_{\rb} G_\a{}^{\b\gb\rb} = 0 .
\end{align}

\begin{align}
    &\Nb \L^\a + \psi^a \Pib^A C_{Aa}{}^\a + 2 w_\b \Pib^A Y_A{}^{\a\b} + \L^\b \dd_{\gb} \Dh_\b{}^{\a\gb} - \psi^a \dd_{\bb} \Vh_a{}^{\bb\a} + 2 w_\b \dd_{\gb} \Zh^{\gb\b\a} \cr 
    &+ \L^\b \Lb^{\gb} \wb_{\rb} S_{\b\gb}{}^{\a\rb} + \L^\b \psib^a \psib^b E_{ab\b}{}^\a + \L^\b \psib^a \wb_{\gb} F_{a\b}{}^{\a\gb} + \Lb^{\bb} \wb_{\gb} \psi^a \Fh_{a\bb}{}^{\gb\a} + \L^\b \wb_{\gb}\wb_{\rb} G_\b{}^{\a\gb\rb}\cr& + 2 w_\b \Lb^{\gb} \wb_{\rb} \Gh_{\gb}{}^{\rb\b\a}+ \psi^a \psib^b \psib^c \Ih_{bca}{}^\a + 2 w_\b \psib^a \psib^b \Jh_{ab}{}^{\b\a} + \psi^a \wb_{\bb} \wb_{\gb} L_a{}^{\a\bb\gb} + 2 w_\b \psib^a \wb_{\gb} \Lh_a{}^{\gb\b\a} \cr 
    &+ \psi^a \psib^b \wb_{\bb} K_{ab}{}^{\a\bb}+ 2 w_\b \wb_{\gb} \wb_{\rb} M^{\a\b\gb\rb} = 0 . 
\end{align}

\begin{align}
    &\Nh \psib_a + \Pi^A \wb_{\ab} \Ch_{Aa}{}^{\ab} - 2 d_\a \psib^b \Uh_{ab}{}^\a - d_\a \wb_{\bb} V_a{}^{\a\bb} + 2 \L^\a w_\b \psib^b E_{ab\a}{}^\b + \L^\a w_\b \wb_{\gb} F_{a\a}{}^{\b\gb} \cr 
    &+ 2 \psi^b \psi^c \psib^d H_{bcad} + \psi^b \psi^c \wb_{\ab} I_{bca}{}^{\ab} - 2 \psi^b w_\a \psib^c \Ih_{acb}{}^\a + 2 w_\a w_\b \psib^b \Jh_{ab}{}^{\a\b} - \psi^b w_\a \wb_{\bb} K_{ba}{}^{\a\bb} \cr 
    &+ w_\a w_\b \wb_{\gb} \Lh_a{}^{\gb\a\b} = 0 .
\end{align}

\begin{align}
    &-\N \wb_{\ab} + d_\b \wb_{\gb} D_{\ab}{}^{\gb\b} + \L^\b w_\g \wb_{\rb} S_{\b\ab}{}^{\g\rb} + \psi^a \psi^b \wb_{\bb} \Eh_{ab\ab}{}^{\bb} \cr 
    &+\psi^a w_\b \wb_{\gb} \Fh_{a\ab}{}^{\gb\b} + w_\b w_\g \wb_{\rb} \Gh_{\ab}{}^{\rb\b\g} = 0 .
\end{align}

\begin{align}
    &\N \Lb^{\ab} + \psib^a \Pi^A \Ch_{Aa}{}^{\ab} + 2 \wb_{\bb} \Pi^A \Yh_A{}^{\bb\ab} + d_\b \Lb^{\gb} D_{\gb}{}^{\ab\b} + d_\b \psib^a V_a{}^{\b\ab} + 2 d_\b \wb_{\gb} Z^{\b\gb\ab} \cr 
    &+\L^\b w_\g \Lb^{\rb} S_{\b\rb}{}^{\g\ab} + \Lb^{\bb} \psi^a \psi^b \Eh_{ab\bb}{}^{\ab} + \L^\b w_\g \psib^a F_{a\b}{}^{\g\ab} + \Lb^{\bb} \psi^a w_\g \Fh_{a\bb}{}^{\ab\g} + 2 \L^\b w_\g \wb_{\rb} G_\b{}^{\g\rb\ab}\cr
    &+ \Lb^{\bb} w_\g w_\r \Gh_{\bb}{}^{\ab\g\r}+\psi^a \psi^b \psib^c I_{abc}{}^{\ab} + 2 \psi^a \psi^b \wb_{\bb} J_{ab}{}^{\ab\bb} + 2 \psi^a w_\b \wb_{\gb} L_a{}^{\b\gb\ab} + \psib^a w_\b w_\g \Lh_a{}^{\ab\b\g} \cr 
    &+ \psi^a w_\b \psib^b K_{ab}{}^{\b\ab}+2 w_\b w_\g \wb_{\rb} M^{\b\g\rb\ab} = 0 .
\end{align}

The procedure to find the equations of motion for the $d_A$'s is more intricate. We start by varying the action with respect to the supercoordinates $Z^P$:
\begin{equation} \label{deltaZP}
    \frac{\delta S}{\delta Z^P}=\frac{1}{2}\eta_{ab}\Big(-\B{\partial}E^aE_P\,^b-E_P\,^a\partial\B{E}^b-\partial Z^M\partial_{[M}E_{P]}\,^a\B{E}^b-\B{\partial}Z^M\partial_{[M}E_{P]}\,^aE^b\Big)+\frac{1}{2}E^A\B{E}^BH_{BAP}
\end{equation}
$$
+\frac{1}{2}\B{\partial}Z^M\partial_{[M}\Omega_{P]ab}\psi^a\psi^b+\frac{1}{2}\Omega_{Pab}\B{\partial}(\psi^a\psi^b)-\Lambda^{\alpha}w_{\beta}\B{\partial}Z^M\partial_{[M}\Omega_{P]\alpha}\,^{\beta}-\Omega_{P\alpha}\,^{\beta}\B{\partial}(\Lambda^{\alpha}w_{\beta})
$$
$$
+\B{\partial}Z^M\partial_{[M}C_{P]a}\,^{\beta}\psi^aw_{\beta}+C_{Pa}\,^{\beta}\B{\partial}(\psi^aw_{\beta})-w_{\alpha}w_{\beta}\B{\partial}Z^M\partial_{[M}Y_{P]}\,^{\alpha\beta}-Y_P\,^{\alpha\beta}\B{\partial}(w_{\alpha}w_{\beta})
$$
$$
+\frac{1}{2}{\partial}Z^M\partial_{[M}\hOmega_{P]ab}\hpsi^a\hpsi^b+\frac{1}{2}\hOmega_{Pab}{\partial}(\hpsi^a\hpsi^b)-\hLambda^{\halpha}\hw_{\hbeta}{\partial}Z^M\partial_{[M}\hOmega_{P]\halpha}\,^{\hbeta}-\hOmega_{P\halpha}\,^{\hbeta}{\partial}(\hLambda^{\halpha}\hw_{\hbeta})
$$
$$
+{\partial}Z^M\partial_{[M}\hC_{P]a}\,^{\hbeta}\hpsi^a\hw_{\hbeta}+\hC_{Pa}\,^{\hbeta}{\partial}(\hpsi^a\hw_{\hbeta})-\hw_{\halpha}\hw_{\hbeta}\partial{Z}^M\partial_{[M}\hY_{P]}\,^{\halpha\hbeta}-\hY_P\,^{\halpha\hbeta}{\partial}(\hw_{\halpha}\hw_{\hbeta})
$$
$$
+\B{\partial}Z^M\partial_{[M}E_{P]}\,^{\alpha}d_{\alpha}+E_P\,^{\alpha}\B{\partial}d_{\alpha}+\partial Z^M\partial_{[M}E_{P]}\,^{\halpha}\hd_{\halpha}+E_P\,^{\halpha}\partial\hd_{\halpha}+\partial_PP^{\alpha\halpha}d_{\alpha}\hd_{\halpha}
$$
$$
-\partial_PU^{\hgamma}_{ab}\psi^a\psi^b\hd_{\hgamma}-\partial_P\hD_{\alpha}^{\beta\hgamma}\Lambda^{\alpha}w_{\beta}\hd_{\hgamma}-\partial_P\hV_a^{\beta\hgamma}\psi^aw_{\beta}\hd_{\hgamma}-\partial_P\HH{Z}^{\alpha\beta\hgamma}w_{\alpha}w_{\beta}\hd_{\hgamma}
$$
$$
-\partial_P\HH{U}^{\alpha}_{bc}d_{\alpha}\hpsi^b\hpsi^c-\partial_PD_{\hbeta}^{\alpha\hgamma}d_{\alpha}\hLambda^{\hbeta}\hw_{\hgamma}+\partial_PV_b^{\alpha\hgamma}d_{\alpha}\hpsi^b\hw_{\hgamma}-\partial_PZ^{\alpha\hbeta\hgamma}d_{\alpha}\hw_{\hbeta}\hw_{\hgamma}
$$
$$
+\psi^a\psi^b\hpsi^c\hpsi^d\partial_PH_{abcd}+\psi^a\psi^b\hLambda^{\hgamma}\hw_{\hdelta}\partial_P\HH{E}_{ab\hgamma}\,^{\hdelta}+\Lambda^{\alpha}w_{\beta}\hpsi^c\hpsi^d\partial_PE_{\alpha cd}^{\beta}+\Lambda^{\alpha}w_{\beta}\hLambda^{\hgamma}\hw_{\hdelta}\partial_PS_{\alpha\hgamma}^{\beta\hdelta}
$$
$$
-\partial_P\HH{I}_{acd}^{\beta}\psi^aw_{\beta}\hpsi^c\hpsi^d-\partial_P\HH{F}_{a\hgamma}^{\beta\hdelta}\psi^aw_{\beta}\hLambda^{\hgamma}\hw_{\hdelta}-\partial_PI_{abc}^{\hdelta}\psi^a\psi^b\hpsi^c\hw_{\hdelta}-\partial_PF^{\beta\hdelta}_{\alpha c}\Lambda^{\alpha}w_{\beta}\hpsi^c\hw_{\hdelta}
$$
$$
+w_{\alpha}w_{\beta}\hpsi^c\hpsi^d\partial_P \HH{J}_{cd}^{\alpha\beta}+w_{\alpha}w_{\beta}\hLambda^{\hgamma}\hw_{\hdelta}\partial_P \HH{G}_{\hgamma}^{\alpha\beta\hdelta}+\psi^a\psi^b\hw_{\hgamma}\hw_{\hdelta}\partial_P J_{ab}^{\hgamma\hdelta}+\Lambda^{\alpha}w_{\beta}\hw_{\hgamma}\hw_{\hdelta}\partial_P G_{\alpha}^{\beta\hgamma\hdelta}
$$
$$
-\partial_P\HH{L}_c^{\alpha\beta\hdelta}w_{\alpha}w_{\beta}\hpsi^c\hw_{\hdelta}-\partial_PL_a^{\beta\hgamma\hdelta}\psi^aw_{\beta}\hw_{\hgamma}\hw_{\hdelta}+\psi^aw_{\beta}\hpsi^c\hw_{\hdelta}\partial_PK_{ac}^{\beta\hdelta}+w_{\alpha}w_{\beta}\hw_{\hgamma}\hw_{\hdelta}\partial_PM^{\alpha\beta\hgamma\hdelta}=0
$$
We then note that
\begin{equation}
    \eta_{ab}{\partial}Z^M\partial_{[M}E_{P]}\,^a\B{E}^b=\eta_{ab}E^A\B{E}^bT_{AP}\,^a+\eta_{ab}E^c\Omega_{Pc}\,^a\B{E}^b-\eta_{ab}E_P\,^cE^A\Omega_{Ac}\,^a\B{E}^b
\end{equation}
such that the first line in (\ref{deltaZP}) can be rewritten as
\begin{equation}
    -\frac{1}{2}\eta_{ab}E_P\,^a\big(\nabla \B{E}^b+\B{\nabla}E^b\big)-\frac{1}{2}\big(E^A\B{E}^a+\B{E}^AE^a\big)T_{AP a}+\frac{1}{2}E^A\B{E}^BH_{BAP}.
\end{equation}
Furthermore, we also rewrite
\begin{equation}
    \B{\partial}Z^M\partial_{[M}E_{P]}\,^{\alpha}d_{\alpha}=\B{\partial}Z^MT_{MP}\,^{\alpha}d_{\alpha}-E_P\,^{\beta}\B{\partial}Z^M\Omega_{M\beta}\,^{\alpha}d_{\alpha}+\Omega_{P\beta}\,^{\alpha}\B{E}^{\beta}d_{\alpha},
\end{equation}
\begin{equation}
    {\partial}Z^M\partial_{[M}E_{P]}\,^{\halpha}\hd_{\halpha}={\partial}Z^M\hT_{MP}\,^{\halpha}\hd_{\halpha}-E_P\,^{\hbeta}\partial Z^M\hOmega_{M\hbeta}\,^{\halpha}\hd_{\halpha}+\hOmega_{P\hbeta}\,^{\halpha}{E}^{\hbeta}\hd_{\halpha}.
\end{equation}

Then, we use the equations of motion for $\B{E}^{\alpha},E^{\halpha}$ and for the worldsheet fields $\psi^a,\hpsi^a,\Lambda^{\alpha},\hLambda^{\halpha},w_{\alpha}$ and $\hw_{\halpha}$ to finally obtain the expression from which we extract the equations of motion for the $d_A$. We arrive at
\newpage
\begin{equation} \label{eomdA}
    -\frac{1}{2}\eta_{ab}E_P\,^a\big(\nabla \B{E}^b+\B{\nabla}E^b\big)-\frac{1}{2}\big(E^A\B{E}^a+\B{E}^AE^a\big)T_{AP a}+\frac{1}{2}E^A\B{E}^BH_{BAP}
\end{equation}
$$
    +\B{E}^AT_{AP}\,^{\alpha}d_{\alpha}+E_P\,^{\alpha}\B{\nabla}d_{\alpha}+E^A\hT_{AP}\,^{\halpha}\hd_{\halpha}+E_P\,^{\halpha}\nabla\hd_{\halpha}
$$
$$
    +\frac{1}{2}\B{E}^AR_{APcd}\psi^c\psi^d-\B{E}^AR_{AP\gamma}\,^{\delta}\Lambda^{\gamma}w_{\delta}+\frac{1}{2}{E}^A\hR_{APcd}\hpsi^c\hpsi^d-\B{E}^A\hR_{AP\hgamma}\,^{\hdelta}\hLambda^{\hgamma}\hw_{\hdelta}
$$
$$
    +\B{E}^A\Big(\nabla_{[A}C_{P]c}\,^{\delta}+T_{AP}\,^{E}C_{Ec}\,^{\delta}\Big)\psi^cw_{\delta}+\B{E}^A\Big(\nabla_{[A}Y_{P]}\,^{\gamma\delta}+T_{AP}\,^EY_E\,^{\gamma\delta}-(-1)^PC_A\,^{e\gamma}C_{Pe}\,^{\delta}\Big)w_{\gamma}w_{\delta}
$$
$$
    +{E}^A\Big(\hnabla_{[A}\hC_{P]c}\,^{\hdelta}+\hT_{AP}\,^{E}\hC_{Ec}\,^{\hdelta}\Big)\hpsi^c\hw_{\hdelta}+{E}^A\Big(\hnabla_{[A}\hY_{P]}\,^{\hgamma\hdelta}+\hT_{AP}\,^E\hY_E\,^{\hgamma\hdelta}-(-1)^P\hC_A\,^{e\hgamma}\hC_{Pe}\,^{\hdelta}\Big)\hw_{\hgamma}\hw_{\hdelta}
$$
$$
+\nabla_PP^{\alpha\halpha}d_{\alpha}\hd_{\halpha}-\nabla_PU_{ab}^{\hgamma}\psi^a\psi^b\hd_{\hgamma}-\nabla_P\hD_{\alpha}^{\beta\hgamma}\Lambda^{\alpha}w_{\beta}\hd_{\hgamma}-\nabla_P\hV_a^{\beta\hgamma}\psi^aw_{\beta}\hd_{\hgamma}-\nabla_P\HH{Z}^{\alpha\beta\hgamma}w_{\alpha}w_{\beta}\hd_{\hgamma}
$$
$$
-\hnabla_P\HH{U}_{bc}^{\alpha}d_{\alpha}\hpsi^b\hpsi^c-\hnabla_PD_{\hbeta}^{\alpha\hgamma}d_{\alpha}\hLambda^{\hbeta}\hw_{\hgamma}+\hnabla_PV_b^{\alpha\hgamma}d_{\alpha}\hpsi^b\hw_{\hgamma}-\hnabla_PZ^{\alpha\hbeta\hgamma}d_{\alpha}\hw_{\hbeta}\hw_{\hgamma}
$$
$$
+\psi^a\psi^b\hpsi^c\hpsi^d\nabla_PH_{abcd}+\psi^a\psi^b\hLambda^{\hgamma}\hw_{\hdelta}\nabla_P\HH{E}_{ab\hgamma}\,^{\hdelta}+\Lambda^{\alpha}w_{\beta}\hpsi^c\hpsi^d\nabla_PE_{\alpha cd}^{\beta}+\Lambda^{\alpha}w_{\beta}\hLambda^{\hgamma}\hw_{\hdelta}\nabla_PS_{\alpha\hgamma}^{\beta\hdelta}
$$
$$
-\nabla_P\HH{I}_{acd}^{\beta}\psi^aw_{\beta}\hpsi^c\hpsi^d-\nabla_P\HH{F}_{a\hgamma}^{\beta\hdelta}\psi^aw_{\beta}\hLambda^{\hgamma}\hw_{\hdelta}-\nabla_PI_{abc}^{\hdelta}\psi^a\psi^b\hpsi^c\hw_{\hdelta}-\nabla_PF^{\beta\hdelta}_{\alpha c}\Lambda^{\alpha}w_{\beta}\hpsi^c\hw_{\hdelta}
$$
$$
+w_{\alpha}w_{\beta}\hpsi^c\hpsi^d\nabla_P \HH{J}_{cd}^{\alpha\beta}+w_{\alpha}w_{\beta}\hLambda^{\hgamma}\hw_{\hdelta}\nabla_P \HH{G}_{\hgamma}^{\alpha\beta\hdelta}+\psi^a\psi^b\hw_{\hgamma}\hw_{\hdelta}\nabla_P J_{ab}^{\hgamma\hdelta}+\Lambda^{\alpha}w_{\beta}\hw_{\hgamma}\hw_{\hdelta}\nabla_P G_{\alpha}^{\beta\hgamma\hdelta}
$$
$$
-\nabla_P\HH{L}_c^{\alpha\beta\hdelta}w_{\alpha}w_{\beta}\hpsi^c\hw_{\hdelta}-\nabla_PL_a^{\beta\hgamma\hdelta}\psi^aw_{\beta}\hw_{\hgamma}\hw_{\hdelta}+\psi^aw_{\beta}\hpsi^c\hw_{\hdelta}\nabla_PK_{ac}^{\beta\hdelta}+w_{\alpha}w_{\beta}\hw_{\hgamma}\hw_{\hdelta}\nabla_PM^{\alpha\beta\hgamma\hdelta}
$$
\begin{equation*}
\begin{split}
    +C_P\,^{a\alpha}\Big(&-2U_{ab}^{\hgamma}\psi^bw_{\alpha}\hd_{\hgamma}+\hV_a^{\beta\hgamma}w_{\alpha}w_{\beta}\hd_{\hgamma}\\
    &-2H_{abcd}w_{\alpha}\psi^b\hpsi^c\hpsi^d-2\HH{E}_{ab\hgamma}\,^{\hdelta}w_{\alpha}\psi^b\hLambda^{\hgamma}\hw_{\hdelta}-2I_{abc}^{\hdelta}w_{\alpha}\psi^b\hpsi^c\hw_{\hdelta}-2J_{ab}^{\hgamma\hdelta}w_{\alpha}\psi^b\hw_{\hgamma}\hw_{\hdelta}\\
    &-\HH{I}_{acd}^{\beta}w_{\alpha}w_{\beta}\hpsi^c\hpsi^d-\HH{F}_{a\hgamma}^{\beta\hdelta}w_{\alpha}w_{\beta}\hLambda^{\hgamma}\hw_{\hdelta}-K_{ac}^{\beta\hdelta}w_{\alpha}w_{\beta}\hpsi^c\hw_{\hdelta}-L_a^{\beta\hgamma\hdelta}w_{\alpha}w_{\beta}\hw_{\hdelta}\hw_{\hgamma}\Big)
\end{split}
\end{equation*}
\begin{equation*}
\begin{split}
    +C_{Pa}\,^{\alpha}\Big(\hD_{\alpha}^{\beta\hgamma}\psi^aw_{\beta}\hd_{\hgamma}+E_{\alpha cd}^{\beta}\psi^aw_{\beta}\hpsi^c\hpsi^d+S_{\alpha\hgamma}^{\beta\hdelta}\psi^aw_{\beta}\hLambda^{\hgamma}\hw_{\hdelta}+F_{\alpha c}^{\beta\hdelta}\psi^aw_{\beta}\hpsi^c\hw_{\hdelta}+G_{\alpha}^{\beta\hgamma\hdelta}\psi^aw_{\beta}\hw_{\hgamma}\hw_{\hdelta}\Big)
\end{split}
\end{equation*}
\begin{equation*}
    +2Y_P\,^{\alpha\gamma}\Big(\hD_{\alpha}^{\beta\hgamma}w_{\beta}w_{\gamma}\hd_{\hgamma}-E_{\alpha cd}^{\beta}w_{\beta}w_{\gamma}\hpsi^c\hpsi^d-S_{\alpha\hgamma}^{\beta\hdelta}w_{\beta}w_{\gamma}\hLambda^{\hgamma}\hw_{\hdelta}+F_{\alpha c}^{\beta\hdelta}w_{\beta}w_{\gamma}\hpsi^c\hw_{\hdelta}-G_{\alpha}^{\beta\hgamma\hdelta}w_{\beta}w_{\gamma}\hw_{\hgamma}\hw_{\hdelta}\Big)
\end{equation*}
\begin{equation*}
\begin{split}
    +\hC_P\,^{a\hbeta}\Big(&2\HH{U}_{ab}^{\gamma}d_{\gamma}\hpsi^a\hpsi^b+V_a^{\gamma\halpha}d_{\gamma}\hw_{\hbeta}\hw_{\hgamma}\\
    &-2H_{cdab}\psi^c\psi^d\hw_{\hbeta}\hpsi^b-2E_{\gamma ab}^{\delta}\Lambda^{\gamma}w_{\delta}\hw_{\hbeta}\hpsi^b+2\HH{I}_{cab}^{\delta}\psi^cw_{\delta}\hpsi^b\hw_{\hbeta}-2\HH{J}_{ab}^{\gamma\delta}w_{\gamma}w_{\delta}\hpsi^b\hw_{\hbeta}\\
    &-I_{cda}^{\halpha}\psi^c\psi^d\hw_{\halpha}\hw_{\hbeta}-F_{\gamma a}^{\delta\halpha}\Lambda^{\gamma}w_{\delta}\hw_{\halpha}\hw_{\hbeta}+K_{ca}^{\gamma\halpha}\psi^cw_{\gamma}\hw_{\halpha}\hw_{\hbeta}-\HH{L}_a^{\gamma\delta\halpha}w_{\gamma}w_{\delta}\hw_{\halpha}\hw_{\hbeta}\Big)
\end{split}
\end{equation*}
\begin{equation*}
    +\hC_{Pa}\,^{\halpha}\Big(-D_{\halpha}^{\gamma\hbeta}d_{\gamma}\hpsi^a\hw_{\hbeta}+\HH{E}_{cd\halpha}\,^{\hbeta}\psi^c\psi^d\hpsi^a\hw_{\hbeta}+S_{\gamma\halpha}^{\delta\hbeta}\Lambda^{\gamma}w_{\delta}\hpsi^a\hw_{\hbeta}-\HH{F}_{c\halpha}^{\delta\hbeta}\psi^cw_{\delta}\hpsi^a\hw_{\hbeta}+\HH{G}_{\halpha}^{\gamma\delta\hbeta}w_{\gamma}w_{\delta}\hpsi^a\hw_{\hbeta}\Big)
\end{equation*}
$$
+2\hY_P\,^{\halpha\hbeta}\Big(D_{\halpha}^{\gamma\hgamma}d_{\gamma}\hw_{\hbeta}\hw_{\hgamma}-\HH{E}_{cd\halpha}\,^{\hgamma}\psi^c\psi^d\hw_{\hbeta}\hw_{\hgamma}-S_{\gamma\halpha}^{\delta\hgamma}\Lambda^{\gamma}w_{\delta}\hw_{\hbeta}\hw_{\hgamma}+\HH{F}_{c\halpha}^{\delta\hgamma}\psi^cw_{\delta}\hw_{\hbeta}\hw_{\hgamma}-\HH{G}_{\halpha}^{\gamma\delta\hgamma}w_{\gamma}w_{\delta}\hw_{\hbeta}\hw_{\hgamma}\Big)=0
$$

In order to find $\B{\partial}d_{\alpha}$ and $\partial\hd_{\halpha}$ from this huge equation, we simply invert the vielbein $E_P\,^{\alpha}$ or $E_P\,^{\halpha}$ by contracting $E_{\rho}\,^P$ or $E_{\hrho}\,^P$ respectively. Finally, in order to find $\B{\partial}d_a$ we need the extra step that is to write
\begin{equation}
    \frac{\B{\partial}E_a+\partial\B{E}_a}{2}=\B{\partial}d_a-\partial_{1}\B{E}_a.
\end{equation}
Then, we should again invert the vielbein $E_P\,^{a}$ with $E_c\,^P$.

\section{$\mathfrak{psu}(2,2|4)$ conventions}\label{app3}
Part of the commutators of the $\mathfrak{psu}(2,2|4)$ superalgebra with $P^{\a\bb}=-\frac12\eta^{\a\bb}$ is given by:
\begin{align}
    &\{T_\a,T_\b\}=\g^{\ua}_{\a\b} T_{\ua},\quad \{T_{\ab},T_{\bb}\}=\g^{\ua}_{\ab\bb} T_{\ua},\cr 
    &[T_{\ua},T_{\bb}]=-\frac12(\eta\g_{\ua})^\a{}_{\bb}T_{\alpha},\cr
    &[T_{\ua},T_\a]=\frac12(\g_{\ua}\eta)_\a{}^{\bb} T_{\bb},\cr
    &[T_a,T_b]=-T_{ab},\quad [T_{a'},T_{b'}]=+T_{a'b'},\cr
    &\{T_\a,T_{\bb}\}=\frac12(\g^{ab}\eta)_{\a\bb} T_{ab} - \frac12 (\g^{a'b'}\eta)_{\a\bb} T_{a'b'} .
\end{align}

We also make use of the non-vanishing supertraces
\begin{equation}
    \mathrm{Str}\big(T_{\ua}T_{\ub}\big)=\eta_{\ua\ub},\quad \mathrm{Str}\big(T_{\alpha}T_{\hbeta}\big)=-2\eta_{\alpha\hbeta},
\end{equation}
\begin{equation}
    \mathrm{Str}(T_{ab}T_{cd})=-\eta_{a[c}\eta_{d]b},\quad \mathrm{Str}\big(T_{a'b'}T_{c'd'}\big)=\eta_{a'[c'}\eta_{d']b'}.
\end{equation}
\end{appendices}

\bibliography{refs}
\nocite{*}

\end{document}